\UseRawInputEncoding
\documentclass[prx,superscriptaddress,amsmath,amssymb,twocolumn,nofootinbib]{revtex4-2}
\usepackage{graphicx,bm,amsmath}
\usepackage[usenames,dvipsnames]{xcolor}
\usepackage[colorlinks,bookmarks=false,citecolor=NavyBlue,linkcolor=Red,urlcolor=blue,
]{hyperref}
\usepackage[percent]{overpic}
\usepackage[bbgreekl]{mathbbol}
\usepackage{xcolor}
\usepackage[normalem]{ulem}
\usepackage{algorithm}
\usepackage{braket}
\usepackage{algpseudocode}
\usepackage{comment}
\usepackage{amsthm}
\usepackage{enumerate}
\usepackage{booktabs}
\usepackage{array}  
\usepackage{lipsum}
\def\doi{http://dx.doi.org/}
\newcommand{\be}{\begin{equation}}
\newcommand{\ee}{\end{equation}}
\newcommand{\bec}{\begin{equation*}}
\newcommand{\eec}{\end{equation*}}
\newcommand{\bea}{\begin{eqnarray}}
\newcommand{\eea}{\end{eqnarray}}

\newcommand{\Tr}{\text{Tr}}   

\newcommand{\titleinfo}{Exploring the Limit of the Lattice-Bisognano-Wichmann Form Describing the Entanglement Hamiltonian: A Quantum Monte Carlo Study}
\begin{document}
\title{\titleinfo}

\author{Siyi Yang}
\affiliation{State Key Laboratory of Surface Physics and Department of Physics, Fudan University, Shanghai 200438, China}
\affiliation{Department of Physics, School of Science and Research Center for Industries of the Future, Westlake University, Hangzhou 310030,  China}
\affiliation{Institute of Natural Sciences, Westlake Institute for Advanced Study, Hangzhou 310024, China}

\author{Yi-Ming Ding}
\email{dingyiming@westlake.edu.cn}
\affiliation{State Key Laboratory of Surface Physics and Department of Physics, Fudan University, Shanghai 200438, China}
\affiliation{Department of Physics, School of Science and Research Center for Industries of the Future, Westlake University, Hangzhou 310030,  China}
\affiliation{Institute of Natural Sciences, Westlake Institute for Advanced Study, Hangzhou 310024, China}

\author{Zheng Yan}
\email{zhengyan@westlake.edu.cn}
\affiliation{Department of Physics, School of Science and Research Center for Industries of the Future, Westlake University, Hangzhou 310030,  China}
\affiliation{Institute of Natural Sciences, Westlake Institute for Advanced Study, Hangzhou 310024, China}

\begin{abstract}
As a powerful theoretical construct, the entanglement Hamiltonian (EH) encapsulates the essential entanglement properties of a quantum many-body system. From the EH, one can extract a variety of entanglement quantities, such as entanglement entropies, negativity, and the entanglement spectrum. However, its general analytical form remains largely unknown.
While the Bisognano-Wichmann theorem gives an exact EH form for Lorentz-invariant field theories, its validity on lattice systems is limited, especially when Lorentz invariance is absent.
In this work, we propose a general scheme based on the lattice-Bisognano-Wichmann (LBW) ansatz and multi-replica-trick quantum Monte Carlo methods to numerically reconstruct the entanglement Hamiltonian in two-dimensional systems and systematically explore its applicability to systems without translational invariance, going beyond the original scope of the primordial Bisognano-Wichmann theorem. Various quantum phases--including gapped and gapless phases, critical points, and phases with either discrete or continuous symmetry breaking--are investigated, demonstrating the versatility of our method in reconstructing entanglement Hamiltonians. Furthermore, we find that when the entanglement boundary of a system is ordinary (i.e., free from surface anomalies), the LBW ansatz provides an accurate approximation well beyond Lorentz-invariant cases. Our work thus establishes a general framework for investigating the analytical structure of entanglement in the complex quantum many-body systems.
\end{abstract}

\maketitle

\section{Introduction}
Entanglement stands as arguably the most fundamentally non-classical feature of quantum systems. It is universally recognized as an indispensable tool for diagnosing and classifying quantum phases of matter. 
The entanglement properties of a pure state $\ket{\psi}$ of a bipartite system $A \cup B$ are encoded in its Schmidt decomposition, $\ket{\psi} = \sum_i \sqrt{\lambda_i} \ket{i_A} \ket{i_B}$, where $\{\lambda_i\}$ are the (squared) Schmidt coefficients. 
Based on the decomposition, the entanglement entropy (EE) is defined as $S=-\Tr_A(\rho_A\ln\rho_A)=-\sum_i\lambda_i\ln\lambda_i$, where $\rho_A = \Tr_B(\ket{\psi}\bra{\psi}) = \sum_i \lambda_i \ket{i_A}\bra{i_A}$ is the reduced density matrix of subsystem $A$~\cite{nielsen2010quantum,PhysRevA.53.2046,plenio2005introduction}.  
Crucially, the scaling of EE serves as a powerful diagnostic for many-body phenomena, including quantum criticality, topological order, and conformal field theories~\cite{PhysRevLett.90.227902, PhysRevLett.96.110404,PhysRevLett.96.110405,RevModPhys.80.517,PasqualeCalabrese_2004,Calabrese_2009,Latorre_2009,RevModPhys.82.277,wang2024probing,wang2025bipartite,ding2025tracking,jiang2024high,zhu2025bipartite}, revealing fundamental aspects of their structure and correlations.

While EE provides a powerful quantification of entanglement, the notion of the entanglement spectrum (ES) was introduced to retain the full distribution of the Schmidt coefficients, offering an alternative and more detailed description of the entanglement structure~\cite{Li2008entangle,XLQi2012,Mao2025,Yan2023,liu2023probing,mao2025detecting,lauchli2012entanglement,lauchli2013operator,LAFLORENCIE20161}. 
By defining the entanglement Hamiltonian (EH) $H_A$ through 
\begin{equation}\label{eq:rho_A}
  \rho_A:= e^{-H_A},
\end{equation}
where we require $\Tr(\rho_A)=1$, the ES of $\rho_A$ is exactly the energy spectrum of $H_A$~\cite{Li2008entangle,RevModPhys.90.045003,Peschel_2011,dalmonte2022entanglement}.
From this perspective, EE is exactly the thermal entropy of an effective canonical system described by $H_A$ at the EH effective inverse temperature $\beta_A=1$~\cite{wang2025sudden,Eisler_2006, PhysRevLett.121.200602, Mendes-Santos_2020}. 
A key application of the ES lies in its power to characterize topological phases~\cite{PhysRevB.81.064439,PhysRevX.6.041033,PhysRevLett.105.080501,PhysRevB.89.195147,PhysRevB.83.245134,PhysRevLett.104.130502,PhysRevB.91.125146,jiang2025identifying}.
Li and Haldane first conjectured that the low-lying ES of the $\nu=5/2$ fractional quantum Hall state closely mirrors the corresponding edge energy spectrum~\cite{Li2008entangle}.
This remarkable connection was soon extended to quantum spin systems~\cite{Didier2010es}, and more generally, a broad correspondence has been established between the ES of (2+1)D gapped topological phases and the spectrum of their (1+1)D edges, particularly when the edge is governed by conformal field theory (CFT)~\cite{XLQi2012}.

This deep connection between EH and quantum many-body physics has motivated extensive studies aiming to derive the explicit functional form of the EH. 
However, this is generally a challenging task.
For lattice systems, exact results are limited to a few special cases, including the EH of the Ising~\cite{https://doi.org/10.1002/andp.19995110203, Peschel_2009} and XYZ chains~\cite{PhysRevLett.58.1395} away from criticality, certain one-dimensional free fermion systems~\cite{Iglói_2010, Eisler_2017,Eisler_2018}, and a handful of other non-generic models. Nevertheless, there is still no general access to obtain an analytic form of EH in a quantum many-body system.

In this work, we build on recently developed multi-replica quantum Monte Carlo (QMC) techniques to propose a scheme for numerically approximating and verifying the functional ansatzes of the EH.
Our focus is on the Bisognano-Wichmann (BW) theorem~\cite{10.1063/1.522605,10.1063/1.522898}, whose field-theoretical insights suggest a specific EH structure for various Lorentz-invariant models~\cite{PhysRevB.86.045117,Dalmonte2018}.
Its lattice counterpart, the so-called lattice-Bisognano-Wichmann (LBW) form, has been numerically demonstrated to provide a good approximation and ansatz for certain translationally invariant lattice systems~\cite{giudici2018bw, PhysRevB.100.155122}.
However, testing and applying the LBW ansatz typically require prior knowledge of the sound velocity (dispersion slope), which is often unavailable, particularly in higher-dimensional settings.
Furthermore, for lattice systems without Lorentz invariance, it remains unclear whether the LBW ansatz of the EH continues to capture even qualitative features of the exact EH. These challenges motivate the present study.

Using the multi-replica trick~\cite{Yan2023}, we simulate the ensemble of EH without requiring prior knowledge of its functional form at various integer inverse temperatures.
This allows us to verify candidate functional forms. By computing related imaginary-time correlations of EH and comparing them with those predicted by the LBW ansatz, we determine the unknown parameter of the functional form and further evaluate the accuracy of the ansatz. We apply our method to the two-dimensional transverse-field Ising model with translational symmetry and the two-dimensional columnar dimerized Heisenberg model without translational symmetry. Our investigation of the LBW ansatz covers not only critical points but also both gapped and gapless phases, providing new insights into the structure of entanglement in these regimes and demonstrating the power of our method as a general tool for studying EH in a broad class of many-body lattice models.

Importantly, we find that the LBW ansatz holds well once the (entanglement) edge of the system is ordinary (i.e., without anomaly), and the presence of Lorentz invariance does not seem to be a necessary condition. It potentially reveals the uncovered deep-correspondence between the research areas of many-body entanglement \cite{RevModPhys.80.517} and surface criticality \cite{binder1990critical}. This discovery actually can also explain the contradictions in the entanglement entropy behaviors recently observed due to the different entanglement splitting schemes~\cite{deng2024diagnosing,d2024entanglement,song2023extracting}. The broadened applicability of the LBW approximation thus opens a powerful new pathway for investigating the entanglement properties of complex many-body systems.

This paper is organized as follows. Sec.~\ref{sec:lbw} provides a brief review of the LBW ansatz. In Sec.~\ref{sec:qmc}, we describe our main methodology for simulating the EH at various integer inverse temperatures, evaluating imaginary-time correlations, and fitting the prefactor used in the LBW form. Sec.~\ref{sec:tfim} presents the results for the two-dimensional transverse-field Ising model, which is translationally invariant, while Sec.~\ref{sec:dhm} discusses the two-dimensional dimerized Heisenberg model, where translational symmetry is explicitly broken. Finally, Sec.~\ref{sec:con} summarizes our conclusions and provides further discussions.


\section{Lattice-Bisognano-Wichmann Entanglement Hamiltonian}\label{sec:lbw}

Consider a ($D+1$)-dimensional relativistic quantum field theory with Hamiltonian density $H(\mathbf{x})$, where $\mathbf{x}=(x_1,x_2,\cdots,x_D)$ is the spatial coordinate and the system has a Lorentz-invariant symmetry. 
According to the Bisognano-Wichmann (BW) theorem~\cite{10.1063/1.522605,10.1063/1.522898,Hislop1982,Brunetti1993}, the entanglement Hamiltonian (or modular Hamiltonian) of its semi-infinite subsystem $A$ ($x_1>0$) under the half-space bipartition is given by:
\begin{equation}\label{eq:sf}
    \tilde{H}_{A} = \frac{2\pi}{c} \int_{\mathbf{x}\in A} d\mathbf{x}\  x_1H(\mathbf{x}),
\end{equation}
where $c$ is the speed of light. 
In this case, the reduced density matrix $\rho_A\propto e^{-\tilde H_A}$ can be viewed as a Gibbs state with space-dependent temperatures. 
Specifically, close to the entangling boundary, the temperature is high, thus dominating the system described by $\tilde H_A$, which is directly connected to the area-law behavior of quantum entanglement for ground states~\cite{RevModPhys.82.277}.
Moreover, by considering the conformal symmetry, the BW theorem can be further generalized to other geometries~\cite{Casini2011, Cardy_2016, Najafi2016,PhysRevD.96.105019}.

In the context of lattice models, the BW theorem can be adapted to provide an ansatz for the EH, known as the lattice-Bisognano-Wichmann entanglement Hamiltonian (LBW-EH), which has been shown to be extremely accurate both numerically and experimentally in many scenarios~\cite{Dalmonte2018,giudici2018bw,PhysRevB.100.155122,Mahdieh2021lbw}.
Specifically, for a one- or two-dimensional lattice model with coupling and on-site terms described by Hamiltonian
\begin{equation}\label{eq:LBW-EH analytice form}
    \begin{split}
    H
    =\Gamma \sum_{x,y,\delta}\bigg[h_{(x,y),(x+\delta,y)} + h_{(x,y),(x,y+\delta)}\bigg]
    +\Theta \sum_{x,y}l_{(x,y)} 
    \end{split},
\end{equation}
where $\delta$ represents the unit vector in the direction of the nearest-neighbor lattice point.
The term $h_{(x,y),(x+\delta,y)}$ denotes the interaction between two nearest-neighbor sites in the horizontal direction, while $h_{(x,y),(x,y+\delta)}$ denotes that in the vertical direction. 
The parameter $\Gamma$ represents the coupling strength.
The term $l_{(x,y)}$ describes an on-site operator at a single lattice site $(x, y)$, with $\Theta$ governing the strength of the transverse or longitudinal field. 
By recasting the BW theorem Eq.~\eqref{eq:sf} on the lattice~\cite{giudici2018bw}, we achieve the LBW-EH, which is 
\begin{equation}\label{eq:LBW-EH}
    \begin{split}
    \tilde{H}_{A}
    =\epsilon_{EH} \bigg\{ & \sum_{x,y,\delta}\bigg[\Gamma_x h_{(x,y),(x+\delta,y)} 
    + \Gamma_y h_{(x,y),(x,y+\delta)}\bigg] \\
    +&\Theta_{x,y}\sum_{x,y}l_{(x,y)} \bigg\},
    \end{split}
\end{equation}
where $\Gamma_x$ and $\Gamma_y$ are the coupling in $x$ and $y$ direction, and $\Theta_{x,y}$ is the on-site term.  
These terms are associated with the distance from the entangling boundary that separates two half-space bipartite subsystems $A$ and $B\equiv \overline{A}$ (environment). 

\begin{figure}[ht!]
\centering
\includegraphics[width=.95\linewidth]{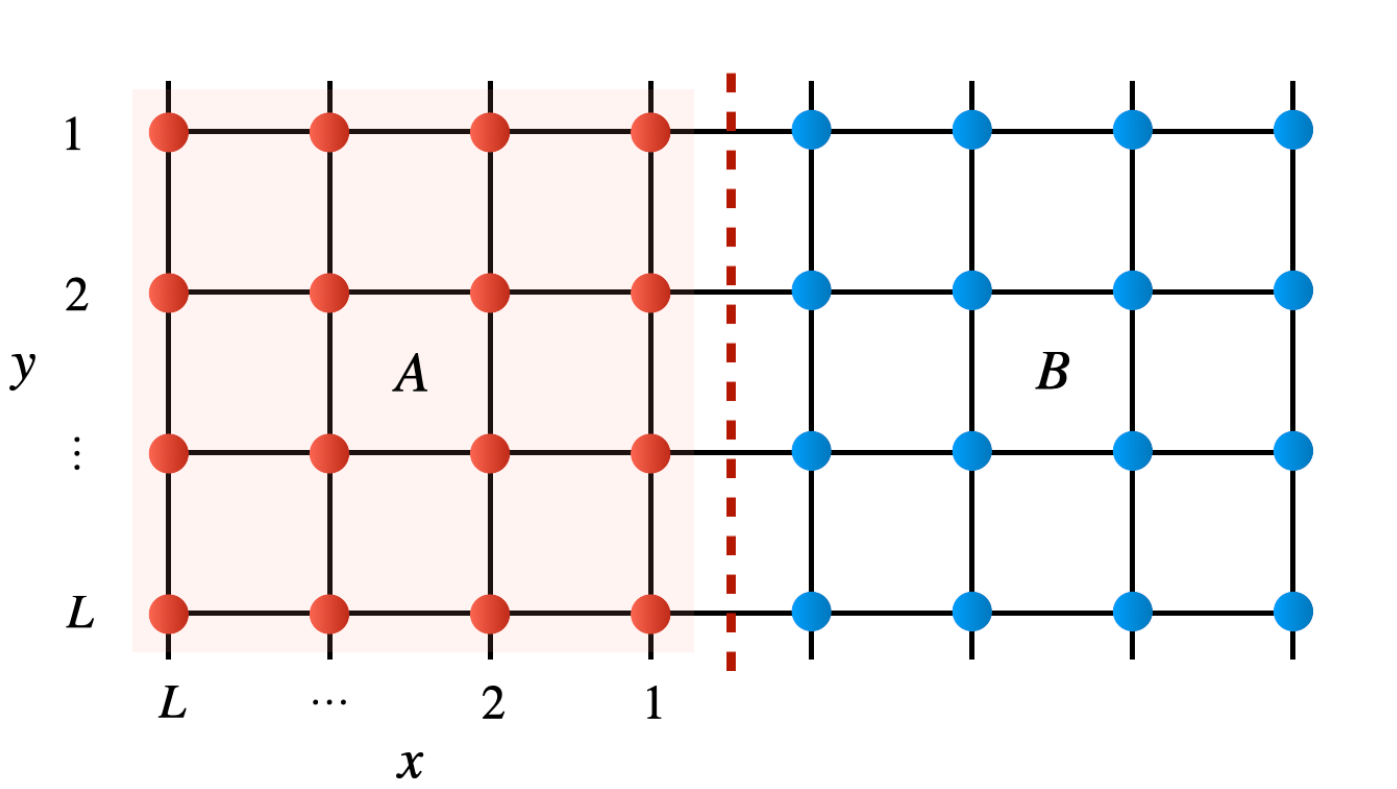}
\caption{A two-dimensional lattice system with cylinder geometry. The half-space bipartite subsystems $A$ and $B$ both have dimensions of $L \times L$. The LBW-EH depends on the distance from the lattice sites and bonds to the boundary that separates the subsystems. The distance from a lattice site to the boundary ranges over $[1/2, L - 1/2]$. The same holds for vertical bonds. Notably, the distance from a horizontal bond to the boundary ranges over $[1,L-1]$, defined as the distance from the center of the horizontal bond to the boundary.} 
\label{fig: BW_config}
\end{figure}

In this work, we focus on two-dimensional systems by considering the cylinder geometry, characterized by open boundary conditions (OBC) along the $x$-axis and periodic boundary conditions (PBC) along the $y$-axis, as illustrated in Fig.~\ref{fig: BW_config}.
In this case, the coupling constants in Eq.~\eqref{eq:LBW-EH} are 
\begin{equation}\label{eq:terms}
    \Gamma_x=x\Gamma,\quad 
    \Gamma_y=\left(x-\frac{1}{2}\right)\Gamma,\quad 
    \Theta_{x,y} = \left(x-\frac{1}{2} \right) \Theta ,
\end{equation}
where $x$ takes values from $1$ to $L$. The parameter $\Gamma_x$ corresponds to the horizontal bonds while $\Gamma_y$ is associated with the vertical bonds. The term $\Theta_{x,y}$ is related to the lattice sites.


In the formulation of LBW-EH, there is a key parameter $\epsilon_{\mathrm{EH}}$, which plays the role of the effective energy scale. Within the framework of low-energy field theory, its explicit expression is given by
\begin{equation}\label{eq:beta_EH}
\epsilon_{\text{EH}} = \frac{2\pi}{v},
\end{equation}
where $v$ is the sound velocity. 
As indicated in this expression, the determination of $\epsilon_{\mathrm{EH}}$ requires prior knowledge of the sound velocity $v$, which is not generally accessible. 

Previous studies suggest that the LBW-EH provides an accurate description of the lattice EH, as long as the low-energy description of the lattice model is well captured by Lorentz-invariant quantum field theory~\cite{Dalmonte2018,giudici2018bw,PhysRevB.100.155122}. 
In this work, we aim to move beyond this paradigm, and further investigate the applicability of the LBW-EH in more scenarios, especially when the system has no translational symmetry  (Sec.~\ref{sec:dhm}).
This will not only extend the scope of applicability of the LBW-EH, but also facilitate future studies to directly use LBW-EH to explore entanglement properties in a broader range of models.


\section{Replicated reduced density matrix and the imaginary-time correlations}\label{sec:qmc}
In this section, we present a universal scheme for studying the LBW-EH within multi-replica-trick QMC methods. Although our focus is on the LBW-EH, the approach is general and can be applied to other ansatzes of the EH. 

Given a functional ansatz of the EH, the reduced density matrix can be regarded as a Gibbs state with respect to this EH. Hence, the ansatz EH can be simulated at various effective inverse temperatures $\beta_A$ using conventional finite-temperature QMC methods.
To assess the validity of the ansatz, we compare physical observables obtained from the ansatz EH [Sec.~\ref{sec:LBW-EH}] with those derived from the exact simulation of the resemble of EH [Sec.~\ref{sec:exact-EH}]. We emphasize that although the exact EH can be simulated via multi-replica-trick QMC methods~\cite{Yan2023,PhysRevB.109.195169}, its analytical form remains unknown and the multi-replica method can only visit the integer effective inverse temperatures $\beta_A$.  

On the other hand, the multi-replica-trick QMC method allows simulating the resemble of the exact EH even though only at integer effective inverse temperatures $\beta_A=n$ ($n=1,2,3,\ldots$)~\cite{Yan2023}, which enables a systematic comparison of observables between the ansatz and the exact EH across different effective inverse temperatures $\beta_A$. Moreover, it provides a way to fit the unknown parameter $\epsilon_{\mathrm{EH}}$ in the LBW ansatz by measuring imaginary-time correlations, as discussed in Sec.~\ref{sec:fit}.

\subsection{LBW entanglement Hamiltonian}\label{sec:LBW-EH}
The approximated reduced density matrix $\tilde \rho_A$, constructed from the LBW entanglement Hamiltonian $\tilde H_A$ given in Eq. \eqref{eq:LBW-EH}, is defined as $\tilde{\rho}_{A} = {e^{-\tilde{H}_A}}/{\tilde{Z}_{A}}$,
where the partition function $\tilde{Z}_{A} = \Tr_A(e^{-\tilde{H}_{A}})$ ensures the normalization of $\tilde{\rho}_{A}$~\footnote{
    Throughout this work, tildes are used to denote LBW approximations of operators obtained from QMC, in contrast to their exact counterparts.
}. 
A general physical observables $ {O}_{A}$ measured in subsystem $A$ under the effective inverse temperatures $\beta_A$ can thus be expressed as 
\begin{equation}\label{eq:BW_O}
 \langle \tilde{O}_{A} \rangle_{\beta_A} 
 \equiv \frac{\Tr_A(e^{-\beta_A\tilde{H}_{A}} {O}_{A})}{\tilde{Z}_{A}}, 
\end{equation}
where the effective inverse temperature ${\beta}_A$ of the LBW-EH is taken to be $1$ in the original definition Eq. \eqref{eq:rho_A}. 
Certainly, the effective inverse temperature ${\beta}_A$ can also take other values as $\tilde\rho_A$ is treated as a Gibbs state with respect to the $\tilde H_A$.
In the limit where the effective inverse temperature $\beta_A$ tends to infinity, the system described by $\tilde H_A$ approaches its ground state. Through measurements of physical observables in this regime, the ground-state properties of the LBW-EH can be extracted.

Note that, compared with the original Hamiltonian of the system, the LBW-EH only modifies the coupling constants, while preserving the interaction structure (see Eq.\eqref{eq:LBW-EH analytice form} and \eqref{eq:LBW-EH}). Therefore, the LBW-EH can be simulated using standard finite-temperature QMC methods, such as the stochastic series expansion (SSE) technique~\cite{PhysRevB.43.5950,PhysRevE.66.046701,PhysRevB.57.10287,PhysRevB.59.R14157,10.1063/1.3518900,PhysRevB.99.165135,PhysRevB.105.184432}, without encountering the sign problem, provided that the original Hamiltonian is free from it. 




\subsection{Exact simulation of the entanglement Hamiltonian}\label{sec:exact-EH}
To verify the reliability of the LBW-EH, a direct comparison with the exact-EH is required. 
Recall that for the exact-EH $H_A$ defined by the reduced density matrix ${\rho}_{A} = e^{-H_A}/{Z}_{A}$, where the effective inverse temperature is set to $\beta_A = 1$, and $Z_A=\Tr_A(\rho_A)$. To investigate the properties of the exact-EH, we need to also simulate it at various effective inverse temperatures $\beta_A$. For the LBW-EH, this is straightforward, as discussed in Sec.~\ref{sec:LBW-EH}. However, for the exact-EH, its analytical form is unknown, making direct simulations impossible.
In this section, we introduce how to simulate the exact-EH using the replica-trick QMC method~\cite{Yan2023, PhysRevB.109.195169}, which can simulate $H_A$ at integer effective inverse temperatures $\beta_A=n$ ($n=1,2,3,\ldots$).



If the effective inverse temperature $\beta_A=1$, the simulation of $\Tr\rho_A$ is exactly the same as simulating the ground state of the original Hamiltonian $H$ if we only consider the degree of freedom in the $A$, since 
\begin{equation}\label{eq:relation}
  \begin{split}
    \Tr(e^{-\beta H})& \propto  \Tr(\rho)\\& =\Tr_A[\Tr_B(\rho)] = \Tr_A(\rho_A)\propto \Tr_A(e^{-H_A}),
  \end{split}
\end{equation}
where $\beta$ is the real inverse temperature of the original system (be careful for the physical inverse temperature $\beta$ and the effective inverse temperature $\beta_A$). 
To make the original system approach its ground state, the physical inverse temperature $\beta$ must sufficiently large in practical simulations.
For a physical observable $O_{A}$ defined on subsystem $A$, its expectation value can be expressed as
\begin{equation}\label{eq:exact_1}
    \langle O_{A} \rangle_{\beta_A=1} = \frac{ \Tr_A[\text{Tr}_B (e^{-\beta H} )O_{A}] } {Z},
\end{equation}
where $Z=\Tr(e^{-\beta H})$.
From this expression, it follows that during the simulation, one must first trace over the environmental degrees of freedom $B$ first, and then perform measurements of physical observables in subsystem $A$. 

\begin{figure}[ht!]
\centering
\includegraphics[width=.6\linewidth]{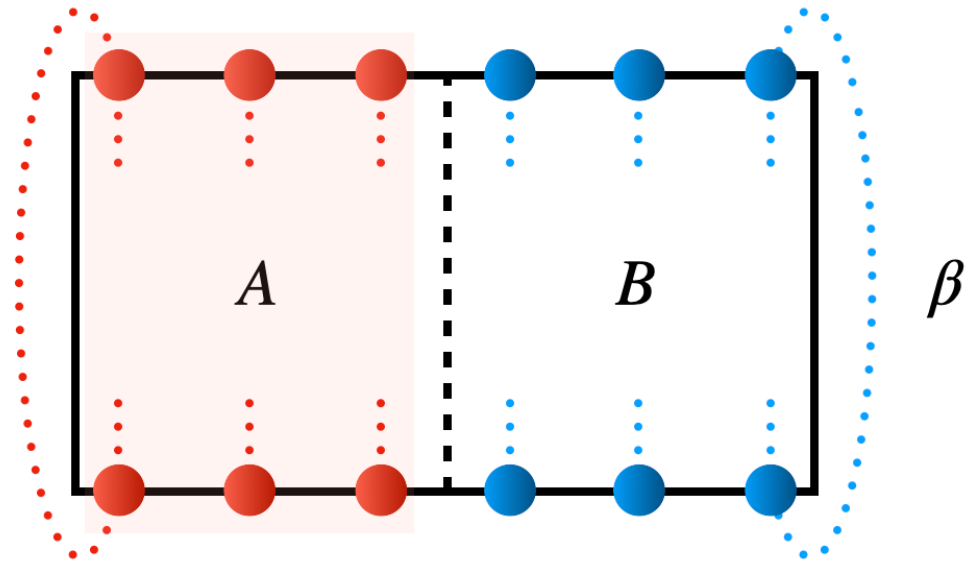}
\caption{
  The path integral representation of exact-EH ${H}_A$ with effective inverse temperature $\beta_A = 1$. The horizontal axis represents the space, and the vertical axis represents the imaginary-time. The original Hamiltonian $H$ with the physical inverse temperature $\beta$ is simulated. Both subsystem $A$ and the environment $B$ are subject to periodic boundary conditions in imaginary-time.} 
\label{fig: QMC_Rep}
\end{figure}

Fig.~\ref{fig: QMC_Rep} illustrates the path integral representation of the ensemble of exact-EH with effective inverse temperature $\beta_A=1$.  
In the path integral representation, the state of the system evolves along the vertical temporal direction. 
For the environment $B$ part, the $\Tr_B$ operation requires that the path must return to its initial state after the imaginary-time $\beta$, thus leading to periodic boundary conditions of $B$ for the imaginary-time direction and remaining $\rho_A\propto \Tr_B(e^{-\beta H})|_{\beta\rightarrow \infty}$.
Similarly, to obtain $\Tr_A(\rho_A)$, we need to trace over the subsystem $A$, and the imaginary-time boundary of subsystem $A$ also satisfies periodic boundary conditions. This corresponds to Eq. \eqref{eq:relation}, which is $\Tr(e^{-\beta H})\propto \Tr_A(e^{-H_A})$.


\begin{figure}[ht!]
\centering
\includegraphics[width=.7\linewidth]{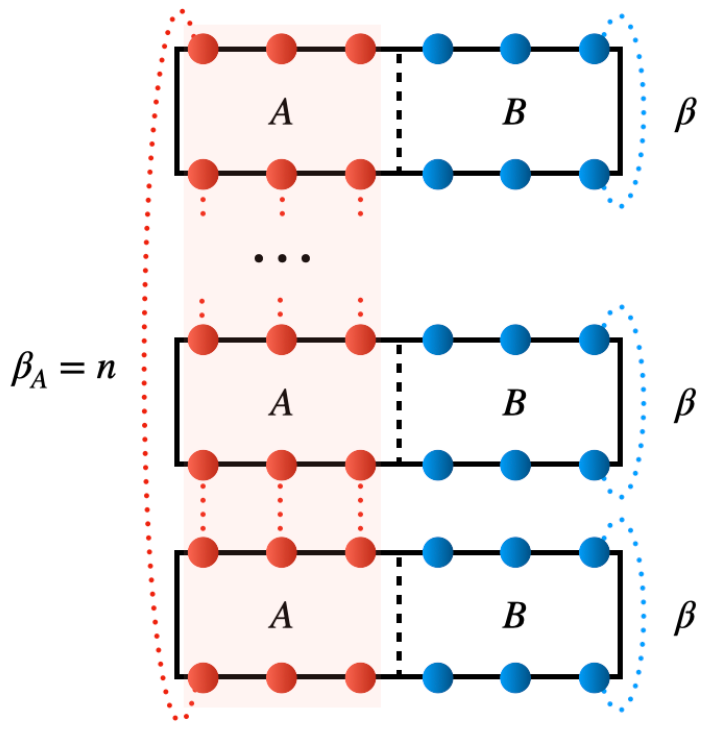}
\caption{The path integral representation of exact-EH with imaginary-time $n$. 
The effective inverse temperature $\beta_A$ equals the number of replicas $n$.
The horizontal axis represents the real-space configuration, while the vertical axis corresponds to imaginary-time.
Each replica is partitioned into subsystem $A$ and environment $B$.
For subsystem $A$, all replicas are interconnected with periodic boundary conditions applied solely between the first and last replica, whereas for environment $B$, each individual replica must independently satisfy periodic boundary conditions.}
\label{fig: QMC_nReps}
\end{figure}

Similarly, if the effective inverse temperature $\beta_A=n>1$, we can extend Eq.~\eqref{eq:relation} to 
\begin{equation}
  \begin{split}
    \Tr(\rho_A^n)&=\Tr_A\{[\Tr_B (\rho)]^n\} \\ &\propto  \Tr_A\{[\Tr_B (e^{-\beta H})]^n\} \propto \Tr_A(e^{-n H_A}),
  \end{split}
\end{equation}
and correspondingly, the expectation value of a physical observable $O_{A}$ defined on subsystem $A$ at $\beta_A=n$ with respect to the exact-EH can be expressed as
\begin{equation}\label{eq:exact_n}
    \langle O_{A} \rangle_{\beta_A=n} = \frac{ \Tr_A[(\text{Tr}_B e^{-\beta H} )^n O_{A}] } {Z_A^{(n)}},
\end{equation}
where $Z_A^{(n)}=\Tr_A[(\text{Tr}_B e^{-\beta H} )^n] $ is a normalization factor.
In Fig.~\ref{fig: QMC_nReps}, we illustrate the path integral representation of the ensemble of exact-EH $Z_A^{(n)}$ with effective inverse temperature $\beta_A=n$, $n$ is an integer since QMC can only simulate integer replicas. Similar to the case of $\beta_A=1$, as we have $n$ replicas of the state, we trace over the environment $B$ for each replica first, and then trace over the subsystem $A$ for the total $n$ replicas of subsystem $A$. Therefore, the length of the total imaginary-time is $n\beta$, where $\beta\to\infty$, which corresponds to $\beta_A=n$ in the ensemble of exact-EH~\cite{PhysRevB.108.075114,wu2023classical,liu2023probing,Mao2025}.
Though it is difficult to generalize this method to non-integer $\beta_A$, Eq.~\eqref{eq:exact_n} still provides a systematic way to study the exact-EH at various integer effective inverse temperatures. This allows us not only to study the finite-temperature properties of the exact-EH~\cite{PhysRevB.109.195169}, but also to compare the physical observables between the LBW-EH and the exact-EH at various effective inverse temperatures.

\subsection{Fitting of the BW  energy scale}\label{sec:fit}

We have established the methodology for simulating both the LBW-EH and the exact-EH using QMC methods, along with techniques for measuring physical observables.
However, although the ansatz of the LBW-EH is given, it contains an unknown parameter $\epsilon_{EH}$ that needs to be determined.
In this section, we introduce a method to fit this parameter by comparing imaginary-time correlations between the LBW-EH and the exact-EH.

Before discussing the EH, we first consider a general Gibbs state $\rho\propto e^{-\beta H}$ for some Hamiltonian $H$.
For the two-dimensional cylinder geometry, we measure the imaginary-time correlation function on a boundary-adjacent line inside subsystem $A$.
Writing lattice sites as $\mathbf{i}=(i_x,i_y)$ and $\mathbf{j}=(j_x,j_y)$, we fix $\mathbf{i}=(1,1)$ and choose $\mathbf{j}=(1,j_y)$ with $j_y=1,2,\dots,L$.
We then define the momentum-resolved imaginary-time correlation function along the periodic boundary direction as
\begin{equation}
C_k(\tau)= \frac{1}{L}\sum_{j_y=1}^{L} C\!\big((1,1),0;(1,j_y),\tau\big)\,e^{-ik(j_y-1)},
\label{eq:Ck_def}
\end{equation}
where
\begin{equation}
C\!\big((1,1),0;(1,j_y),\tau\big)
=
\left\langle \sigma^z_{(1,j_y)}(\tau)\sigma^z_{(1,1)}(0)\right\rangle,
\end{equation}
and $k=2\pi n/L$ with $n=0,1,\dots,L-1$.

For convenience, we define 
\begin{equation}
O_k=\frac{1}{L}\sum_{j_y=1}^{L} e^{-ik(j_y-1)} \sigma^z_{(1,j_y)},
\end{equation}
so that
\begin{equation}
C_k(\tau)=\langle O_k(\tau)\,\sigma^z_{(1,1)}(0)\rangle,
\end{equation}
where $O_k(\tau)=e^{\tau H}O_k e^{-\tau H}$.
Under the eigenbasis of the Hamiltonian, $H|m\rangle=E_m|m\rangle$, we have
\begin{equation}
\begin{split}
C_k(\tau)
=& \frac{1}{Z}\sum_{m}\langle m|e^{-\beta H}e^{\tau H}O_k e^{-\tau H}\sigma^z_{(1,1)}|m\rangle \\
=& \frac{1}{Z}\sum_{mn} \langle m|e^{-\beta H}e^{\tau H}O_k e^{-\tau H}|n\rangle
\langle n|\sigma^z_{(1,1)}|m\rangle \\
=& \frac{1}{Z}\sum_{mn} e^{-\beta E_m} e^{-\tau(E_n-E_m)}
\langle m|O_k|n\rangle \langle n|\sigma^z_{(1,1)}|m\rangle ,
\end{split}
\end{equation}
where $E_m$ and $E_n$ are the eigenvalues corresponding to the eigenstates $|m\rangle$ and $|n\rangle$ of $H$, respectively.
For convenience, we write $A^{(k)}_{mn}=\langle m|O_k|n\rangle \langle n|\sigma^z_{(1,1)}|m\rangle$, thus
\begin{equation}
\begin{split}
C_k(\tau)
=& \frac{1}{Z}\sum_{mn} e^{-\beta E_m} e^{-\tau(E_n-E_m)} A^{(k)}_{mn} \\
=& \frac{1}{Z} e^{-\beta E_0}\sum_{mn} e^{-\beta(E_m-E_0)} e^{-\tau(E_n-E_m)} A^{(k)}_{mn},
\end{split}
\end{equation}
where $E_0$ is the ground-state energy of $H$.
When $\beta\to\infty$ and $\tau$ is finite, the terms with $m\neq 0$ can be ignored, the system
approaches its ground state, and we have
\begin{equation}
C_k(\tau)=\sum_n e^{-\tau(E_n-E_0)} A^{(k)}_{0n}.
\end{equation}

If the equal-time constant contribution $A^{(k)}_{00}\neq 0$, it can be subtracted from the correlation function, i.e., by redefining $C_k(\tau)\rightarrow C_k(\tau)-A^{(k)}_{00}$.
Therefore, without loss of generality, we assume $A^{(k)}_{00}=0$ in the following discussion.
In this case, when $\tau$ is sufficiently large, the summation is dominated by the first nonzero term, thus
\begin{equation}
C_k(\tau)\sim A^{(k)}_{0n_k^*} e^{-\tau(E_{n_k^*}-E_0)},
\end{equation}
where $n_k^*$ denotes the index of the first nonzero term in the summation, typically corresponding to the lowest-energy state with nonzero spectral weight in the selected momentum channel $k$.
Taking the logarithm of both sides yields
\begin{equation}
\log \left[C_k(\tau)\right] \sim -(E_{n_k^*}-E_0)\tau,
\label{eq:logCk_linear}
\end{equation}
which exhibits linear dependence on $\tau$.

By replacing $H$ with the exact-EH $H_A$ and the LBW-EH $\tilde{H}_A$, we can similarly define the corresponding momentum-resolved imaginary-time correlation functions $C^{H_A}_{k}(\tau)$ and $C^{\tilde{H}_A}_{k}(\tau)$, respectively.
For 
\begin{equation}\label{eq:fitting}
\frac{\log \left[ C_{H_{A},k}(\tau)\right]}
{\log \left[ C_{\tilde{H}_{A},k}(\tau)\right]}
=
\frac{E_{n_k^*}-E_{0}}{\tilde{E}_{n_k^*}-\tilde{E}_{0}},
\end{equation}
if the LBW-EH provides a complete description of the exact-EH and the correct overall prefactor is used, then the slope ratio given by the right-hand side of Eq.~\eqref{eq:fitting} should be unity.

In practice, the overall prefactor is not known a priori, but we can simulate LBW-EH with a fixed convention for this prefactor, and then determine the missing rescaling factor by comparing the large-$\tau$ behavior of the two logarithmic correlation functions, according to Eq.~\eqref{eq:fitting}. 



Practically, as the imaginary-time correlation function decays very fast, we compare the two logarithmic correlation functions in the larger-$\tau$ regime where a clear linear relation is observed.
In our QMC implementation, the factor $2\pi$ has already been included in $\tilde{H}_A$.
Therefore, with the convention $\epsilon_{\mathrm{EH}}=2\pi/v$, Eq.~\eqref{eq:fitting} yields the sound velocity $v$ directly.
Once $v$ is obtained, the physical LBW scale $\epsilon_{\mathrm{EH}}$ follows immediately from Eq.~\eqref{eq:beta_EH}.


We remark that we specifically focus on the boundary-adjacent region because it typically carries the dominant contribution to the imaginary-time signal and therefore provides the cleanest and most stable input for extracting the overall LBW scale from the large-$\tau$ decay.
While operators close to the boundary may in principle be more sensitive to short-distance (UV) lattice effects, correlators further away are often substantially weaker and exhibit noisier large-$\tau$ behavior, making the scale extraction less reliable in practice.



\section{Transverse-field Ising model}\label{sec:tfim}
\subsection{LBW ansatz}

To verify the scheme outlined above, we first apply it to the two-dimensional transverse-field Ising model (TFIM), whose Hamiltonian is given by 
\begin{equation}\label{eq:TFIM}
    \begin{split}
    H = -J \sum_{\langle \mathbf{i},\mathbf{j} \rangle}  
    \sigma_{\mathbf{i}}^z \sigma_{\mathbf{j}}^z 
    - h \sum_{\mathbf{i}} \sigma_{\mathbf{i}}^x
    \end{split},
\end{equation}
where $J$ denotes the nearest-neighbor spin-spin coupling strength, $h$ is the strength of the transverse magnetic field, and $\langle \mathbf{i},\mathbf{j} \rangle$ represents the nearest-neighbor pair. This model has two distinct phases: the ferromagnetic (FM) phase and the paramagnetic (PM) phase, separated by a quantum critical point (QCP) at $h =3.04438(2)$~\cite{PhysRevE.66.066110}.
The FM phase and PM phase are both gapped, as their low-energy excitations require a finite energy cost to create. At the QCP, the system exhibits gapless excitations.

By bringing the original Hamiltonian Eq.~\eqref{eq:TFIM} into the LBW-EH ansatz Eq.~\eqref{eq:LBW-EH}, we achieve its LBW-EH form as
\begin{equation}\label{eq:TFIM-LBW}
    \begin{split}
    \tilde{H}_{A}
    = & \epsilon_{\text{EH}} 
    \bigg\{ \sum_{x,y,\delta}
    \bigg[x J\sigma_{(x,y)}^z \sigma_{(x+\delta,y)}^z \\
    + & \left(x-\frac{1}{2} \right) J \sigma_{(x,y)}^z \sigma_{(x,y+\delta)}^z \bigg] 
    - \sum_{x,y} \left(x-\frac{1}{2} \right) h\sigma_{(x,y)}^x \bigg\},
    \end{split}
\end{equation}
for the full system on a cylinder. 
From the formulation of the LBW-EH, the dependence of its terms on the distance to the boundary, which separates the system from the environment, can be clearly discerned. 
The first term in the expression corresponds to horizontal bonds, and the distance from the center of the bond to the boundary is $x$. The second and third terms represent vertical bonds and lattice sites respectively, both located at a distance of $(x-{1}/{2})$ from the boundary. These spatial relationships are clearly illustrated in the Fig. \ref{fig: BW_config}.


\subsection{Fitting the LBW-EH}
We begin by studying the EH at the QCP. The QCP of the two-dimensional TFIM is located at
$h = 3.04438(2)$~\cite{PhysRevE.66.066110}. As discussed above, fixing the overall energy scale
$\epsilon_{\mathrm{EH}}$ of the LBW-EH requires the sound velocity $v$.
For the $d=2$ TFIM at the QCP, an independent continuous-time Monte Carlo Renormalization Group (MCRG) study~\cite{PhysRevB.102.014456} estimated the sound velocity in the interval $(3.40,3.42)$, which we take as a reference value for comparison. 
In that approach, continuous-time configurations are
mapped to an effective $(2+1)$-dimensional discrete lattice 
, and $v$ is identified by enforcing large-scale space--time isotropy.

\begin{figure}[ht!]
\centering
\includegraphics[width=0.95\linewidth]{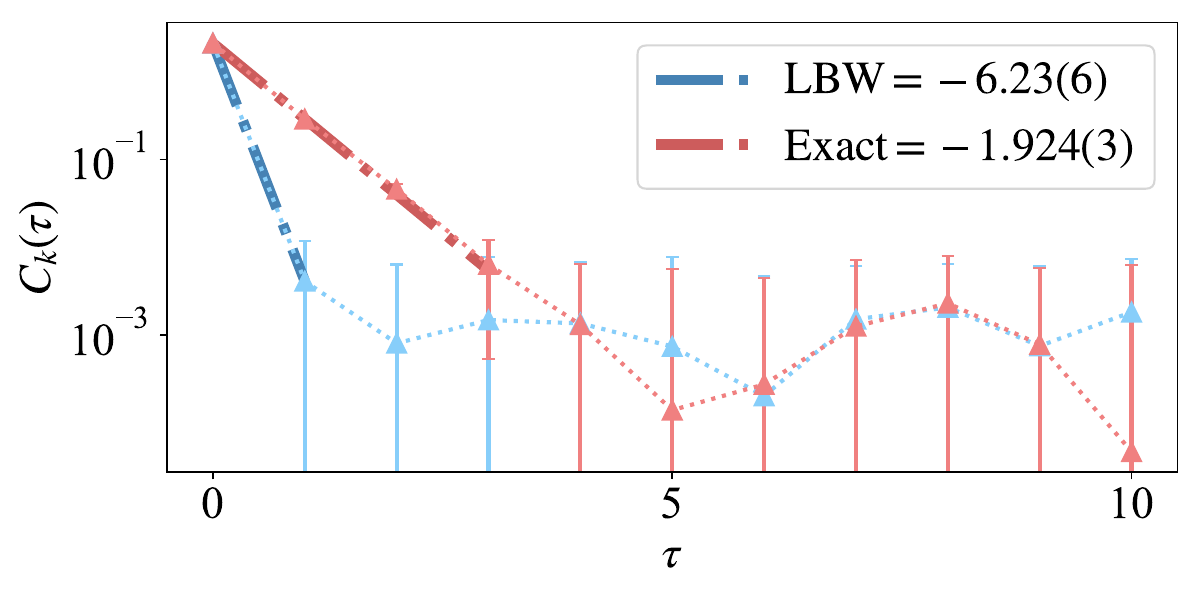}
\caption{
The imaginary-time correlation $C_k(\tau)$ of $16\times16$ LBW-EH and $32\times16$ exact-EH with maximum imaginary-time $\tau =50$ at QCP of two-dimensional TFIM.
Measurements are performed along the boundary-adjacent line inside subsystem $A$ and Fourier transformed along the periodic boundary direction in the uniform momentum channel $k=0$.
A linear fit to the large-$\tau$ regime of $C_k(\tau)$ yields slopes $-6.23(6)$ of LBW-EH data using points 1-2 and $-1.924(3)$ of exact-EH data using points 1-4, giving a sound velocity $v=3.24(3)$.
}
\label{fig:TFIM_QCP_velocity}
\end{figure}

In addition, we obtain an independent estimate of the sound velocity $v$ directly from the imaginary-time boundary correlators using the fitting procedure in Sec.~\ref{sec:fit}.
Specifically, we fit the $\tau$ linear regime of $C_k(\tau)$ for both the LBW-EH and the exact-EH at the original system size $32\times16$ in Fig.~\ref{fig:TFIM_QCP_velocity}, and obtain $v=3.24(3)$ under our LBW-EH normalization convention.
This value is consistent with the MCRG estimate within a few percent.
We attribute the remaining deviation mainly to finite-size effects and systematic uncertainties associated with the choice of the fitting window in $\tau$ and the discretization choices inherent to finite-$L$ data.


Using the same method, we also determine the sound velocities in both the gapped FM phase ($h=1$) and the gapped PM phase ($h=5$), which are summarized in Table \ref{table: TFIM_velocities}. 
Once the velocity parameter $v$ is determined, the value of $\epsilon_{\text{EH}}$ can be directly obtained through the relation Eq.~\eqref{eq:beta_EH}. 

\begin{table}[htbp]
  \centering
  \caption{Fitted velocities $v$ of two-dimensional TFIM at $L=16$.}
  \label{table: TFIM_velocities}
  \begin{tabular}{ccc} 
    \toprule
   & $h$  &  $v$ \\ 
    \midrule
    FM phase &   $1$   & $0.997(6)$\\ 
    QCP &   $3.04438(2)$   & $3.24(3)$ \\ 
    PM phase &   $5$ &     $1.66(8)$ \\ 
    \bottomrule
  \end{tabular}
\end{table}


The slope-ratio method relies on the existence of a large-$\tau$ regime in which the momentum-resolved boundary correlator is dominated by a single mode. 
For the datasets considered here, the extracted velocity remains stable under reasonable variations of the fitting window within that regime.
Detailed robustness tests and discussions of the momentum choices are presented in Appendix~\ref{app:robustness}. 

In the gapped FM and PM phases, the available linear regime becomes narrower due to the rapid decay of the correlator, so the fitted velocities should be regarded as approximate estimates.
More generally, a clearly developed single-exponential regime is not guaranteed for all models or parameter regimes, so the fitting procedure should be understood as a controlled approximation for matching the exact EH to a physically motivated ansatz. Even when the asymptotic linear regime is not sharply developed, the fit still defines an effective decay scale that provides a meaningful quantitative estimate.

\subsection{Validation via equal-time correlation functions}
After we fix the parameter $\epsilon_{\mathrm{EH}}$ and obtain the complete LBW-EH, we proceed to evaluate its accuracy by comparing physical observables obtained from the LBW-EH ansatz and from the exact-EH.
We focus on equal-time spin correlations, which are directly determined by the EH eigenstates and therefore provide a sensitive benchmark of the ansatz.
We define the correlation function as
\begin{equation}\label{eq:cc}
    C(\mathbf{r}) = \big\langle \sigma^z_{\mathbf{i}}\,\sigma^z_{\mathbf{i}+\mathbf{r}} \big\rangle, 
\end{equation}
where $\mathbf{i}=(i_x,i_y)$ labels a lattice site in subsystem $A$ and $\mathbf{r}=(r_x,r_y)$ is the separation vector.
In the two-dimensional cylinder geometry (periodic along $y$), we measure correlations on the boundary-adjacent line inside $A$ by choosing $i_x=1$ and taking the separation along the entanglement boundary as $r_x=0$.
Concretely, we take $\mathbf{i}=(1,i_y)$ and $\mathbf{r}=(0,r_y)$ so that $\mathbf{i}+\mathbf{r}=(1,i_y+r_y)$ (mod $L$) along the periodic $y$ direction.
In practice, we average over all boundary positions $i_y=1,\dots,L$ to improve statistics and report the along-boundary correlator as a function of the boundary separation $r_y$, which we define as
\begin{equation}
C(r_y)= \frac{1}{L}\sum_{i_y=1}^{L}\big\langle \sigma^z_{(1,i_y)}\,\sigma^z_{(1,i_y+r_y)} \big\rangle,
\label{eq:cry_def}
\end{equation}
where $i_y+r_y$ is understood modulo $L$ due to periodic boundary conditions along $y$.
A key advantage of the two-dimensional setup is that it allows us to probe correlations both along and perpendicular to the entanglement boundary.
In what follows we emphasize the along-boundary correlations, which carry the dominant boundary signal and provide the most direct and discriminating test of the LBW-EH ansatz in our geometry.
In the present work, we validate the LBW-EH ansatz primarily through low-energy two-point correlators along the entanglement boundary, which provide a direct and sensitive benchmark.

\begin{figure}[ht!]
\centering
\includegraphics[width=0.95\linewidth]{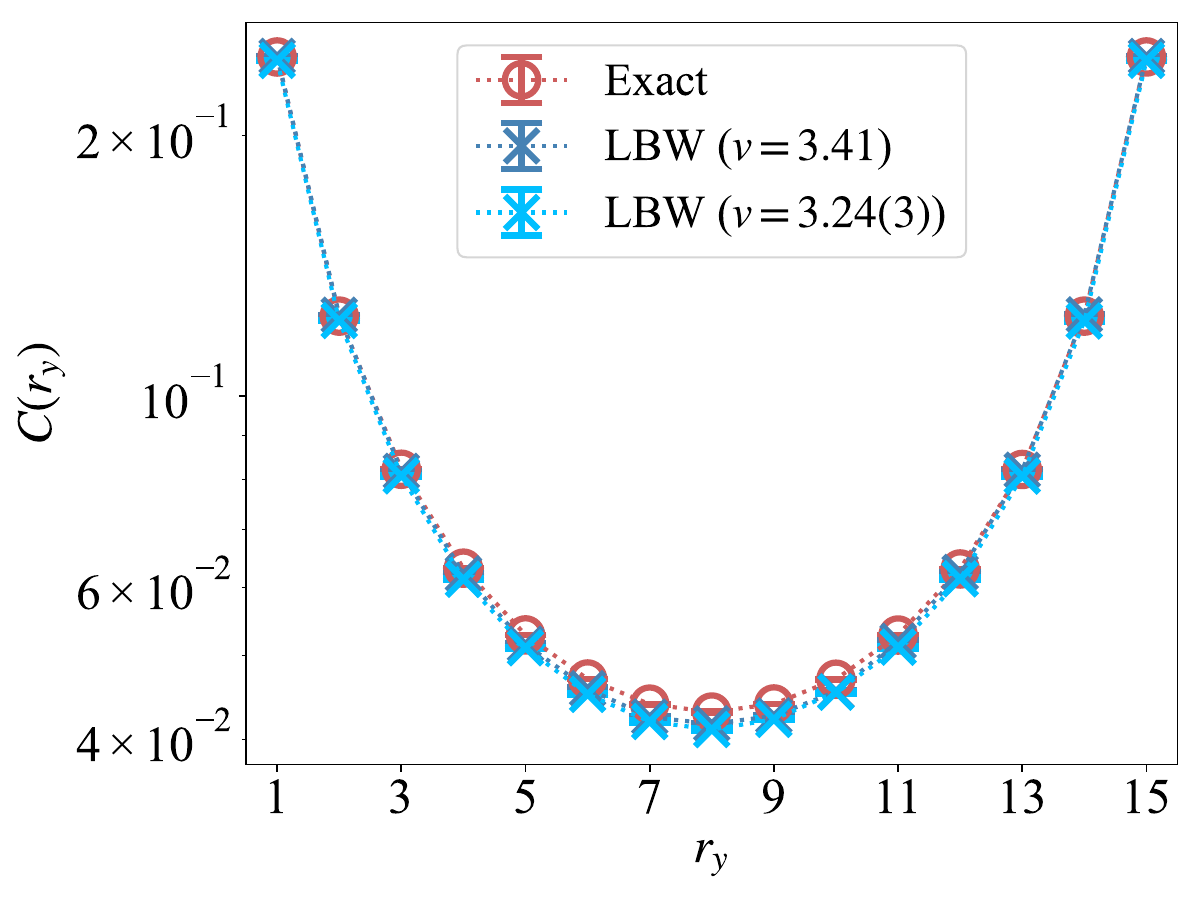}
\caption{
Equal-time spin correlation functions at the QCP of the two-dimensional TFIM comparing the $16\times16$ LBW-EH and the $32\times16$ exact-EH, evaluated at effective inverse temperature $\beta_A=1$.
The along-boundary correlator $C(r_y)$ is measured within subsystem $A$ on the boundary-adjacent line, with separations taken along the periodic boundary direction as defined in the text.
The exact-EH results are compared with LBW-EH predictions using two independently determined velocities, $v=3.41$ (from MCRG) and $v=3.24(3)$ (from the imaginary-time correlator fit).
}
\label{fig:TFIM_QCP_Rep}
\end{figure}

We now discuss the equal-time correlation function at the QCP. The QMC results for the LBW-EH and the exact-EH at effective inverse temperature $\beta_A=1$ are shown in Fig.~\ref{fig:TFIM_QCP_Rep}. 
We simulate the LBW-EH with MCRG given velocity and imaginary-time correlation approximated velocity, compared with the exact-EH results. 
First, we evaluate the quality of the imaginary-time correlation velocity fitting. For the results of the correlation function obtained from the LBW-EH, using the midpoint value $v=3.41$ from the MCRG interval and the velocity result $v=3.24(3)$ derived from our imaginary-time correlation fitting, the two correlation function curves almost completely overlap. 
This demonstrates that the imaginary-time correlation method provides a good fitting result, proving it to be a reliable approach to extract the velocity.
Next, we compare the correlation functions of the LBW-EH and the exact-EH. 
We apply the logarithmic scale to the correlation function results to clearly visualize the discrepancies, then we find that the values of the two correlation functions differ only slightly at large $r_y$ with PBC. When $r_y$ is small, the values coincide.
Therefore, we conclude that the LBW-EH ansatz provides a good functional form for the two-dimensional TFIM at QCP.

\begin{figure}[ht!]
\centering
\includegraphics[width=0.95\linewidth]{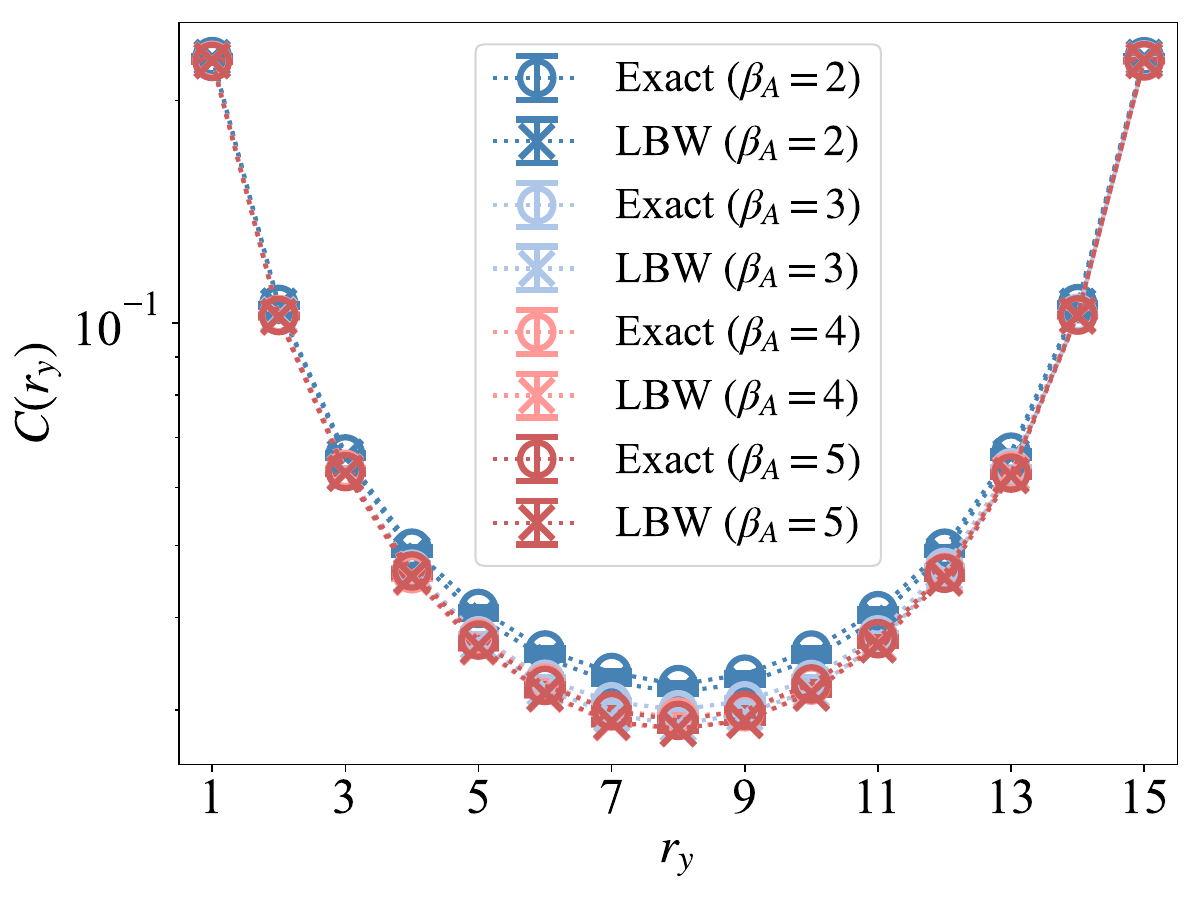}
\caption{
Equal-time spin correlation functions at the QCP of the two-dimensional TFIM for larger effective inverse temperatures $\beta_A$.
We compare the $16\times16$ LBW-EH results at effective inverse temperature $\beta_A$ with the $32\times16$ exact-EH results obtained from the multi-replica construction at the corresponding replica number $n=\beta_A$.
The along-boundary correlator $C(r_y)$ is measured within subsystem $A$ on the boundary-adjacent line, with separations taken along the periodic boundary direction as defined in the text.
Data for $\beta_A=2,3,4,5$ show the progressive approach toward the EH ground-state behavior with increasing $\beta_A$.
}
\label{fig:TFIM_QCP_nRep}
\end{figure}

We also simulated the LBW-EH and exact-EH with higher effective inverse temperatures $\beta_A$, which brings the system closer to the ground state of the EH. 
For the LBW-EH, the effective inverse temperature ${\beta}_A>1$ is used as the imaginary-time to construct the imaginary-time path integral in QMC simulation. 
For the exact-EH, we employ the multi-replica-trick QMC methods~\cite{PhysRevB.109.195169,wu2023classical}, where $\beta_A$ effectively corresponds to an imaginary-time path composed of $n$ replicas. 
For the physical Hamiltonian within each replica, the physical inverse temperature $\beta$ is taken to be proportional to the system size in order to approximate the ground state of the real system. 
Note that a larger effective inverse temperature $\beta_A$ corresponds to a state closer to the ground state of the EH.
The correlation function results of LBW-EH and Exact-EH with higher effective inverse temperatures $\beta_A$ at QCP are shown in Fig.~\ref{fig:TFIM_QCP_nRep}. 
The correlation function results of the LBW-EH and the exact-EH exhibit highly consistent characteristics across different temperatures, with the correlations converging at higher inverse temperatures. This demonstrates that at the QCP, the LBW-EH ansatz also offers a reliable functional form of the EH when the system approaches the ground state.

\begin{figure*}[ht!]
\centering
\begin{overpic}[width=0.45\linewidth]{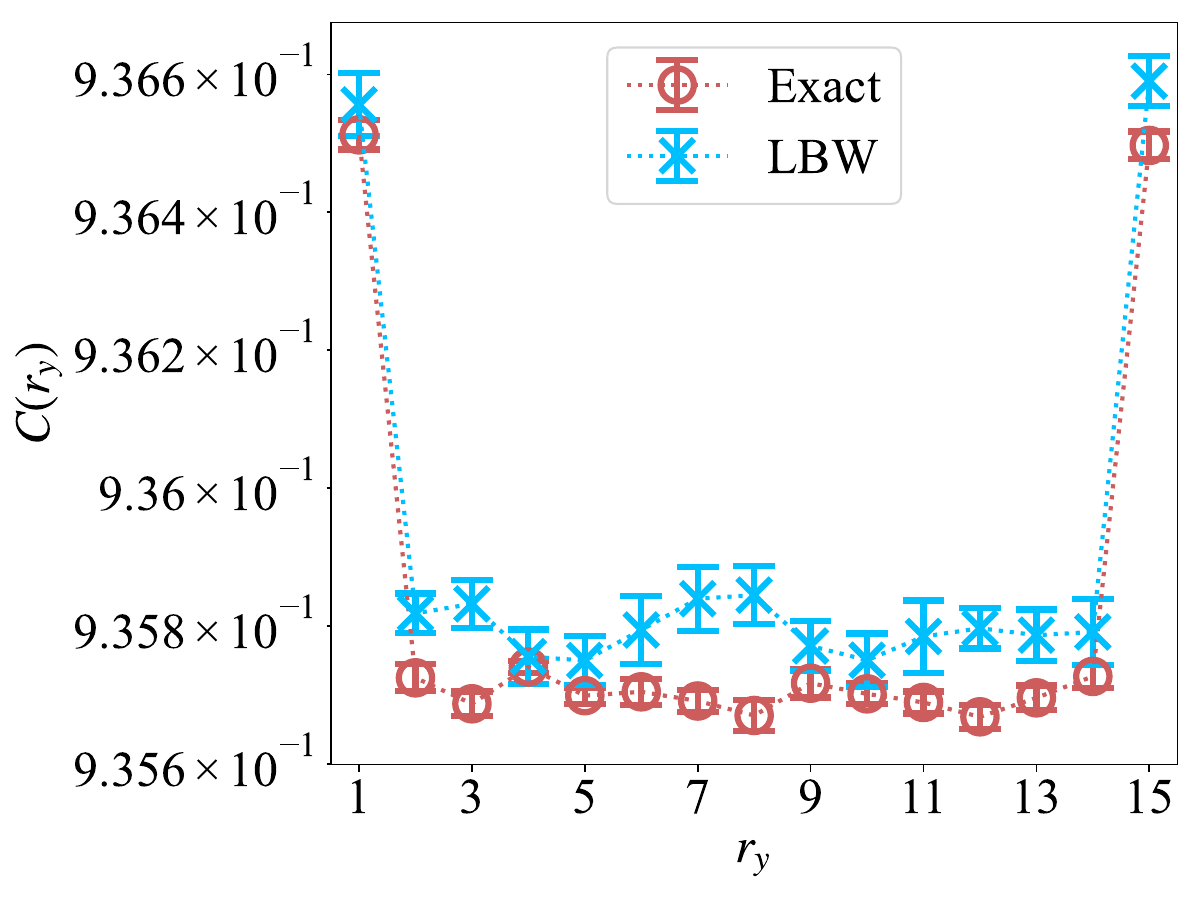}
\put (0,75) {{\textbf{(a)}}}
\end{overpic}
\begin{overpic}[width=0.45\linewidth]{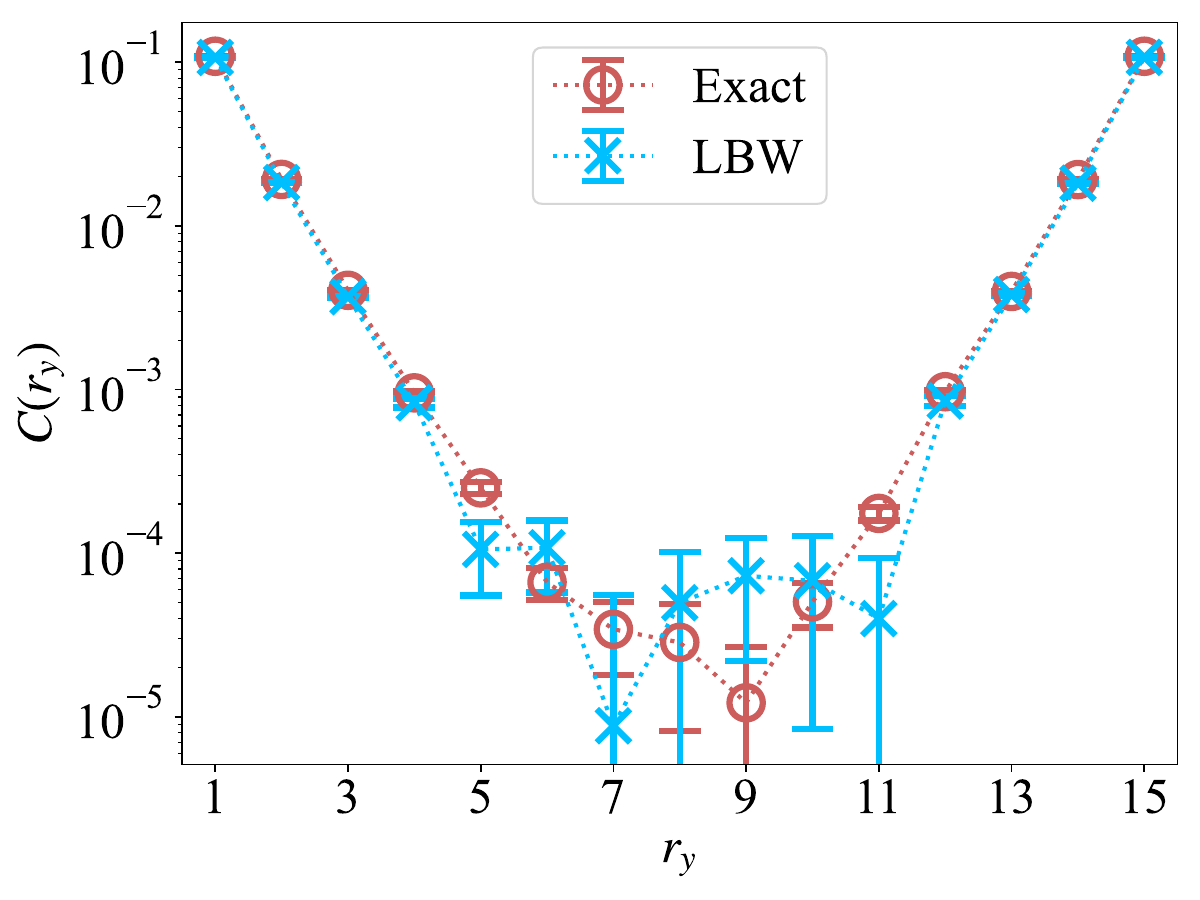}
\put (0,75) {{\textbf{(b)}}}
\end{overpic}
\begin{overpic}[width=0.45\linewidth]{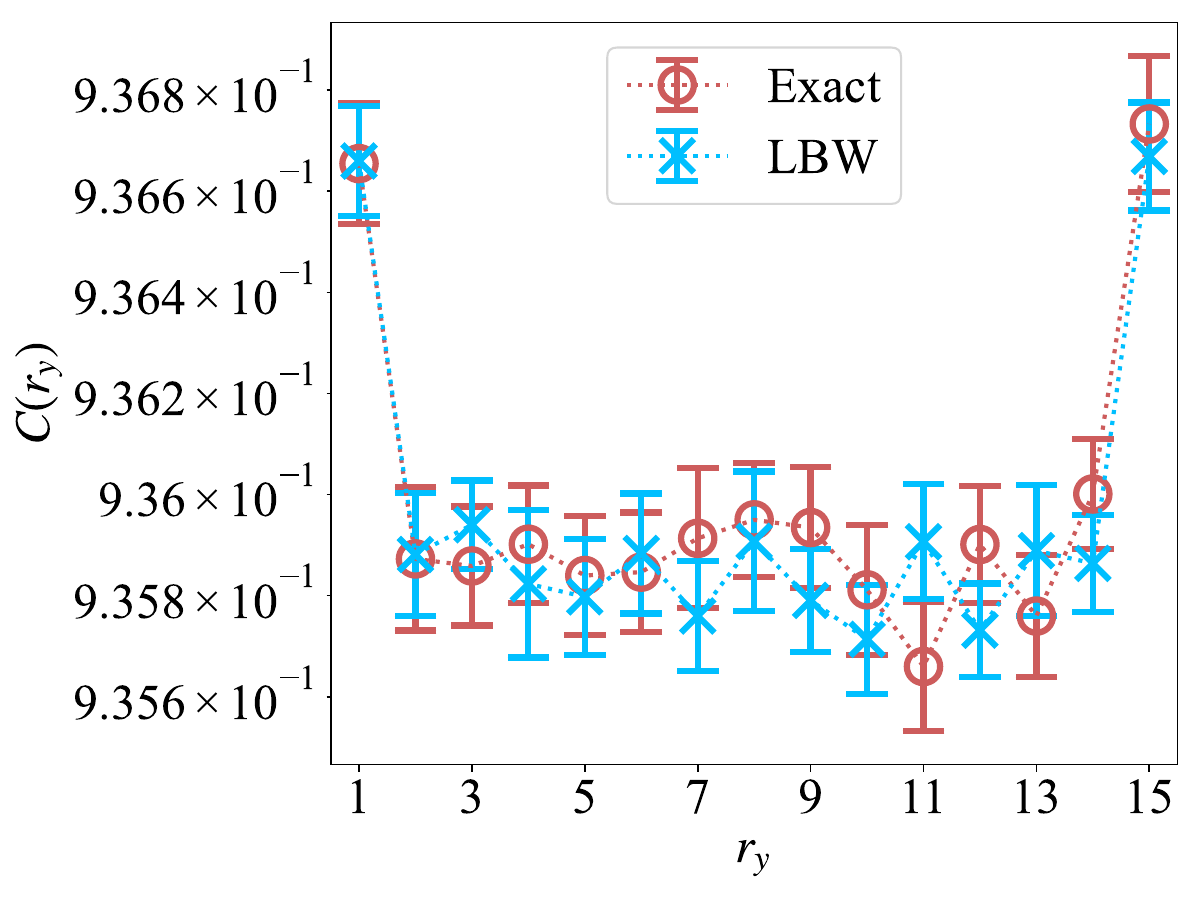}
\put (0,75) {{\textbf{(c)}}}
\end{overpic}
\begin{overpic}[width=0.45\linewidth]{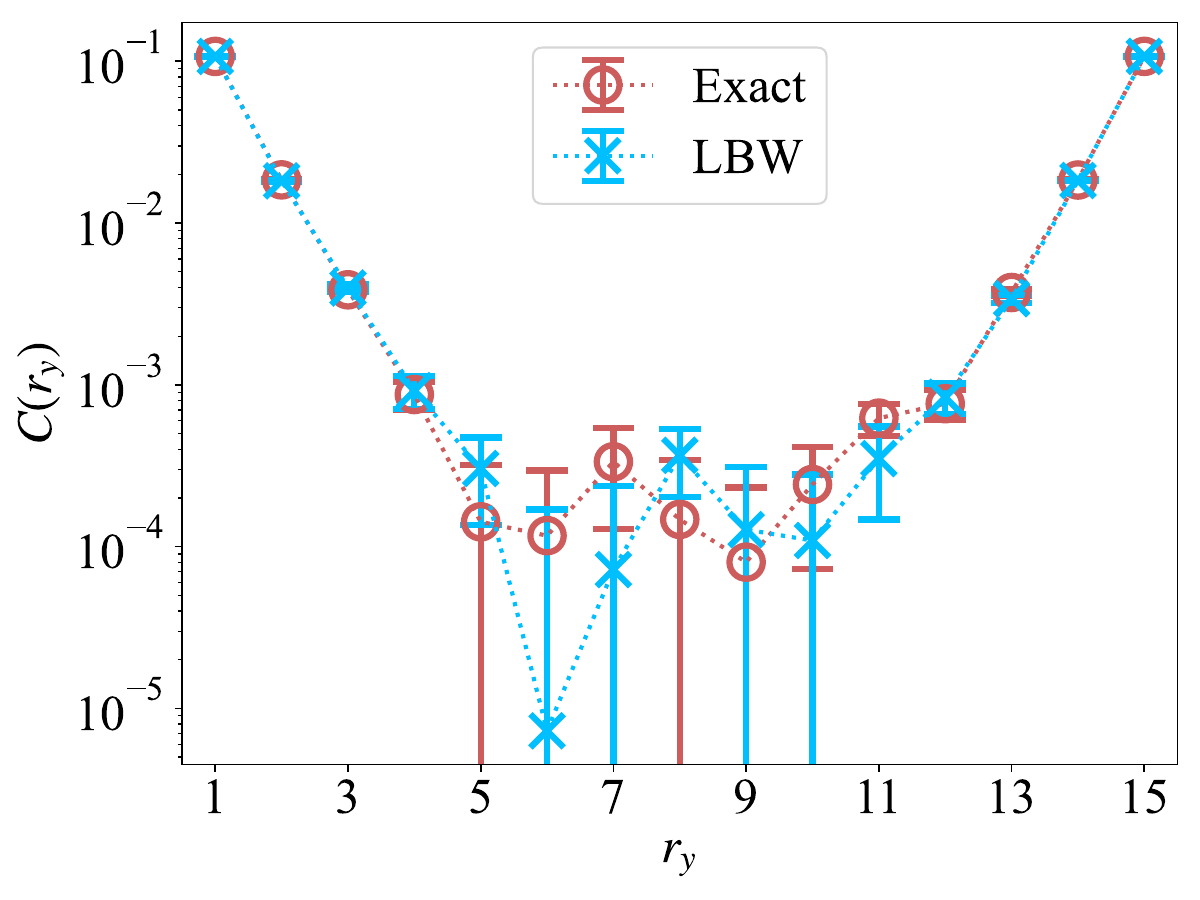}
\put (0,75) {{\textbf{(d)}}}
\end{overpic}
\caption{
Equal-time spin correlation functions of the two-dimensional TFIM away from the QCP, comparing the $16\times16$ LBW-EH and the $32\times16$ exact-EH at effective inverse temperatures $\beta_A=1$ and $5$.
The along-boundary correlator $C(r_y)$ is measured within subsystem $A$ on the boundary-adjacent line, with separations taken along the periodic boundary direction as defined in the text.
(a) FM phase with $h=1$ and $\beta_A=1$.
(b) PM phase with $h=5$ and $\beta_A=1$.
(c) FM phase with $h=1$ and $\beta_A=5$.
(d) PM phase with $h=5$ and $\beta_A=5$.
} 
\label{fig: TFIM_PM_FM}
\end{figure*}

Then we discuss the correlation functions in FM phase ($h=1$) and PM phase ($h=5$), which are presented in Fig.~\ref{fig: TFIM_PM_FM}. 
For FM phase at $h=1$, the correlation functions of LBW-EH and exact-EH do not completely coincide in Fig.~\ref{fig: TFIM_PM_FM}(a). However, it should be noted that at this parameter value, the measured correlation functions at different distances are very close to each other. Moreover, the plotted correlation functions are presented on a logarithmic scale, and the actual numerical difference between the two correlation functions is on the order of $10^{-4}$, which is indeed a very small discrepancy. Additionally, the trends of both correlation functions are similar. The nearest-neighbor correlation function is significantly larger than those at other distances, while the results at other distances are comparable. 

Therefore, we can conclude that in the FM phase, the LBW-EH ansatz provides a good approximation. 
For the PM phase at $h=5$, the first few data points of the LBW-EH and exact-EH correlation functions coincide, as shown in Fig.~\ref{fig: TFIM_PM_FM}(b). However, obtaining accurate correlation functions for the intermediate data points is challenging for both the LBW-EH and the exact-EH, making it impossible to compare the results in this region. Nevertheless, the overlapping data points are sufficient to demonstrate that the LBW-EH ansatz also provides a good approximation in the PM phase. 
Therefore, even in the gapped phases of a two-dimensional system with translational invariance, the parameter $\epsilon_{\mathrm{EH}}$ in the functional form of LBW-EH ansatz can be obtained by fitting imaginary-time correlation functions, and the correlation function results from QMC simulations show that the LBW-EH ansatz provides a reliable functional form.

We further simulate LBW-EH and exact-EH in FM phase and PM phase with effective inverse temperatures greater than $1$ to access the ground state. Through the increase of the effective inverse temperature $\beta_A$ for the LBW-EH in QMC simulation and use of the replica-trick QMC methods $\beta_A = n$ for the exact-EH, the finite-temperature properties of the EH are measured. The results with large effective inverse temperatures $\beta_A =5$ are shown in Fig.~\ref{fig: TFIM_PM_FM}(c) and (d). 
For the FM phase at $h=1$, the correlation functions coincide, although with error bars. This uncertainty arises from the significant computational cost required to achieve high precision in this parameter. Nevertheless, the correlation values are on the order of $10^{-1}$ and the errors are on the order of $10^{-4}$, and the agreement can be considered valid. 
For the PM phase at $h=5$, the correlation functions show excellent agreement for the first four measurable points, indicating that the LBW-EH provides a good approximation in this regime. 
Therefore, we conclude that the functional form of the LBW-EH remains applicable in both gapped FM and PM phases when approaching the ground state.

For the critical point of the two-dimensional TFIM, which exhibits translational invariance, its low-energy behavior can be described by a Lorentz-invariant quantum field theory~\cite{dalmonte2022entanglement}. 
By comparing the correlation functions obtained from the LBW-EH ansatz and the exact EH, across different effective inverse temperatures, as well as in the gapped FM phase, gapped PM phase, and at the QCP, the results show close or exact agreement. Therefore, the numerical results from QMC simulations support the conclusion that the LBW-EH ansatz provides a reliable functional form in translationally invariant systems.


\section{Dimerized Heisenberg model}\label{sec:dhm}
\subsection{LBW ansatz without translational invariance}

We have discussed translationally invariant systems and demonstrated through QMC simulations that the LBW-EH ansatz provides a reliable functional form even in gapped phases in the previous section. Moreover, we are more interested in whether the LBW-EH ansatz remains valid in systems without translational invariance, as extending the applicability of the LBW-EH functional form holds significant importance.

\begin{figure*}[ht!]
\centering
\begin{overpic}[width=0.32\linewidth]  {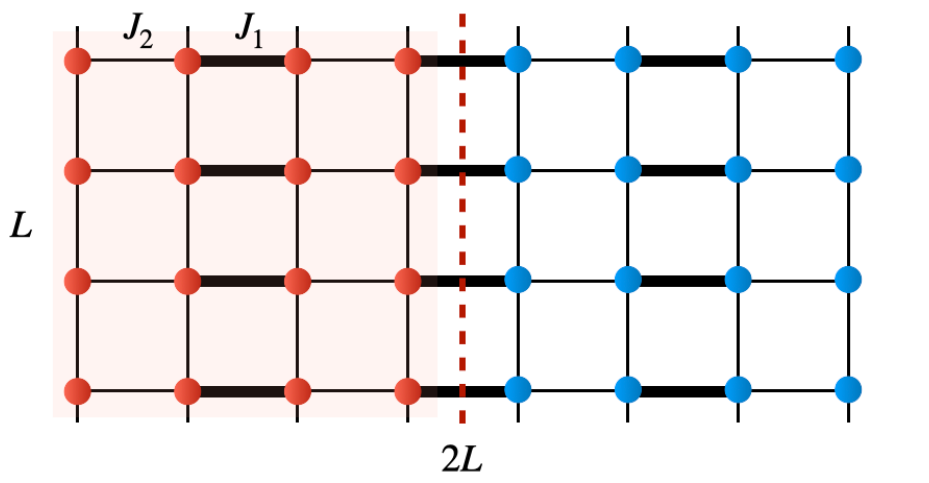} \label{fig: strong}
\put (-3,55) {{\textbf{(a)}}}
\end{overpic}
\begin{overpic}[width=0.32\linewidth]{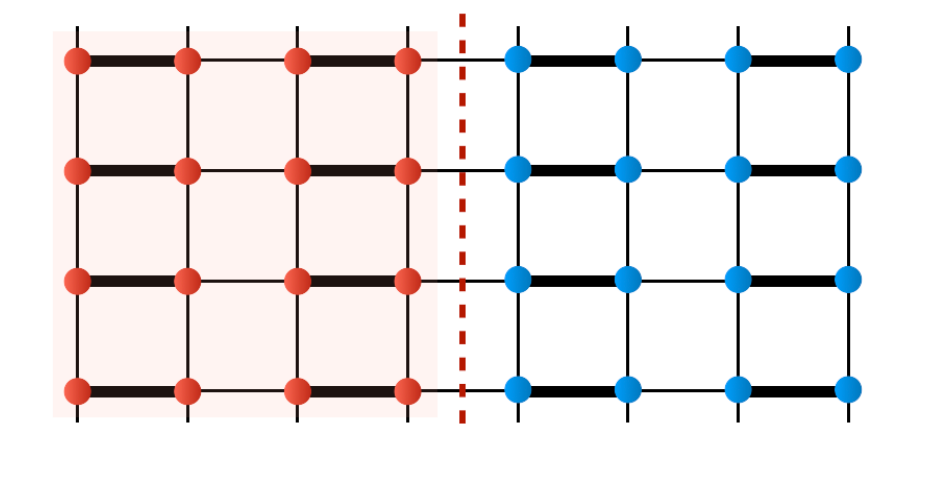}
\put (-3,55) {{\textbf{(b)}}}
\end{overpic}
\begin{overpic}[width=0.32\linewidth]{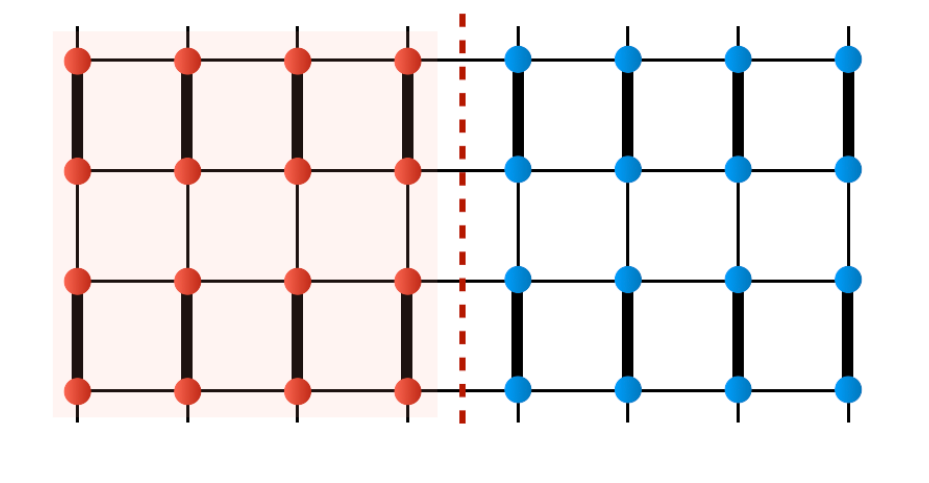}
\put (-3,55) {{\textbf{(c)}}}
\end{overpic}
\caption{The configurations of the two-dimensional dimerized Heisenberg model with the strong-bonds $J_1$ and the weak-bonds $J_2$. (a) Strong and weak-bonds are arranged alternately in the horizontal direction, and the boundary separating the system and the environment cuts through the horizontal strong-bonds. (b) Strong and weak-bonds are alternately arranged along the horizontal direction, and the boundary cuts through the horizontal weak-bonds. (c) Strong and weak-bonds are alternately arranged along the vertical direction, and the boundary cuts through the horizontal weak-bonds.} 
\label{fig:J1J2_config}
\end{figure*}

For the system without translational invariance, we consider the two-dimensional columnar dimerized Heisenberg model whose Hamiltonian is given by
\begin{equation}\label{eq:DHM-H}
    H = J_1 \sum_{\langle \mathbf{i}\mathbf{j} \rangle} \vec{S}_\mathbf{i} \cdot \vec{S}_\mathbf{j} + J_2 \sum_{{\langle \mathbf{i}\mathbf{j} \rangle}'} \vec{S}_\mathbf{i} \cdot \vec{S}_\mathbf{j},
\end{equation}
where $\vec{S}_\mathbf{i} = (S_\mathbf{i}^x,S_\mathbf{i}^y,S_\mathbf{i}^z)$ is the spin-1/2 operator on site $\mathbf{i}$, and $\langle \mathbf{i}\mathbf{j} \rangle$ and ${\langle \mathbf{i}\mathbf{j} \rangle}'$ denote different nearest-neighbor pairs on the lattice. $J_1$ and $J_2$ are the coupling strengths of strong and weak-bonds respectively (Fig. \ref{fig:J1J2_config}). The ratio of these couplings is defined as $J_r = J_1/J_2$, called the dimerization strength.

The most interesting feature of this model is the quantum phase transition that occurs as the dimerization strength $J_r$ is tuned.
The system resides at the Heisenberg limit with translational invariance when $J_r=1$. As the dimerization strength increases, it reaches a QCP at $J_r = 1.90951(1)$~\cite{PhysRevLett.121.117202} from a N\'eel order. Beyond this point, the system goes into the dimer phase. Therefore, a comprehensive investigation of the various phases and points-the Heisenberg limit, N\'eel ordered phase, QCP and the dimer phase-in this two-dimensional dimerized Heisenberg model is essential.

The first step is to derive the LBW-EH functional form of this model for QMC simulations. Based on the original Hamiltonian in Eq.~\eqref{eq:DHM-H} and the functional ansatz for the LBW-EH given in Eq.~\eqref{eq:LBW-EH}, we obtain the LBW-EH functional form of the two-dimensional dimerized Heisenberg model with cylinder geometry is
\begin{widetext}
\begin{equation}\label{eq:DHM-LBW}
\begin{split} \tilde{H}_{A} = \epsilon_{\text{EH}} \bigg\{   \sum_{x,y,\delta} & \bigg[x J_1\vec{S}_{(x,y)} \cdot \vec{S}_{(x+\delta,y)} + \left(x-\frac{1}{2} \right) J_1\vec{S}_{(x,y)}  \cdot \vec{S}_{(x,y+\delta)} \bigg] \\ +  \sum_{x',y',\delta} & \bigg[x' J_2\vec{S}_{(x',y')}  \cdot \vec{S}_{(x'+\delta,y')}  + \left(x'-\frac{1}{2} \right) J_2\vec{S}_{(x',y')}   \cdot \vec{S}_{(x',y'+\delta)} \bigg] \bigg\}. \end{split}
\end{equation}
\end{widetext}
Each term is related to the distance to the boundary separating the system and the environment. For horizontal bonds, the distance is $x$, while for vertical bonds, the distance is $(x - 1/2)$.

In addition to the breaking of the translational symmetry, this model can give different half-space bipartition configurations when dividing the system into system $A$ and environment $B$,
which are generally categorized into three types, as shown in Fig.~\ref{fig:J1J2_config}:
\begin{itemize}
    \item The strong-bond case in Fig.~\ref{fig:J1J2_config}(a) features alternating strong and weak-bonds along the horizontal direction, with all vertical bonds being weak. The boundary separating the system and the environment cuts vertically through the midpoints of the strong horizontal bonds;
    \item In the weak-bond case shown in Fig.~\ref{fig:J1J2_config}(b), the strong and weak-bonds also alternate along the horizontal direction. However, the boundary separating the system and the environment cuts vertically through the midpoints of the weak horizontal bonds;
    \item The vertical case in Fig.~\ref{fig:J1J2_config}(c) features alternating strong and weak-bonds along the vertical direction, while all horizontal bonds remain weak. Consequently, the boundary separating the system and the environment also cuts vertically through the horizontal weak-bonds. 
\end{itemize}
Different partitioning methods may yield distinct entanglement information, thus each of these three configurations must be discussed individually.

We note that cutting strong-bonds (Fig.~\ref{fig:J1J2_config}(a)) will introduce an effective dangling spin chain with Lieb-Schultz-Mattis anomaly \cite{cheng2023lieb} on the edge while the other cuts (Fig.~\ref{fig:J1J2_config}(b) and (c)) will not. The extra gapless surface mode can hence affect the surface critical behaviors, which has been carefully studied~\cite{Ding2018}. It inspires us to pay attention to whether the edge effect in entanglement cut has a potential connection to LBW ansatz.

\subsection{The strong-bond case of bipartition}
We first consider the configuration with horizontally cut strong-bonds, as shown in Fig.~\ref{fig:J1J2_config}(a). 
To complete the LBW-EH ansatz, we determine the sound velocity $v$ using the imaginary-time fitting procedure described in Sec.~\ref{sec:fit}.
We start from the Heisenberg limit $J_r=1$.
We measure the boundary-adjacent correlator, Fourier transform it along the periodic direction, and fit the large-$\tau$ linear regime of $C_k(\tau)$ in the staggered channel $k=\pi$.

\begin{figure}[ht!]
\centering
\includegraphics[width=0.95\linewidth]{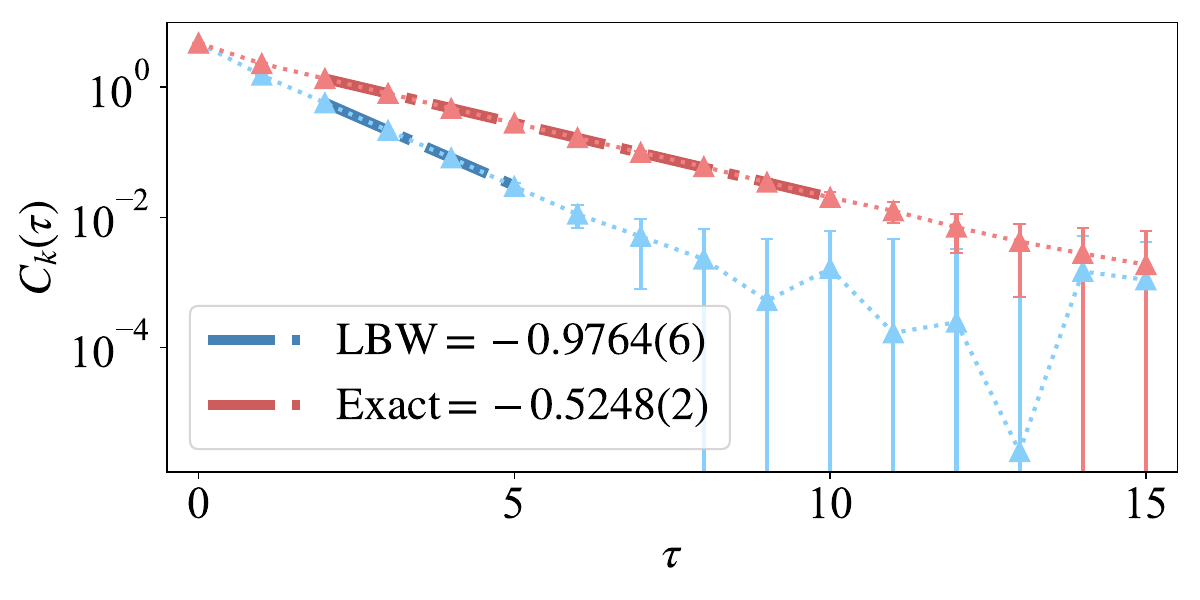}
\caption{
The imaginary-time correlation results for the two-dimensional dimerized Heisenberg model with the configuration in Fig.~\ref{fig:J1J2_config}(a) at the Heisenberg limit $J_r = 1$. 
Measurements are performed for the $16\times16$ LBW-EH and the $32\times16$ exact-EH with maximum imaginary-time $\tau =50$ along the boundary-adjacent line inside subsystem $A$, and Fourier transformed along the periodic boundary direction in the staggered momentum channel $k=\pi$.
The linear fits to $C_k(\tau)$ are performed over points 3--6 for the LBW-EH data and points 3--11 for the exact-EH data, yielding slopes $-0.9764(6)$ and $-0.5248(2)$, respectively. The resulting sound velocity is $v=1.860(1)$.
}
\label{fig:Strong_J1J2_Jr_1_velocity}
\end{figure}

From the fitted slopes $-0.9764(6)$ of LBW-EH and $-0.5248(2)$ of exact-EH in Fig.~\ref{fig:Strong_J1J2_Jr_1_velocity}, we obtain $v=1.860(1)$ under our normalization convention.
This value is in reasonable agreement with the independent estimate value of $1.657(2)$ reported in Ref.~\cite{PhysRevLett.80.1742}. 
The remaining difference is attributable to finite-size effects and differences in estimation procedures.

We repeat the same analysis to fit the sound velocities $v$ in the N\'eel order phase ($J_r=1.5$), at the QCP ($J_r=1.90951(1)$~\cite{PhysRevLett.121.117202}), and in the dimer phase ($J_r=3$). 
The velocities from fitting the different phases and points are summarized in TABLE \ref{table:J1J2_velocities}. 

\begin{table}[htbp]
  \centering
  \caption{Fitted velocities $v$ of two-dimensional dimerized Heisenberg model at $L=16$.}
  \label{table:J1J2_velocities}
  \begin{tabular}{ccc} 
    \toprule
    Phase & $J_r$  &  $v$ \\ 
    \midrule
    Heisenberg limit & $1$ & $1.860(1)$\\ 
    N\'eel ordered phase & $1.5$ & $2.337(1)$ \\ 
    QCP & $1.90951(1)$ & $2.697(1)$ \\ 
    Dimer phase &  $3$  & $4.111(5)$ \\ 
    \bottomrule
  \end{tabular}
\end{table}


The slope extraction is performed in the large-$\tau$ regime where $C_k(\tau)$ is well described by a single dominant linear slope. 
The extracted velocity is stable under reasonable variations of the fitting window. 
Detailed robustness checks across different phases are presented in Appendix~\ref{app:robustness}. 
The momentum-channel choice is physically motivated and stable. 
The dominant boundary signal for this model is captured in the staggered sector $k=\pi$, while other channels carry substantially less low-energy weight and lead to noisier large-$\tau$ data without changing the central value within uncertainties.
Additional larger-size checks of the fitted velocity are presented in Appendix~\ref{app:large_size}.

\begin{figure*}[ht!]
\centering
\begin{overpic}[width=0.45\linewidth]{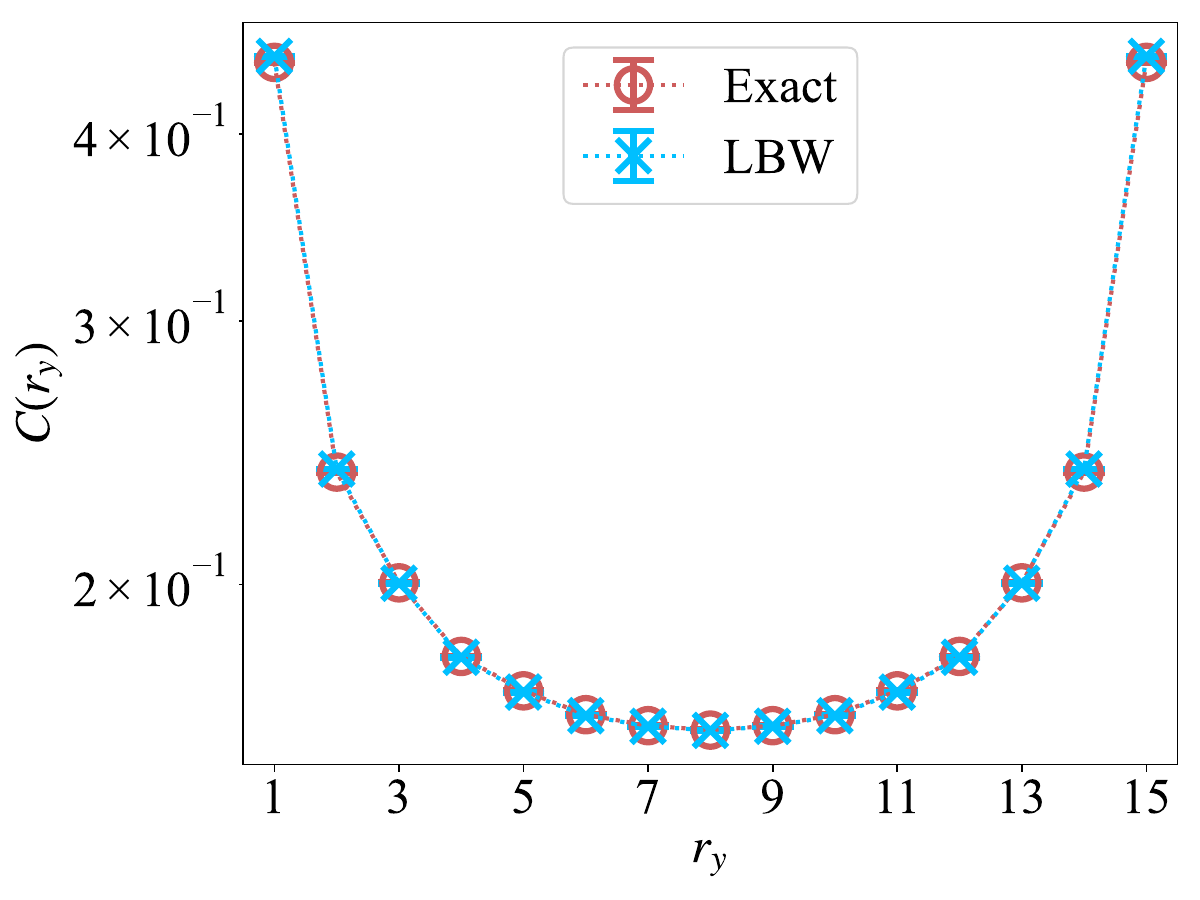}
\put (0,75) {{\textbf{(a)}}}
\end{overpic}
\begin{overpic}[width=0.45\linewidth]{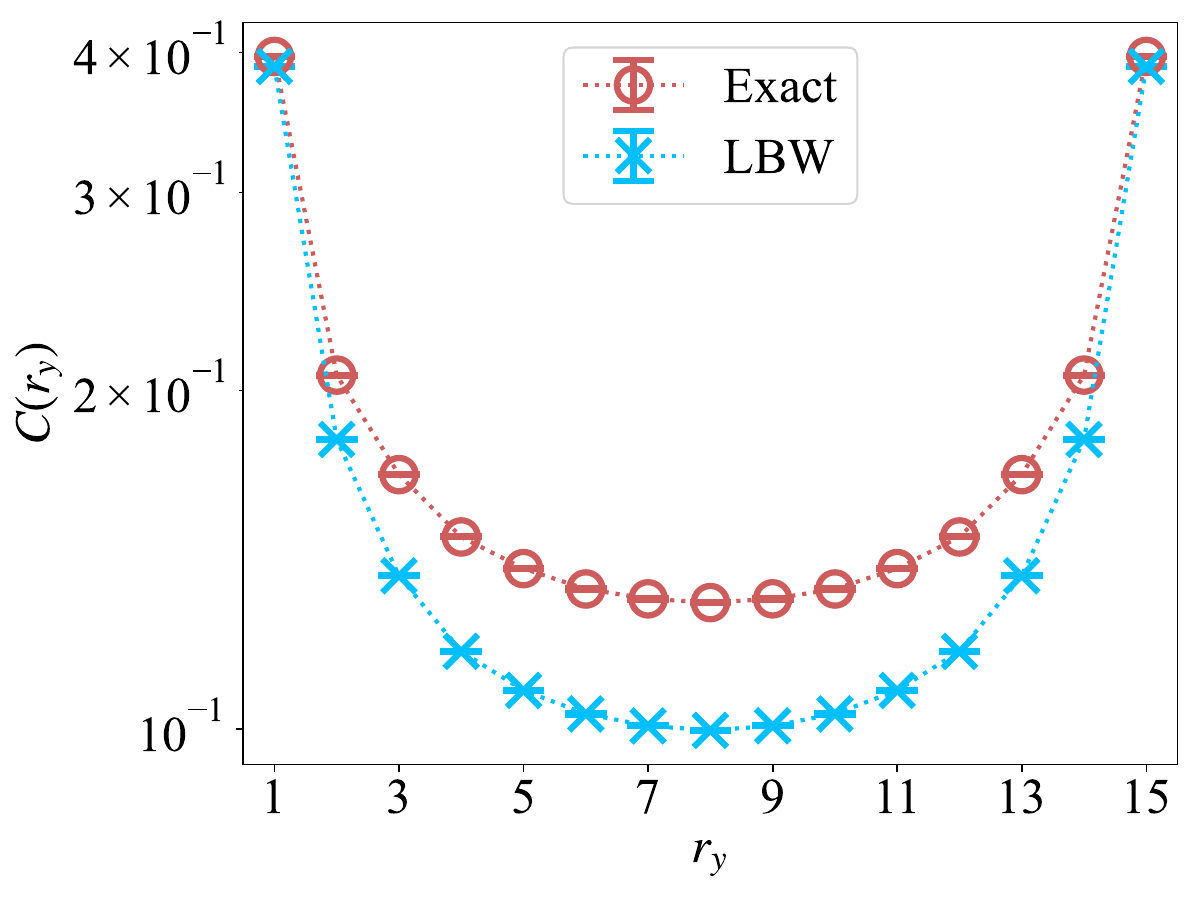}
\put (0,75) {{\textbf{(b)}}}
\end{overpic}
\begin{overpic}[width=0.45\linewidth]{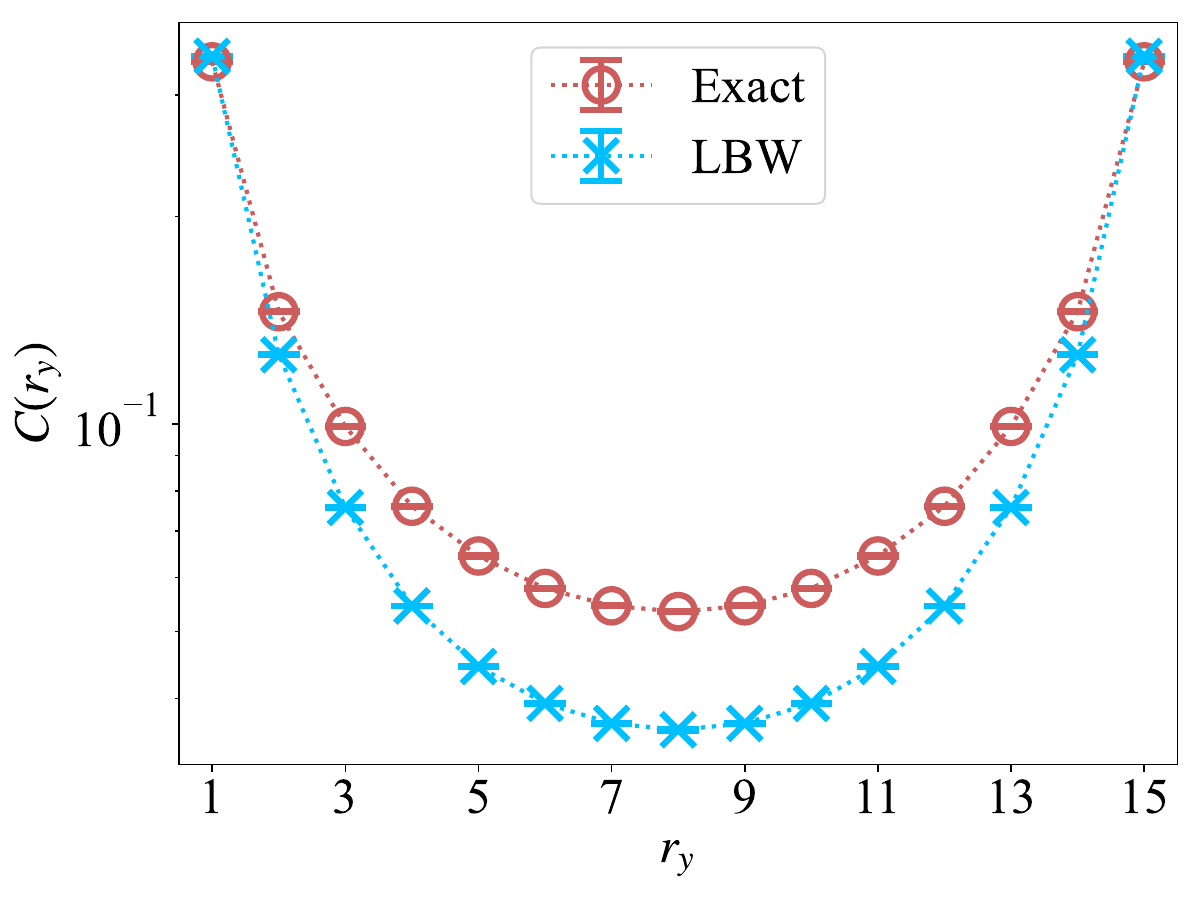}
\put (0,75) {{\textbf{(c)}}}
\end{overpic}
\begin{overpic}[width=0.45\linewidth]{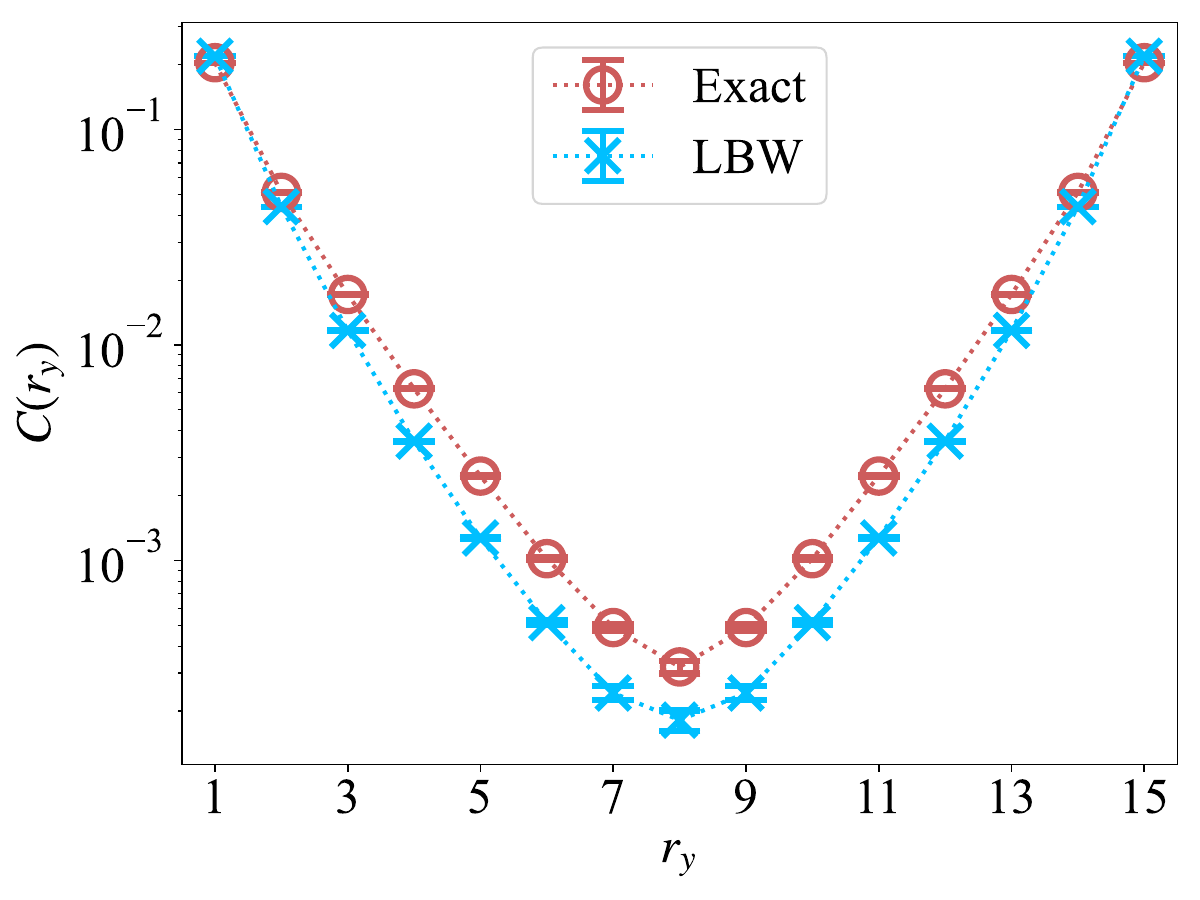}
\put (0,75) {{\textbf{(d)}}}
\end{overpic}
\caption{
Equal-time spin correlation functions of the two-dimensional dimerized Heisenberg model for the strong-bond bipartition shown in Fig.~\ref{fig:J1J2_config}(a), comparing the $16\times16$ LBW-EH and the $32\times16$ exact-EH at effective inverse temperature $\beta_A=1$.
The along-boundary correlator $C(r_y)$ is measured within subsystem $A$ on the boundary-adjacent line, with separations taken along the periodic boundary direction as defined in the text.
(a) Heisenberg limit $J_r=1$.
(b) N\'eel phase $J_r=1.5$.
(c) QCP $J_r=1.90951(1)$.
(d) Dimer phase $J_r=3$.
} 
\label{fig:strong_rep_1}
\end{figure*}

With $v$ (and thus $\epsilon_{\mathrm{EH}}$) fixed, we next benchmark the LBW-EH ansatz by comparing equal-time correlation functions computed in the LBW-EH ensemble and in the exact-EH ensemble at the same effective inverse temperature $\beta_A=1$.
We use the along-boundary $\sigma^z$ correlator $C(r_y)$ defined in Eq.~\eqref{eq:cry_def}.
In the two-dimensional cylinder geometry, $C(r_y)$ is evaluated within subsystem $A$ on the boundary-adjacent line ($i_x=1$), with separations taken along the periodic boundary direction and averaged over boundary positions.
We focus on these along-boundary correlations because they carry the dominant boundary signal and provide the most direct and discriminating test of the LBW-EH ansatz in our geometry.

The correlation function results for the Heisenberg limit, N\'eel ordered phase, QCP, and the dimer phase are shown in Fig.~\ref{fig:strong_rep_1}. 
At the Heisenberg limit with $J_r=1$, the correlation function results of LBW-EH and exact-EH almost completely coincide. The two-dimensional dimerized Heisenberg model at this point actually possesses translational invariance, indicating that the LBW-EH approximation performs exceptionally well in such translationally invariant systems.

When the system is in the N\'eel order phase at $J_r=1.5$, the correlation function results of LBW-EH and exact-EH exhibit consistent trends, but there is a clear separation between the two sets of data. Note that the correlation functions are presented on a logarithmic scale. By examining the actual numerical values, at the distance $r_y=8$, which corresponds to the farthest point due to PBC in y-axis, the correlation function value for LBW-EH is $0.0997(1)$, while that for exact-EH is $0.1295(2)$. The absolute difference between the two is $0.0298(2)$. It must be acknowledged that there is a noticeable discrepancy between these values. Aside from the nearest-neighbor point, the performance of LBW-EH ansatz is not particularly strong in this regime.

A similar behavior is observed at the QCP where $J_r=1.90951(1)$. At this point, the absolute difference between the two correlation functions at the farthest distance $r_y=8$ is $0.0174(1)$, which is slightly smaller than the value in the N\'eel order phase.
Finally, when the system is in the dimer phase at $J_r=3$, the discrepancy still exists between the correlation function results of LBW-EH and exact-EH. However, the difference is now an order of magnitude smaller than those observed in the N\'eel order phase and at the QCP.
Aside from the fact that the velocity fitting from imaginary-time correlations is an approximation, which may introduce certain errors in the LBW-EH simulation, we must acknowledge that, except at the Heisenberg limit where the system is translationally invariant, the LBW-EH ansatz does not fully coincide with the exact-EH beyond the nearest-neighbor points when performing the bipartition at strong-bonds.

Moreover, we measure the correlation functions of LBW-EH and exact-EH for larger system sizes to observe whether these differences would disappear.
Across the N\'eel order phase, QCP, and dimer phases, the numerical discrepancy between the correlation functions of the LBW-EH and the exact-EH decreases as the system size increases.
Nevertheless, these discrepancies do not vanish completely. Within the system sizes we have investigated, we have not observed a scenario in which the correlation functions of LBW-EH and exact-EH fully coincide.
We therefore conclude that, within the system sizes investigated, the strong-bonds bipartition does not support the LBW-EH as an equally accurate functional form away from the translationally invariant case.
This conclusion remains unchanged when the system size is increased to $L=32$ (see Appendix~\ref{app:large_size} for the corresponding larger-size comparisons).

\begin{figure*}[ht!]
\centering
\begin{overpic}[width=0.45\linewidth]{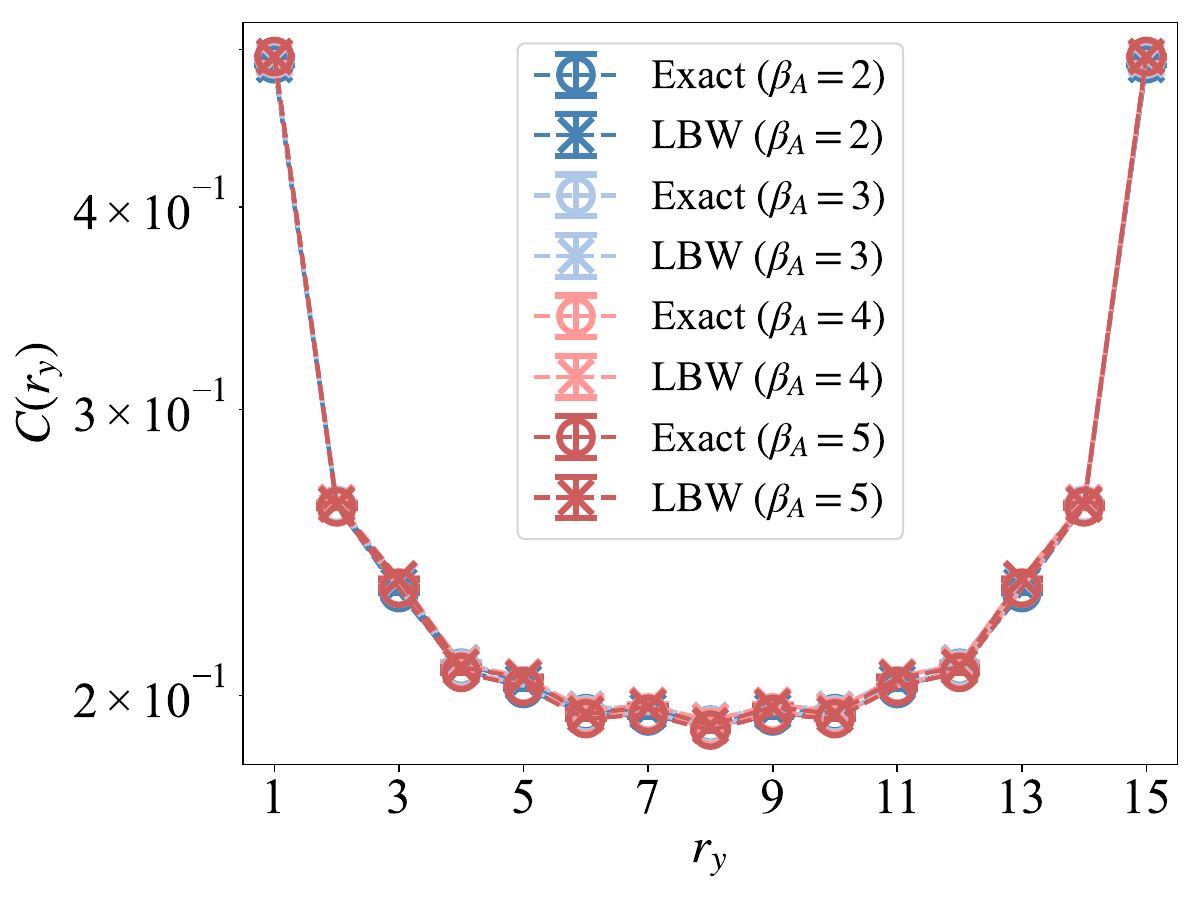}
\put (0,75) {{\textbf{(a)}}}
\end{overpic}
\begin{overpic}[width=0.45\linewidth]{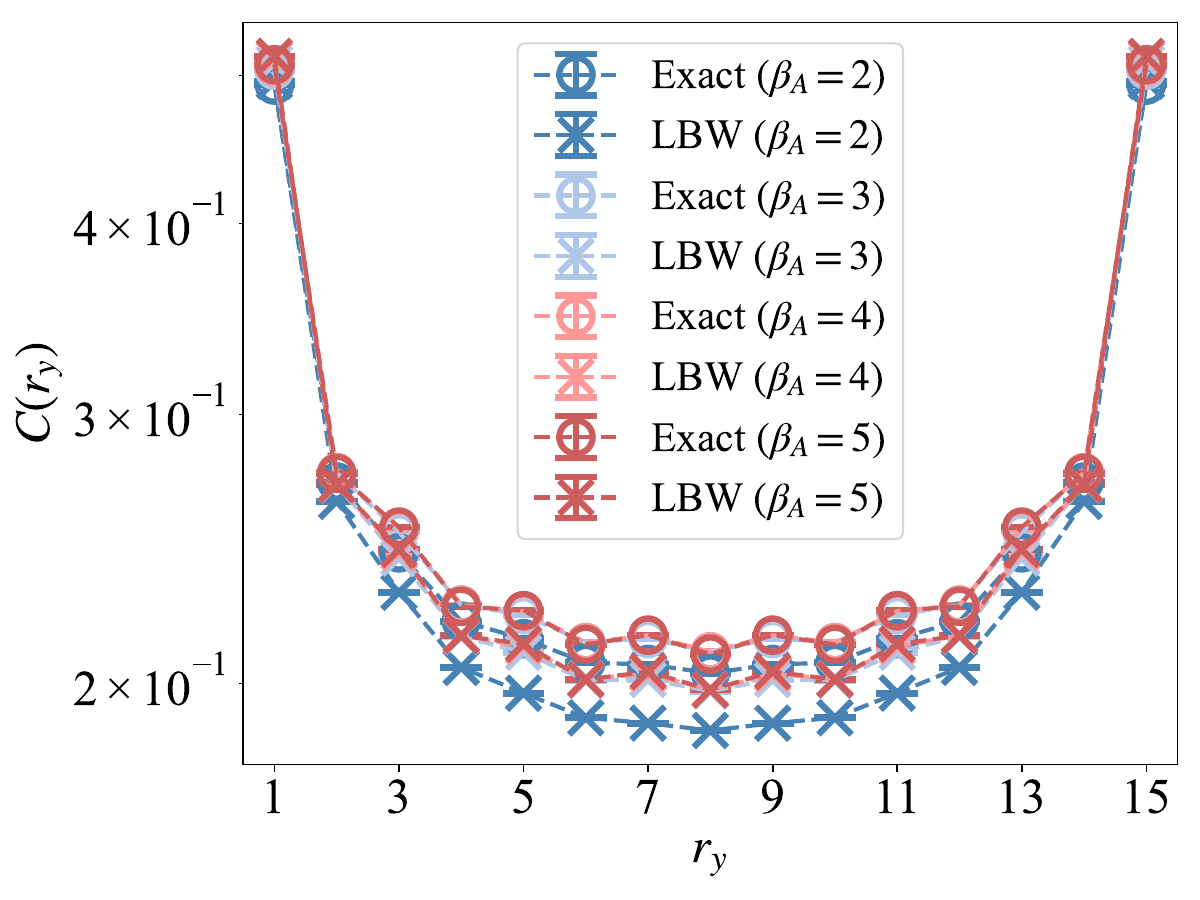}
\put (0,75) {{\textbf{(b)}}}
\end{overpic}
\begin{overpic}[width=0.45\linewidth]{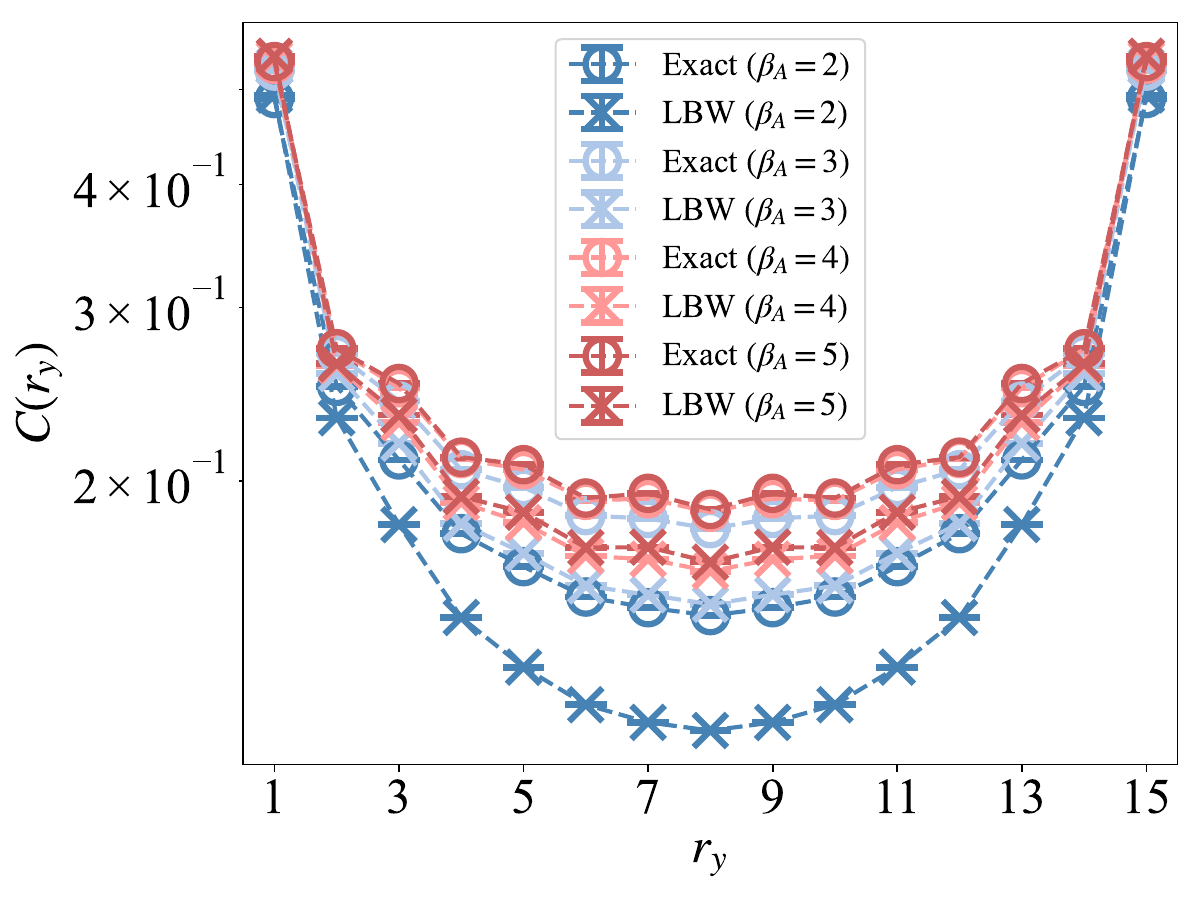}
\put (0,75) {{\textbf{(c)}}}
\end{overpic}
\begin{overpic}[width=0.45\linewidth]{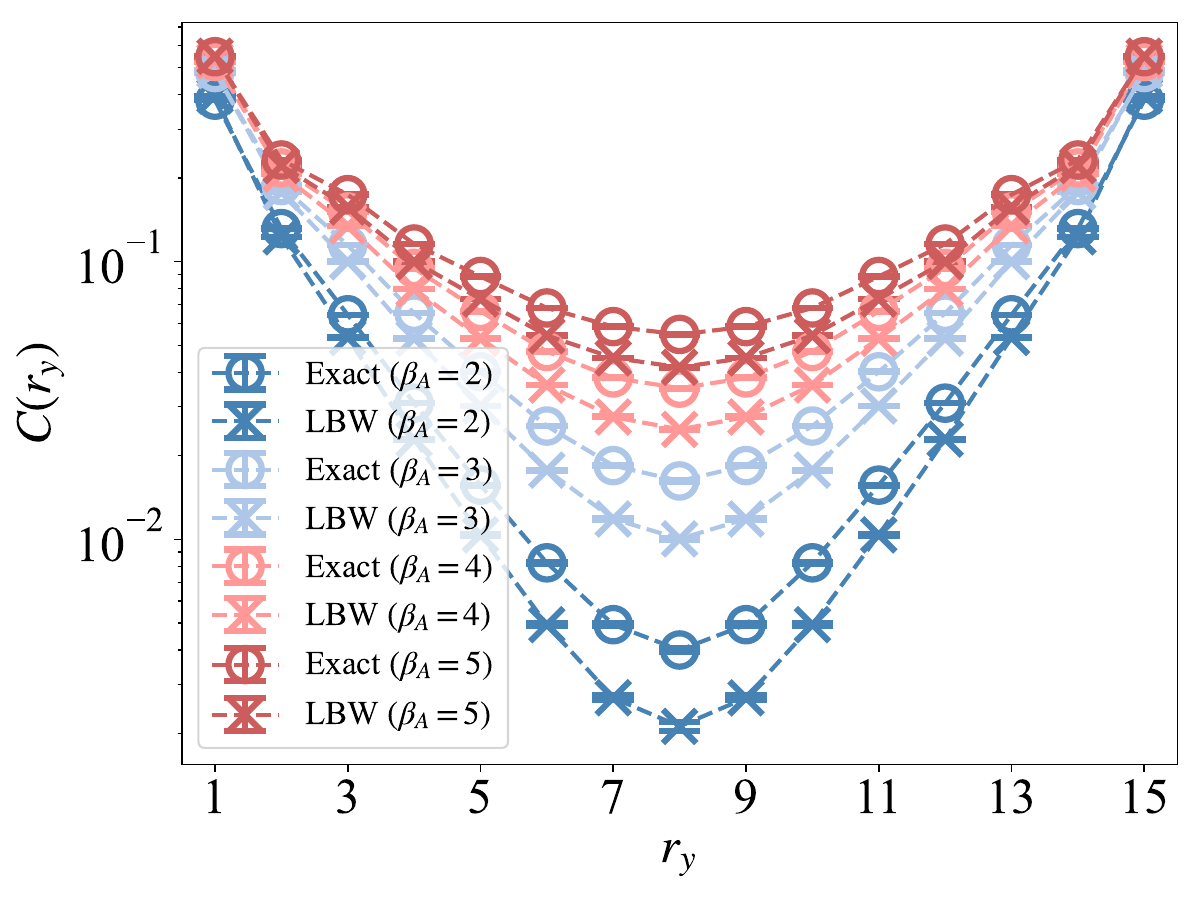}
\put (0,75) {{\textbf{(d)}}}
\end{overpic}
\caption{
Equal-time spin correlation functions of the two-dimensional dimerized Heisenberg model for the strong-bonds bipartition shown in Fig.~\ref{fig:J1J2_config}(a), comparing the $16\times16$ LBW-EH and the $32\times16$ exact-EH for larger effective inverse temperatures $\beta_A=2,3,4,5$.
The LBW-EH data are obtained by simulating the $16\times16$ ansatz Hamiltonian at inverse temperature $\beta_A$, while the exact-EH data are obtained using the multi-replica construction where $\beta_A=n$.
The along-boundary correlator $C(r_y)$ is measured within subsystem $A$ on the boundary-adjacent line, with separations taken along the periodic boundary direction as defined in the text.
(a) Heisenberg limit $J_r=1$.
(b) N\'eel phase $J_r=1.5$.
(c) QCP $J_r=1.90951(1)$.
(d) Dimer phase $J_r=3$.
}
\label{fig:strong_reps_n}
\end{figure*}

In the end, we investigate LBW-EH ansatz and exact-EH with larger effective inverse temperature $\beta_A$. 
For the QMC simulation, the effective inverse temperature $\beta_A$ is regarded as imaginary-time in the path integral representation, and the replica-trick QMC method should be applied where the effective inverse temperature $\beta_A$ is set equal to the number of replicas $n$. In each replica, the true Hamiltonian is simulated with the physical inverse temperature $\beta$, which scales proportional to the system size approaching the ground state of the real system.
The correlation functions for the Heisenberg limit, N\'eel ordered phase, QCP, and the dimer phase of LBW-EH and exact-EH are all shown in Fig.~\ref{fig:strong_reps_n}.
At the Heisenberg limit $J_r=1$, the correlation functions of LBW-EH and exact-EH coincide completely across different finite temperatures. However, in the N\'eel ordered phase, at the QCP, and in the dimer phase, discrepancies persist between the correlation functions of LBW-EH and exact-EH at various finite temperatures. 
When we perform the bipartition at strong-bonds in the two-dimensional dimerized Heisenberg model, even as the system approaches the ground state of the EH, the LBW-EH ansatz fails to provide a good functional form due to the absence of translational invariance.


\begin{figure*}[ht!]
\centering
\begin{overpic}[width=0.45\linewidth]{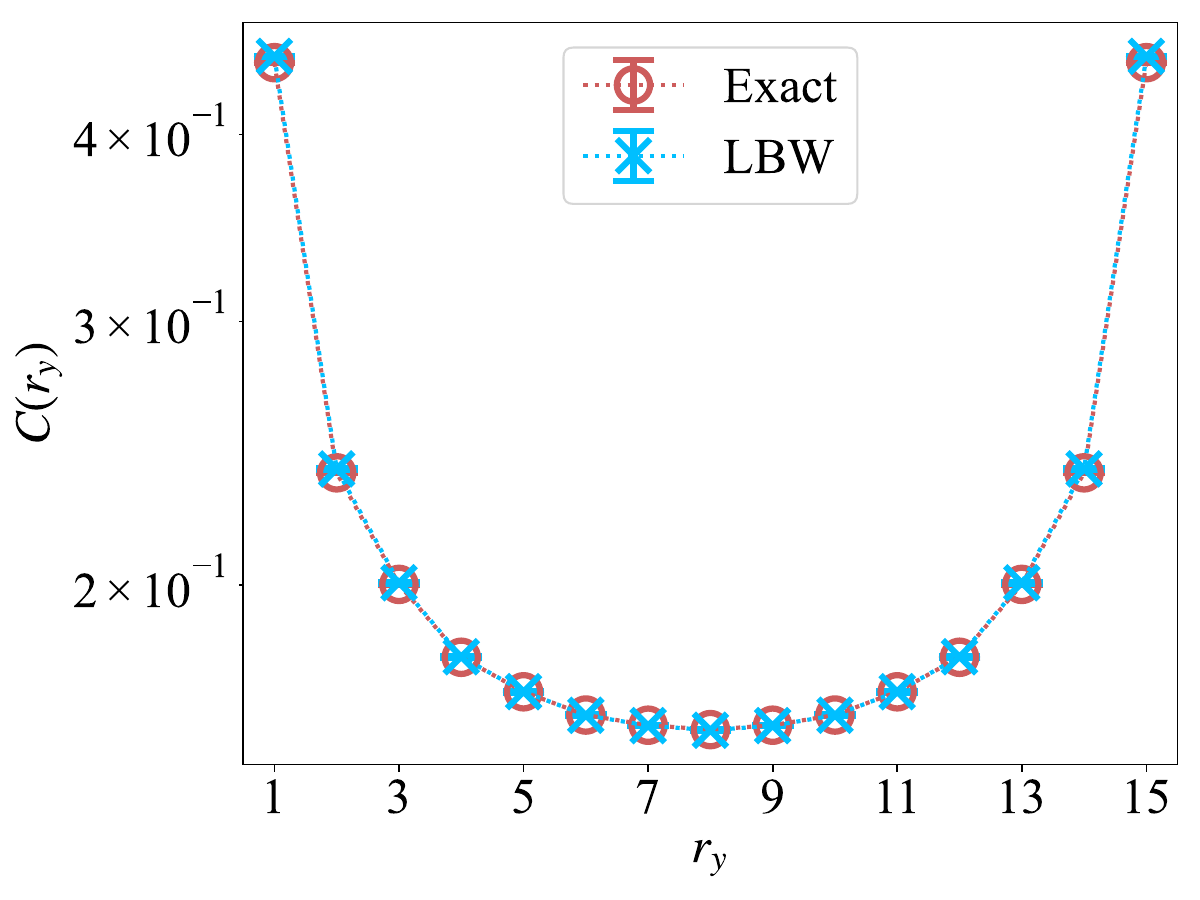}
\put (0,75) {{\textbf{(a)}}}
\end{overpic}
\begin{overpic}[width=0.45\linewidth]{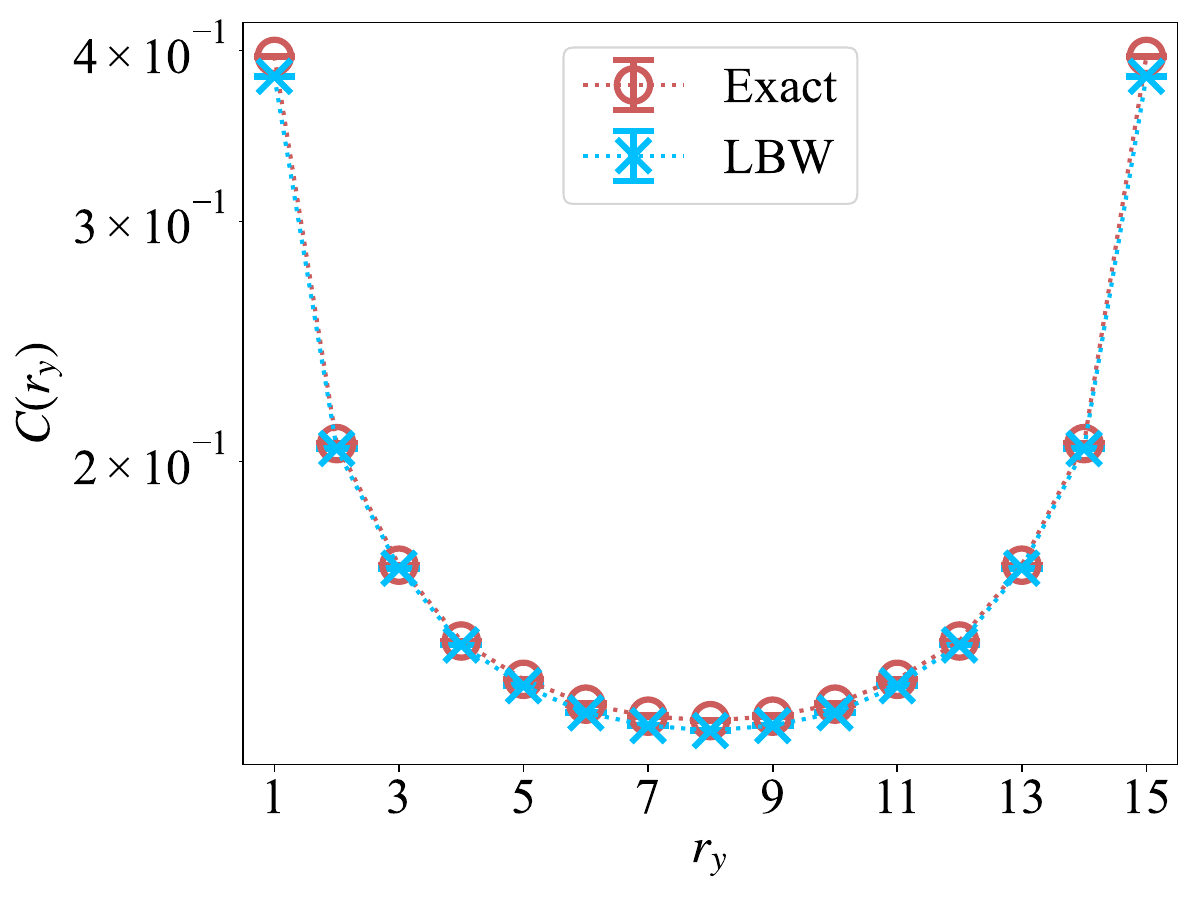}
\put (0,75) {{\textbf{(b)}}}
\end{overpic}
\begin{overpic}[width=0.45\linewidth]{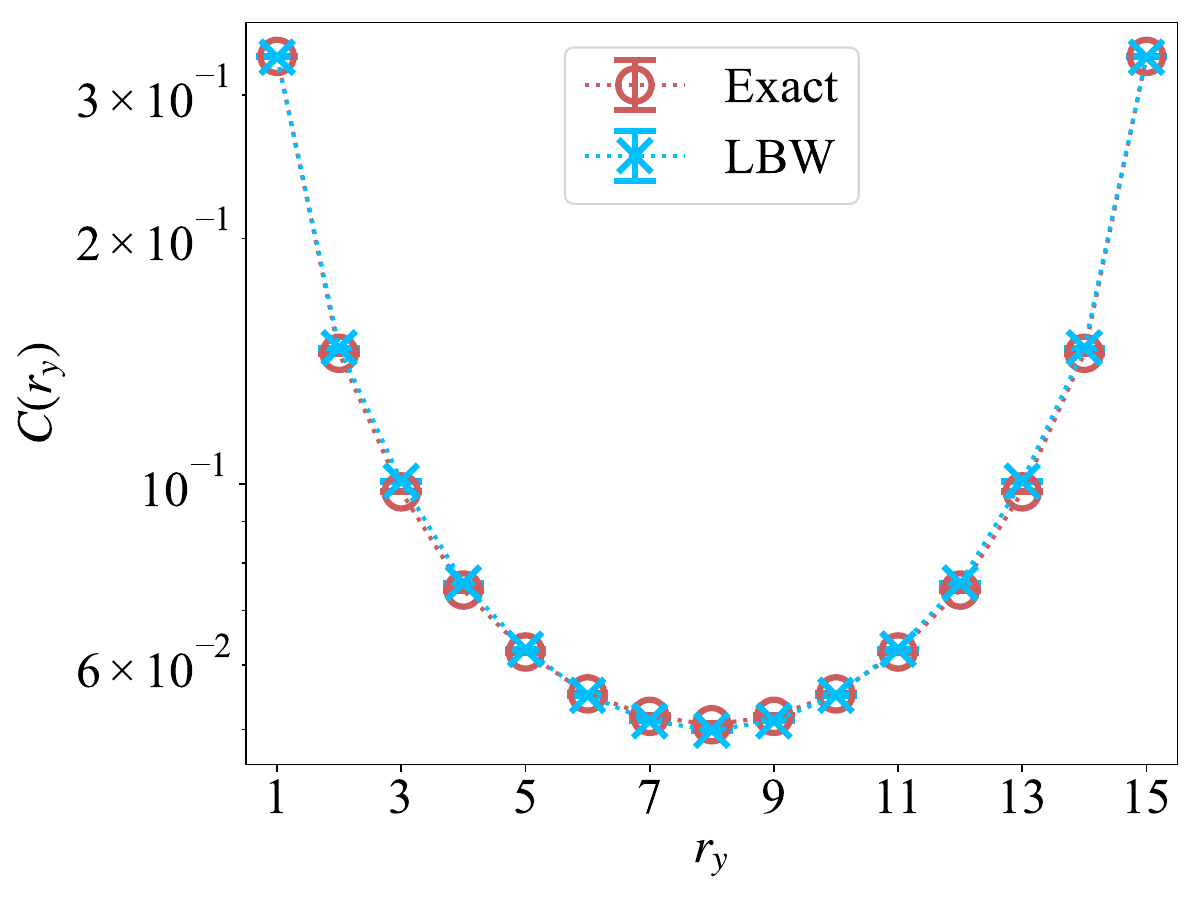}
\put (0,75) {{\textbf{(c)}}}
\end{overpic}
\begin{overpic}[width=0.45\linewidth]{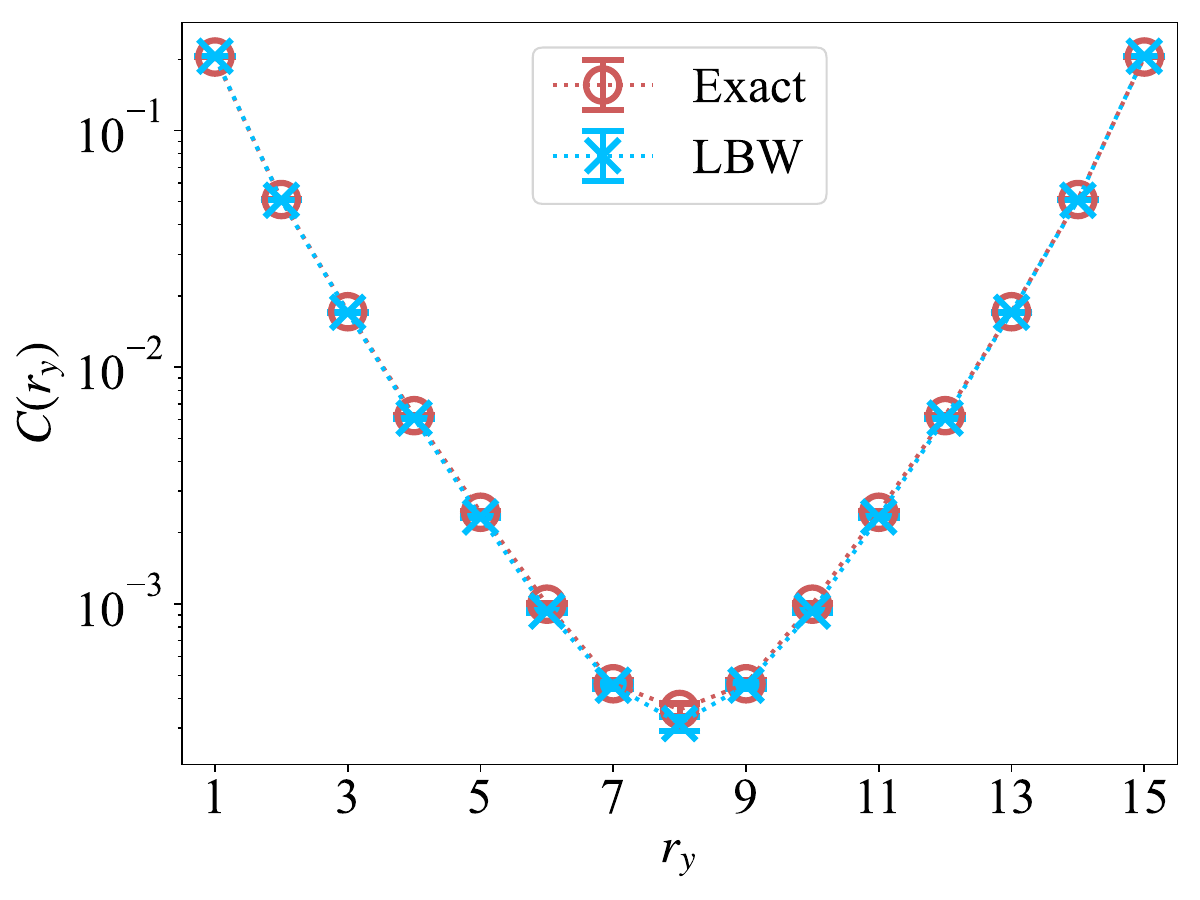}
\put (0,75) {{\textbf{(d)}}}
\end{overpic}
\caption{
Equal-time spin correlation functions of the two-dimensional dimerized Heisenberg model for the horizontal weak-bonds bipartition shown in Fig.~\ref{fig:J1J2_config}(b), comparing the $16\times16$ LBW-EH and the $32\times16$ exact-EH at effective inverse temperature $\beta_A=1$.
The along-boundary correlator $C(r_y)$ is measured within subsystem $A$ on the boundary-adjacent line, with separations taken along the periodic boundary direction as defined in the text.
(a) Heisenberg limit $J_r=1$.
(b) N\'eel phase $J_r=1.5$.
(c) QCP $J_r=1.90951(1)$.
(d) Dimer phase $J_r=3$.
} 
\label{fig:Weak_J1J2}
\end{figure*}

\subsection{The weak-bond case of bipartition}
We now discuss the second case of bipartition at the weak-bonds cutting horizontal weak-bonds, as illustrated in Fig.~\ref{fig:J1J2_config}(b).
The QMC simulations of LBW-EH and exact-EH with effective inverse temperature $\beta_A = 1$ are conducted for the Heisenberg limit, N\'eel ordered phase, QCP, and the dimer phase. The methodological details of the simulations are consistent with the previous descriptions. 

The correlation results in various phases and points are shown in Fig.~\ref{fig:Weak_J1J2}. 
Surprisingly, the correlation functions of the LBW-EH and the exact-EH almost completely coincide, whether in the Heisenberg limit or in the N\'eel ordered phase, at the QCP, and in the dimer phase.
Note that we use the logarithmic scale for the correlation functions, which magnifies any discrepancies.
The deviation at the nearest-neighbor point in the N\'eel ordered phase is attributed to the precision limit of the imaginary-time correlation fitting velocity. 
As for the most distant point at the distance $r_y=8$ in the dimer phase, the minor discrepancies arise from incomplete measurement accuracy. Nevertheless, the computational resources used here were smaller than those for the strong-bonds case, yet we obtained results with good data quality.
When the system is bipartitioned at the horizontal weak-bonds, we obtain results that differ from those at the horizontal strong-bonds case. For this weak-bonds bipartition in the two-dimensional dimerized Heisenberg model, within the system sizes investigated, the LBW-EH provides a good functional form across the Heisenberg limit, N\'eel ordered phase, QCP, and the dimer phase, despite the lack of translational invariance in the system.

\subsection{The vertical case of bipartition}
Compared with the weak-bond case, although the third bipartition is also performed at the weak-bonds, strong and weak-bonds alternate along the vertical direction and all bonds in the horizontal direction are weak in this case. The only feasible way to partition the system and the environment is by cutting the horizontal weak-bonds, as illustrated in Fig.~\ref{fig:J1J2_config}(c). 

\begin{figure*}[ht!]
\centering
\begin{overpic}[width=0.45\linewidth]{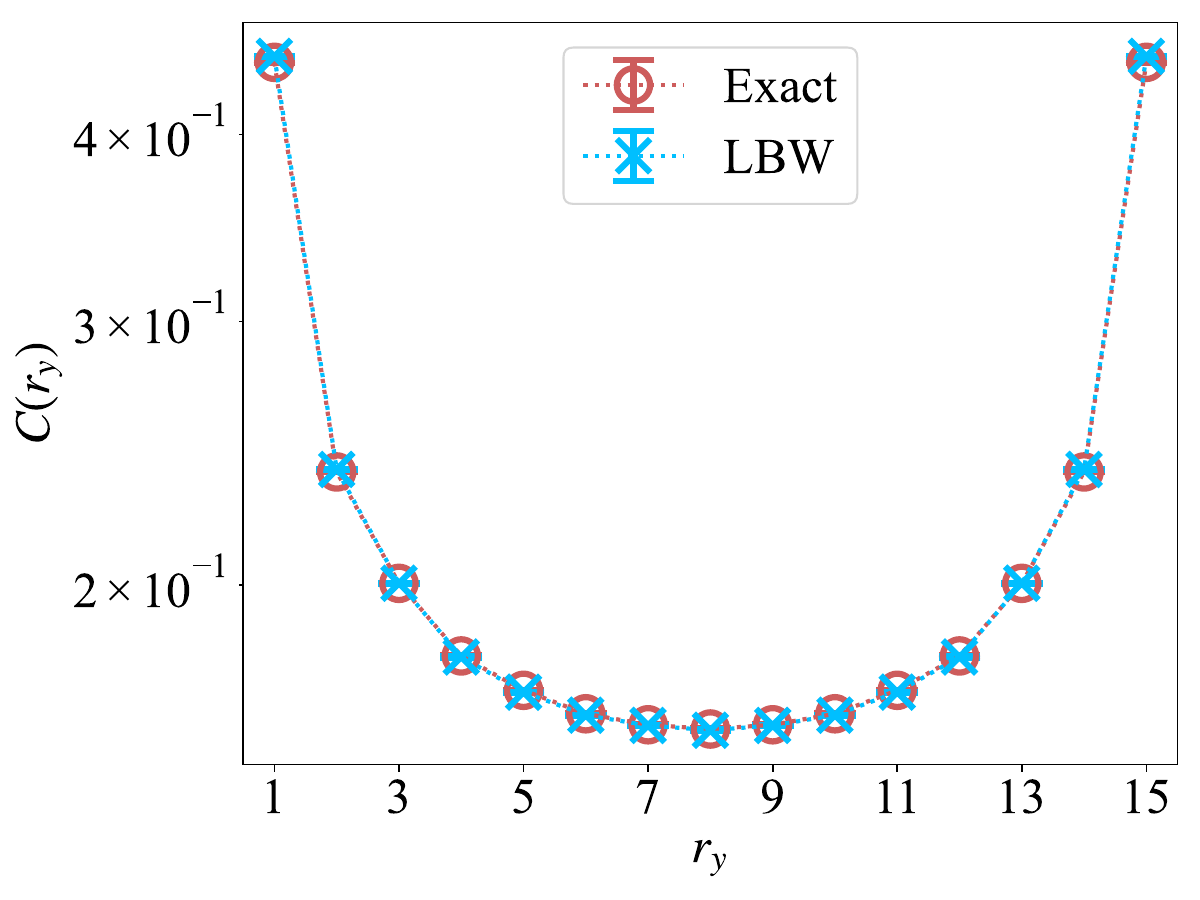}
\put (0,75) {{\textbf{(a)}}}
\end{overpic}
\begin{overpic}[width=0.45\linewidth]{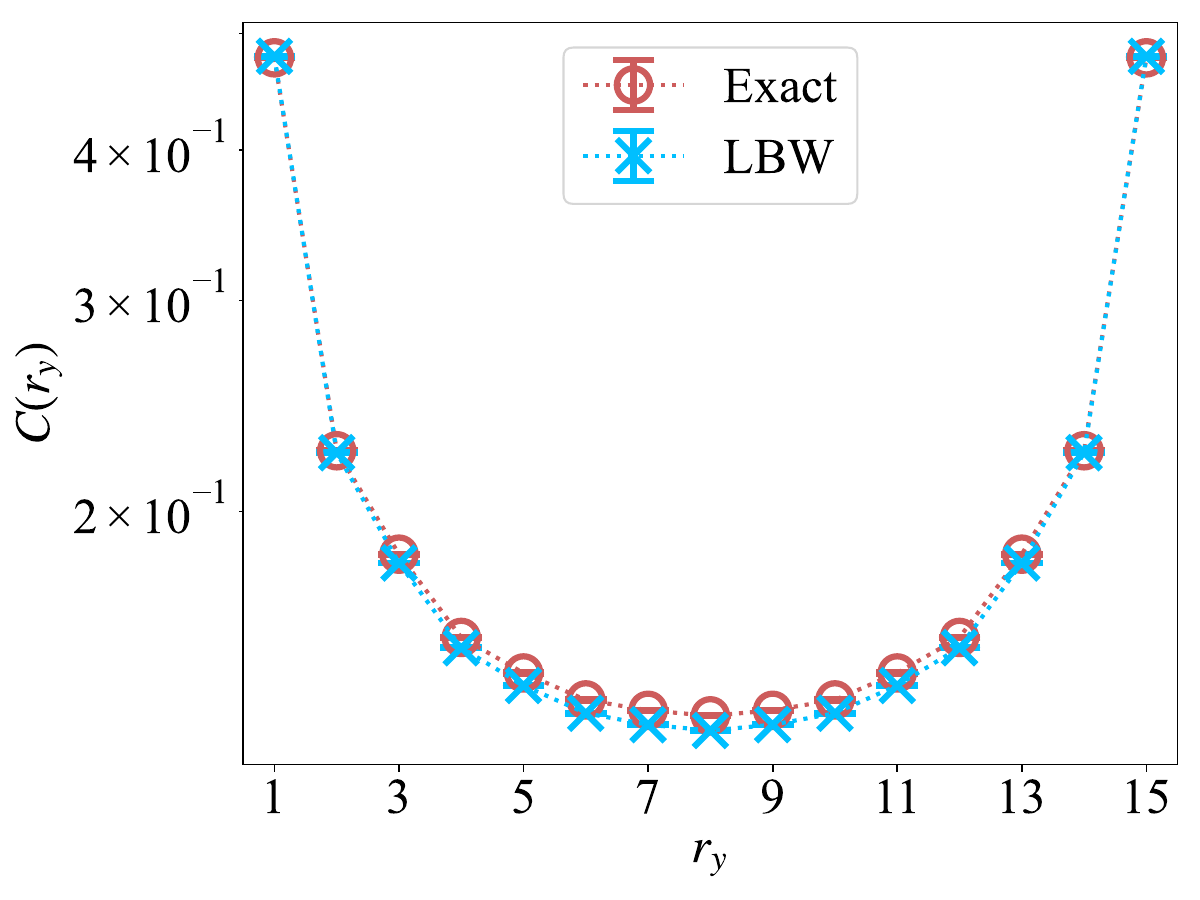}
\put (0,75) {{\textbf{(b)}}}
\end{overpic}
\begin{overpic}[width=0.45\linewidth]{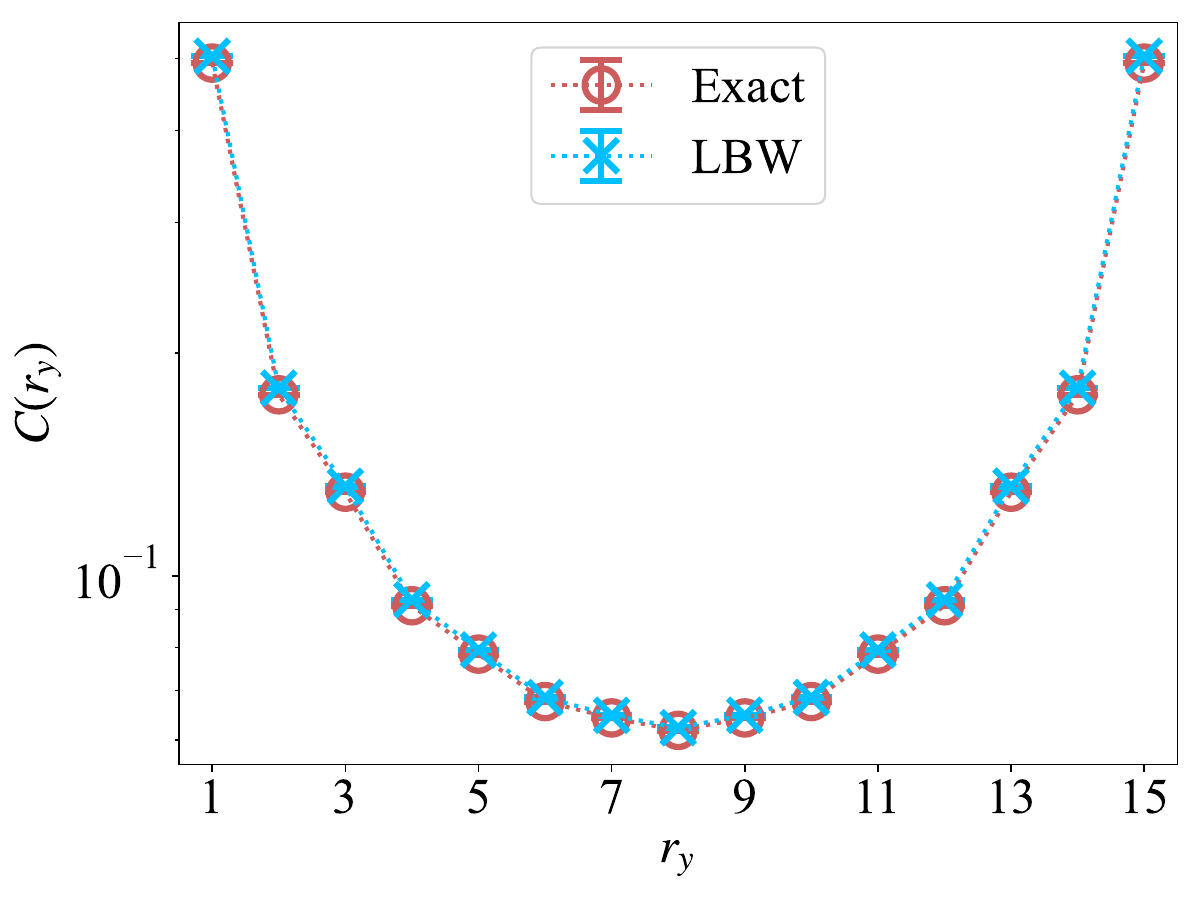}
\put (0,75) {{\textbf{(c)}}}
\end{overpic}
\begin{overpic}[width=0.45\linewidth]{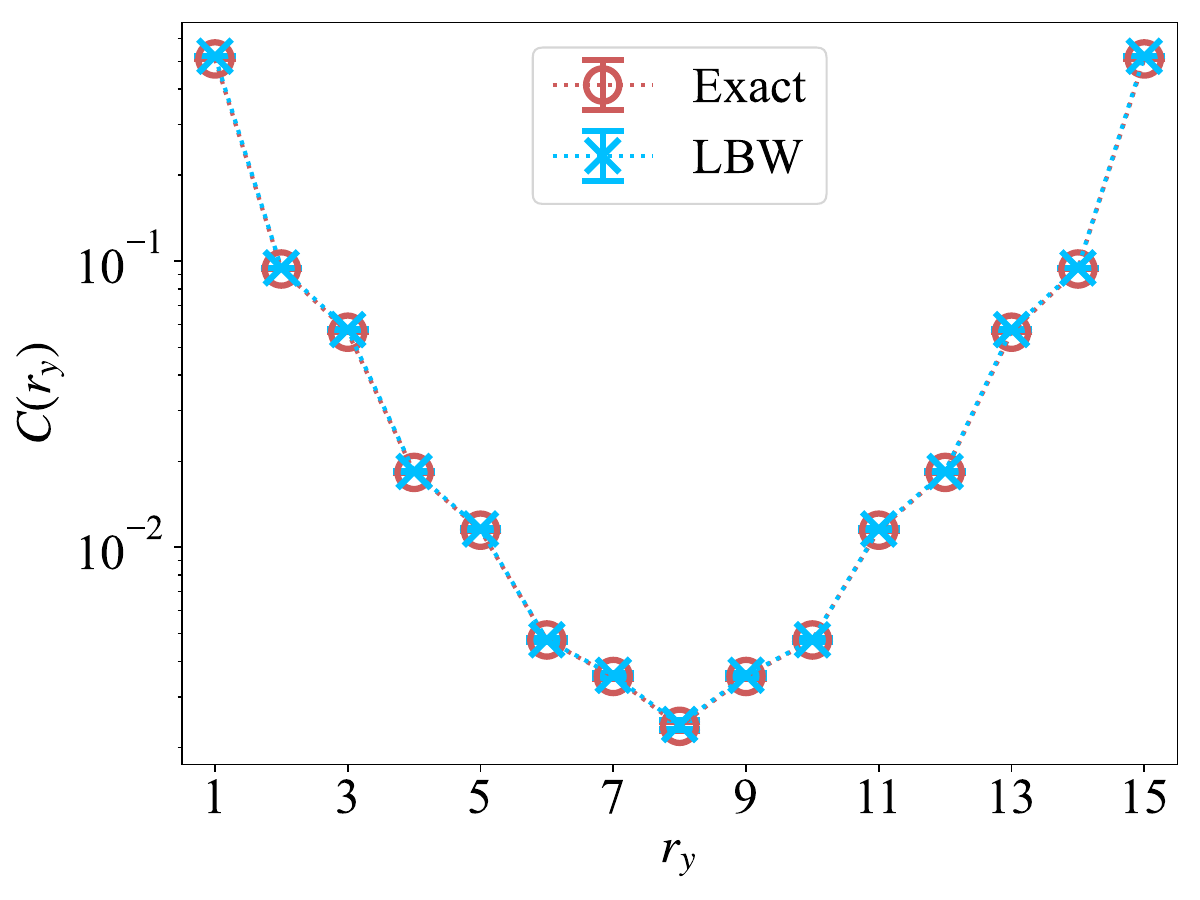}
\put (0,75) {{\textbf{(d)}}}
\end{overpic}
\caption{
Equal-time spin correlation functions of the two-dimensional dimerized Heisenberg model for the vertical weak-bonds bipartition shown in Fig.~\ref{fig:J1J2_config}(c), comparing the $16\times16$ LBW-EH and the $32\times16$ exact-EH at effective inverse temperature $\beta_A=1$.
The along-boundary correlator $C(r_y)$ is measured within subsystem $A$ on the boundary-adjacent line, with separations taken along the periodic boundary direction as defined in the text.
(a) Heisenberg limit $J_r=1$.
(b) N\'eel phase $J_r=1.5$.
(c) QCP $J_r=1.90951(1)$.
(d) Dimer phase $J_r=3$.
}
\label{fig:WeakVertical_J1J2}
\end{figure*}

To distinguish this case from the previous weak-bonds scenario, we refer to it as bipartitioning at the vertical weak-bonds. It is crucial to note that the cut is not literally made along the vertical bonds in this instance. 
Similarly, we perform QMC simulations for both LBW-EH and exact-EH at the effective inverse temperature $\beta_A = 1$, and measure the correlation functions at the Heisenberg limit, in the N\'eel ordered phase, at the QCP, and in the dimer phase. The correlation functions for various phases and points are shown in Fig.~\ref{fig:WeakVertical_J1J2}. 
The correlation functions of the LBW-EH and the exact-EH are in perfect agreement across the Heisenberg limit, N\'eel ordered phase, QCP, and the dimer phase, with a very minor discrepancy in N\'eel ordered phase when the distance $\mathbf{r}$ is large.
Consequently, for the two-dimensional dimerized Heisenberg model with the bipartition at vertical weak-bonds, within the system sizes investigated, the LBW-EH ansatz provides an excellent functional form across various phases and points, despite that the system lacks translational invariance.

To summarize, for two-dimensional dimerized Heisenberg model, we performed QMC simulations for LBW-EH and exact-EH with bipartitions at both strong and weak-bonds. 
The correlation functions lead to distinct conclusions, although the system lacks translational invariance in all these cases. 
This is because the system favors a dimerized state with entanglement dominated by strong-bonds when $J_r>1$. Thus, the bipartition at strong-bonds severely disturbs the original configuration by introduce an effective dangling spin chain on the edge, in contrast to the weak-bonds bipartition which minimizes this effect. The dangling spin chain provides a Lieb-Schultz-Mattis anomaly on the surface, which strongly modify the entanglement property.
Using the language in the field of surface criticality, the edge without extra gapless boundary mode is ordinary. An ordinary surface criticality purely reflects the information of bulk criticality. In our case, cutting weak-bonds is an ordinary splitting.
Therefore, we conclude that for the ordinary cut, the LBW-EH can provide a reliable functional form, even in the absence of Lorentz invariance.
The larger-size data in Appendix~\ref{app:large_size} further support this interpretation by showing that the strong-bond cut continues to deviate from the exact-EH, whereas the two weak-bond cuts remain in near-perfect agreement at $L=32$.


\section{Discussion and Conclusion}\label{sec:con}
At the methodological level, we have proposed a systematic approach to explore the applicability of the LBW-EH approximation in two-dimensional quantum many-body systems. This scheme involves determining the key parameter $\epsilon_{\text{EH}}$ in the LBW-EH form through fitting the sound velocity $v$ using imaginary-time correlation methods. We then employ QMC methods to simulate both LBW-EH and exact-EH, and evaluate the quality of the LBW-EH approximation by comparing their respective correlation functions. Importantly, our approach allows us to investigate the LBW-EH approximation at various effective temperatures of EH, extending beyond previous studies that were limited to comparisons at a single finite temperature that the effective inverse temperature $\beta_A=1$.

For demonstration, we first consider the two-dimensional TFIM as a representative of translationally invariant systems within the requirement of the LBW-EH approximation. Notably, in both the FM and PM gapped phases and at the QCP, we find that the LBW-EH ansatz as well as our approach performs well.

Next, we explore the two-dimensional dimerized Heisenberg model. This model has no translational invariance when the coupling ratio $J_r \ne 1$.
Moreover, under half-space bipartition, this model has three distinct ways to separate the system from the environment, which correspond to the ordinary/special surface at the bulk critical point~\cite{Ding2018,zhu2025bipartite}. 
For cuts along strong horizontal bonds, except at the isotropic limit, there are discrepancies between the correlation functions of LBW-EH and exact-EH across different phases and critical points. 
Even when increasing the effective inverse temperature to approach the ground state, the results do not fully coincide. 
However, for cuts along weak horizontal bonds and weak vertical bonds, we observe nearly perfect agreement between the correlation functions of LBW-EH and exact-EH in all phases, indicating an excellent approximation by LBW-EH. 

At the physical level, in our understanding, it is because the cut along weak-bonds gives out an ordinary boundary while cutting strong-bonds introduces a dangling spin chain with Lieb-Schultz-Mattis anomaly, which contributes an extra gapless edge mode in the entanglement Hamiltonian.
Here we conclude that the LBW form provides a good description of the entanglement Hamiltonian when the edge is ordinary (cutting weak-bonds only in our case), even the system loses the Lorentz-invariance. 
In the previous studies of the surface criticality~\cite{Ding2018,zhang2017unconventional,Binder1983phase,binder1990critical,wang2024surface}, the authors found that only the ordinary cut purely reflects the bulk criticality on the surface, otherwise, the extra gapless edge mode would also affect the critical behaviors on the surface. We think similar physics can also happen in the entanglement boundary instead of a real physical edge.
Another numerical evidence is that the EE behaviors in the columnar dimerized Heisenberg model are consistent with field theory only when the entanglement boundary avoids all the dimers (i.e., an ordinary boundary)~\cite{zhao2022scaling}. Similar case also happens in the scaling behaviors of disorder operators (a nonlocal measure similar to entanglement entropy) in the columnar dimerized Heisenberg model~\cite{wang2022scaling}. 
As the Ref.~\cite{liu2024measuring} shows, the behaviors of the disorder operator on the edge of a special surface criticality reflect the information containing the bulk (2+1) D $O(3)$ criticality and the gapless Luttinger liquid on the boundary. 

In addition to pointing out that LBW can be extended to non-Lorentz invariant situations, but the extra boundary effect needs to be avoided,
importantly, we note that our approach for fitting and studying the EH ansatz is not limited to the LBW-EH discussed in this paper. 
It would be interesting to apply the same scheme to other EH ansatz forms and to more general subregion geometries, which constitutes a natural next step toward a more complete and geometry-resolved characterization of entanglement Hamiltonians.
The scheme opens an access to obtain the full information of EH.

\section{Acknowledgement} 
We thank the helpful discussions with Marcello Dalmonte and Bin-Bin Mao. 
The work is supported by the Scientific Research Project (No.WU2024B027) and the Start-up Funding of Westlake University. 
The authors thank the high-performance computing centers of Westlake University and the Beijing PARATERA Tech Co., Ltd. for providing HPC resources.

\bibliographystyle{apsrev4-1}
\bibliography{ref}

\clearpage
\onecolumngrid
\appendix
\section{Robustness of the correlator-based scale fitting}
\label{app:robustness}

The correlator-based extraction of the overall LBW scale relies on the existence of a large-$\tau$ regime in which the momentum-resolved boundary correlator $C_k(\tau)$ is dominated by a single low-energy mode, so that $C_k(\tau)$ is approximately linear in $\tau$. 
In our analysis, for each dataset and each system size, we therefore first identify the candidate large-$\tau$ linear regime individually and then test the stability of the extracted slope by shifting, shrinking, or modestly expanding the fitting window within that regime. 
Whenever such a regime is clearly identifiable, we find that the extracted velocity remains stable within statistical uncertainty.

For the two-dimensional TFIM, we focus the robustness test on the gapless QCP, where a clearer large-$\tau$ linear regime is present. 
Using the fit shown in Fig.~\ref{fig:TFIM_QCP_velocity}, the original fit to the exact-EH correlator gives a slope of $-1.924(3)$, corresponding to $v=3.24(3)$. 
If we shrink the fitting window to the two most linear $\tau$ points, the slope becomes $-1.915(8)$ and the resulting velocity is $v=3.25(4)$. 
These two estimates agree within statistical uncertainty, showing that once the fit is restricted to the asymptotic $\tau$ linear regime, the extracted velocity is stable.
Both the FM and PM phases are gapped, and the imaginary-time correlator decays rapidly. 
In these cases, the available linear regime can be very short, which limits the quantitative precision of the extracted scale. 
We therefore regard the fitted scale in the gapped phases as an approximate estimate. 
Nevertheless, as shown by the subsequent equal-time correlation comparisons in the main text, the resulting LBW-EH still reproduces the overall behavior of the exact-EH correlator reasonably well, providing a meaningful consistency check of the extracted scale. 

For the two-dimensional dimerized Heisenberg model, we perform the same stability test across different phases. 
At the Heisenberg limit as shown in Fig.~\ref{fig:Strong_J1J2_Jr_1_velocity}, the original fit yields a slope of $-0.9764(6)$ for the LBW-EH correlator and $-0.5248(2)$ for the exact-EH correlator, resulting in $v=1.860(1)$. 
If we shrink the fitting window, fitting points 3--5 for the LBW-EH correlator and points 3--8 for the exact-EH correlator yields slopes of $-0.9752(4)$ and $-0.5251(2)$, respectively, giving $v=1.858(1)$. 
If we instead modestly expand the fitting window, fitting points 3--7 for the LBW correlator and points 3--12 for the exact-EH correlator yields slopes of $-0.9767(5)$ and $-0.5248(2)$, respectively, resulting in $v=1.861(1)$. 
These values agree within statistical uncertainty. 
We have carried out analogous checks in the N\'eel phase at $J_r=1.5$, at the QCP $J_r=1.90951(1)$, and in the dimer phase at $J_r=3$, obtaining stable velocities $v=2.337(1)$, $v=2.697(1)$, and $v=4.111(5)$, respectively. 
As long as the fit is restricted to the large-$\tau$ linear regime, the extracted velocity remains robust within statistical uncertainties.

Regarding the dependence on the momentum channel, our fitting is always performed after Fourier transforming the boundary-adjacent correlator along the periodic boundary direction, so that the scale is extracted in a fixed boundary momentum sector $k$. 
The physically relevant choice of $k$ is model dependent and is guided by which channel carries the dominant low-energy spectral weight or exhibits the cleanest single-exponential decay in imaginary time. 
For the two-dimensional TFIM, the uniform channel $k=0$ provides the clearest signal because the dominant low-energy fluctuations are uniform along the boundary. 
For the two-dimensional dimerized Heisenberg model, the staggered channel $k=\pi$ is the appropriate choice, since the dominant spin correlations along the boundary are staggered. 
We have verified numerically that these channels provide the most stable large-$\tau$ fitting behavior for the corresponding models.

We note, however, that the existence of a clear single-exponential regime—manifested as an approximately linear window in $\ln C_k(\tau)$—is not guaranteed for all models or parameter regimes. 
Our procedure should therefore be viewed as a controlled approximation aimed at matching the exact EH with a physically motivated ansatz. 
Since the theoretical derivation is based on the large-$\tau$ limit, we always prioritize fitting within the asymptotic linear regime whenever such a regime can be identified. 
Even when the asymptotic regime is not sharply developed, the fitting still yields an effective decay scale that provides a meaningful quantitative estimate. 
Moreover, the stability of the fitted results under reasonable variations of the fitting window shows that, whenever a linear regime is present, the extracted velocity, and hence the overall LBW scale $\epsilon_{EH}$, remains robust within statistical uncertainties.

\section{Larger-size finite-size check}
\label{app:large_size}

In Sec.~\ref{sec:dhm}, the main nontrivial conclusion of this work is established in the two-dimensional dimerized Heisenberg model, where translational invariance is absent and the performance of the LBW-EH ansatz depends sensitively on the cut geometry. 
The three inequivalent half-space bipartitions are illustrated in Fig.~\ref{fig:J1J2_config}, and the main-text results show that the strong-bond cut shown in Fig.~\ref{fig:J1J2_config}(a) behaves qualitatively differently from the two weak-bond cuts shown in Fig.~\ref{fig:J1J2_config}(b) and (c). 
For this reason, the most meaningful finite-size test is to revisit this model at larger system size and check whether the same cut-dependent conclusions remain valid. 
The purpose of this appendix is therefore not to repeat the theoretical setup already given in Sec.~\ref{sec:dhm}, but to provide additional large-size evidence showing that the conclusions reported at $L=16$ remain stable when $L$ is increased.


As discussed in Sec.~\ref{sec:dhm}, the strong-bond bipartition requires fixing the overall LBW scale through the imaginary-time fitting procedure of Sec.~\ref{sec:fit}. 
At $L=16$, this was done by measuring the boundary-adjacent correlator, Fourier transforming it along the periodic boundary direction in the staggered momentum channel $k=\pi$, and fitting the large-$\tau$ linear regime of $C_k(\tau)$ for the strong-bond geometry shown in Fig.~\ref{fig:J1J2_config}(a). 
The resulting velocities for the Heisenberg limit, N\'eel phase, QCP, and dimer phase were summarized in Table~\ref{table:J1J2_velocities}. 
To examine the size dependence of this scale extraction, we repeat the same procedure at $L=32$. 
The LBW-EH is simulated on a $32\times32$ subsystem, while the exact-EH is obtained from the corresponding $64\times32$ bipartitioned system. 
As in the main text, the correlator is measured along the boundary-adjacent line inside subsystem $A$, Fourier transformed along the periodic direction in the staggered channel $k=\pi$, and fitted in the large-$\tau$ linear regime. 
The fitted velocities are collected in Table~\ref{table:J1J2_velocities_L32_appendixB}.
\begin{table}[htbp]
  \centering
  \caption{Fitted velocities $v$ of the two-dimensional dimerized Heisenberg model at $L=32$ for the strong-bond bipartition.}
  \label{table:J1J2_velocities_L32_appendixB}
  \begin{tabular}{ccc}
    \toprule
    Phase & $J_r$ & $v$ \\
    \midrule
    Heisenberg limit & $1$ & $1.9063(4)$ \\
    N\'eel ordered phase & $1.5$ & $2.377(1)$ \\
    QCP & $1.90951(1)$ & $2.675(2)$ \\
    Dimer phase & $3$ & $4.03(1)$ \\
    \bottomrule
  \end{tabular}
\end{table}

Compared with the $L=16$ values reported in Table~\ref{table:J1J2_velocities}, these $L=32$ results remain quantitatively close across all four representative points. 
The same phase dependence of the fitted scale is preserved from the Heisenberg limit to the dimer phase, which shows that the overall LBW scale used in the strong-bond analysis is not an accidental consequence of the smaller size. 
This larger-$L$ check therefore supports the stability of the velocity extraction employed in the main-text comparisons.

\vspace{0.1cm}
\begin{figure}[ht!]
\centering
\begin{overpic}[width=0.45\linewidth]{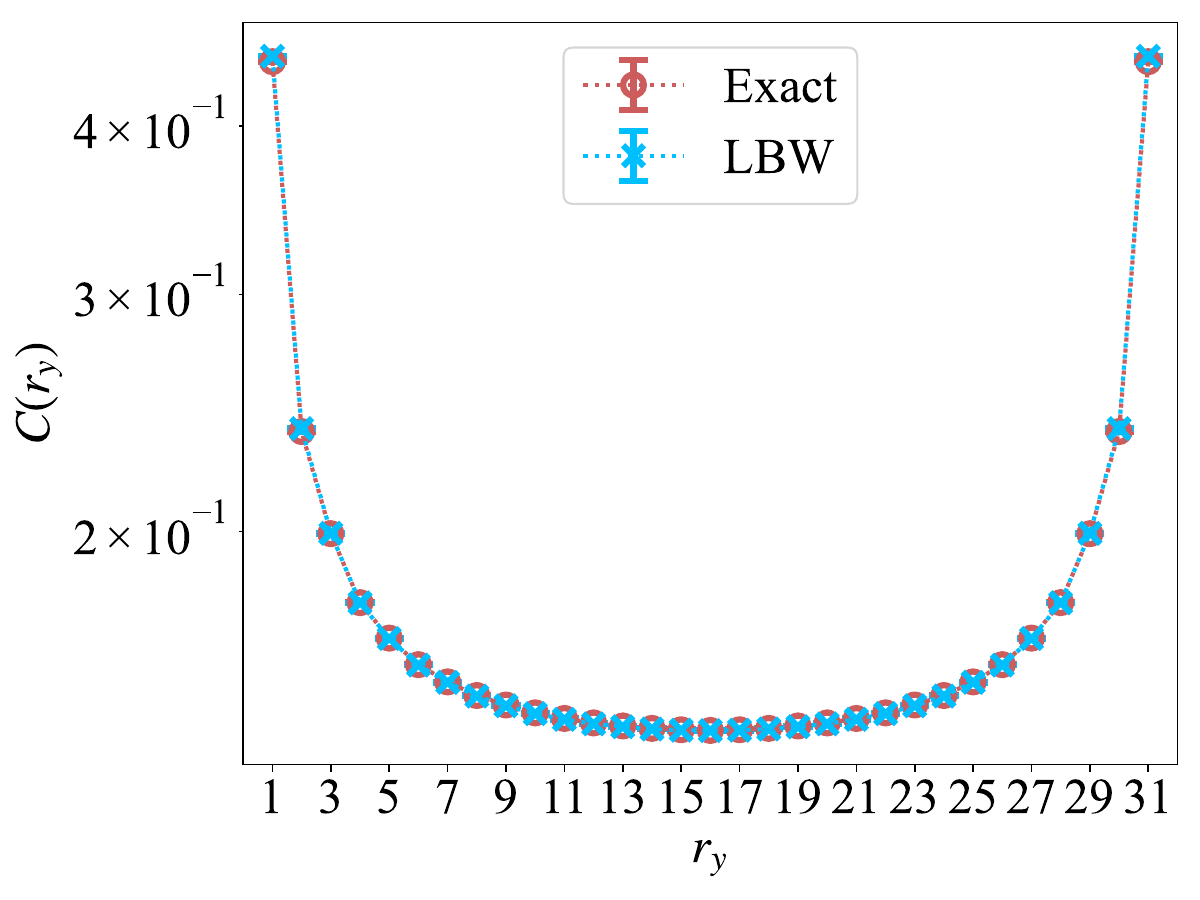}
\put (0,75) {{\textbf{(a)}}}
\end{overpic}
\begin{overpic}[width=0.45\linewidth]{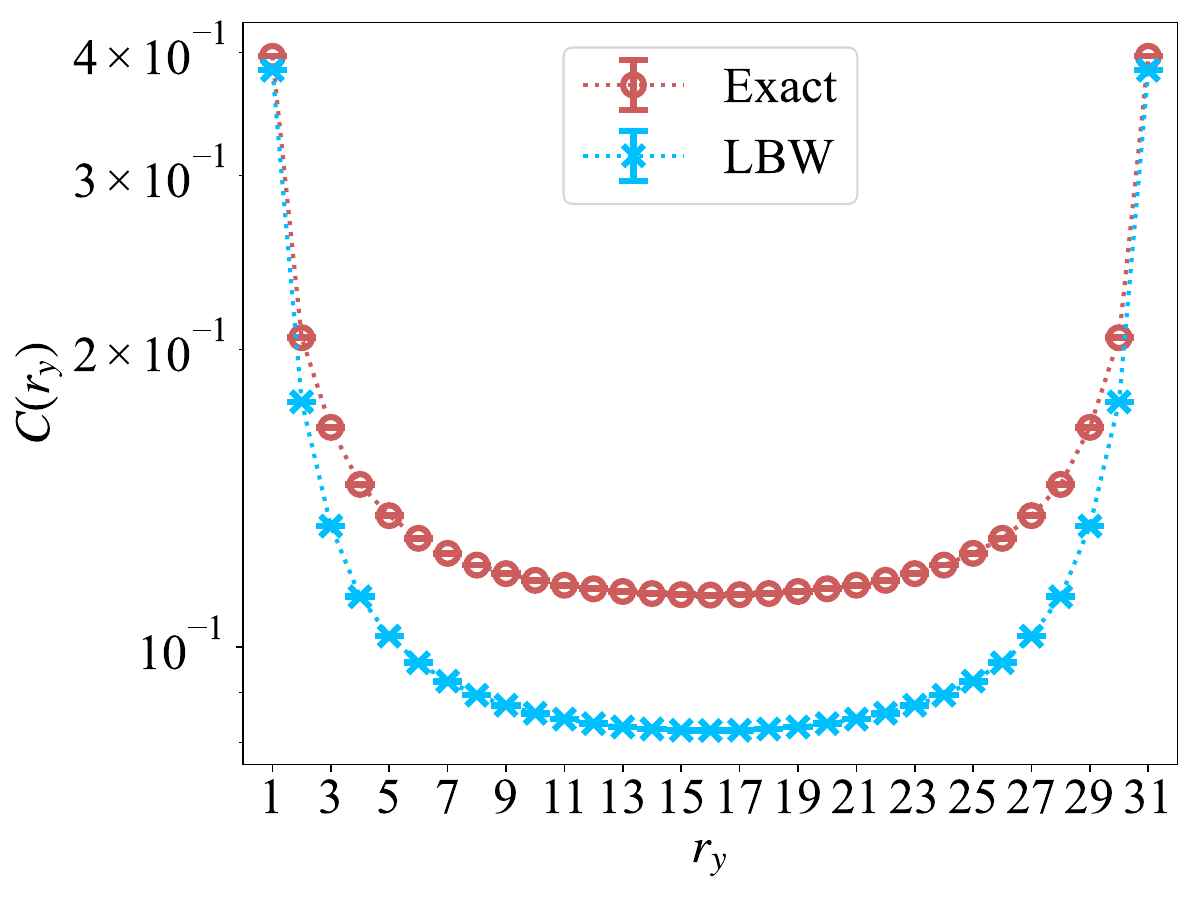}
\put (0,75) {{\textbf{(b)}}}
\end{overpic}
\begin{overpic}[width=0.45\linewidth]{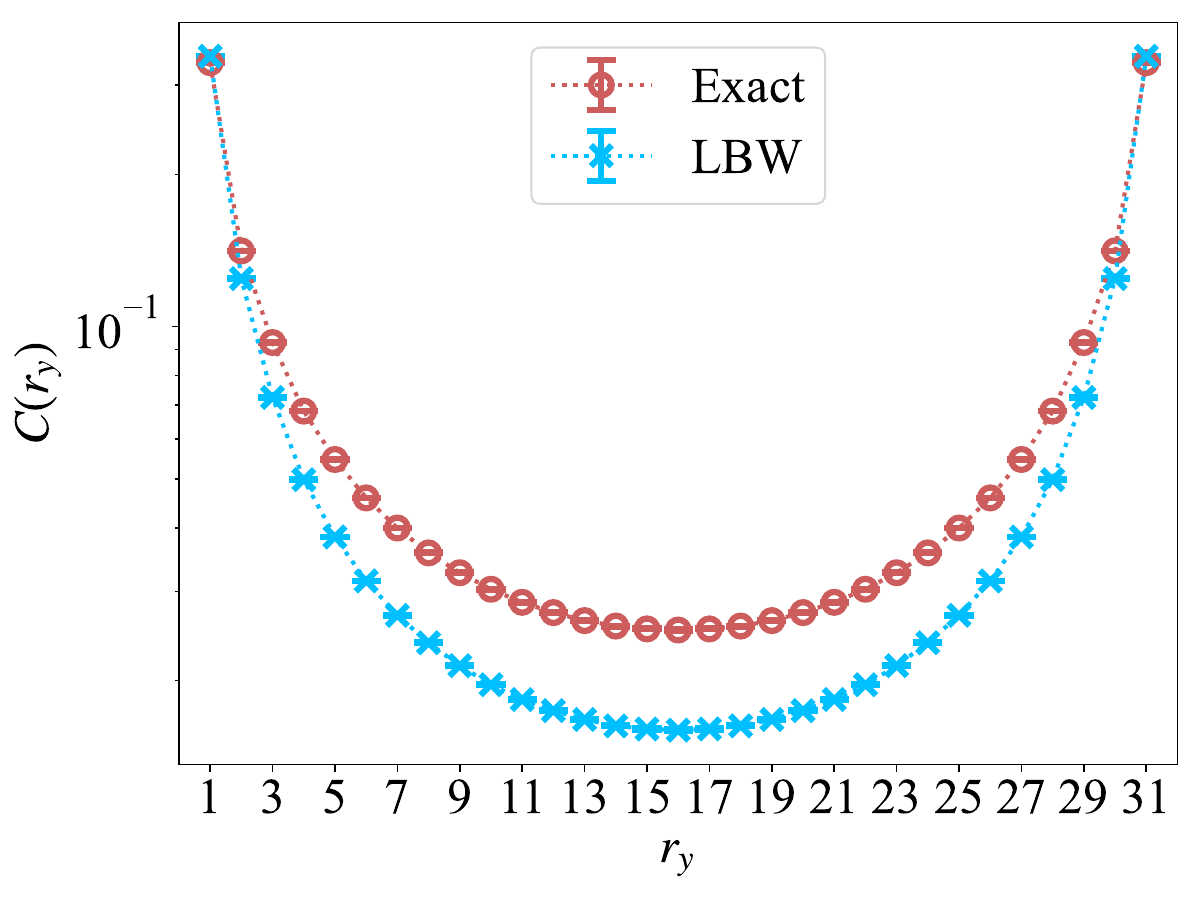}
\put (0,75) {{\textbf{(c)}}}
\end{overpic}
\begin{overpic}[width=0.45\linewidth]{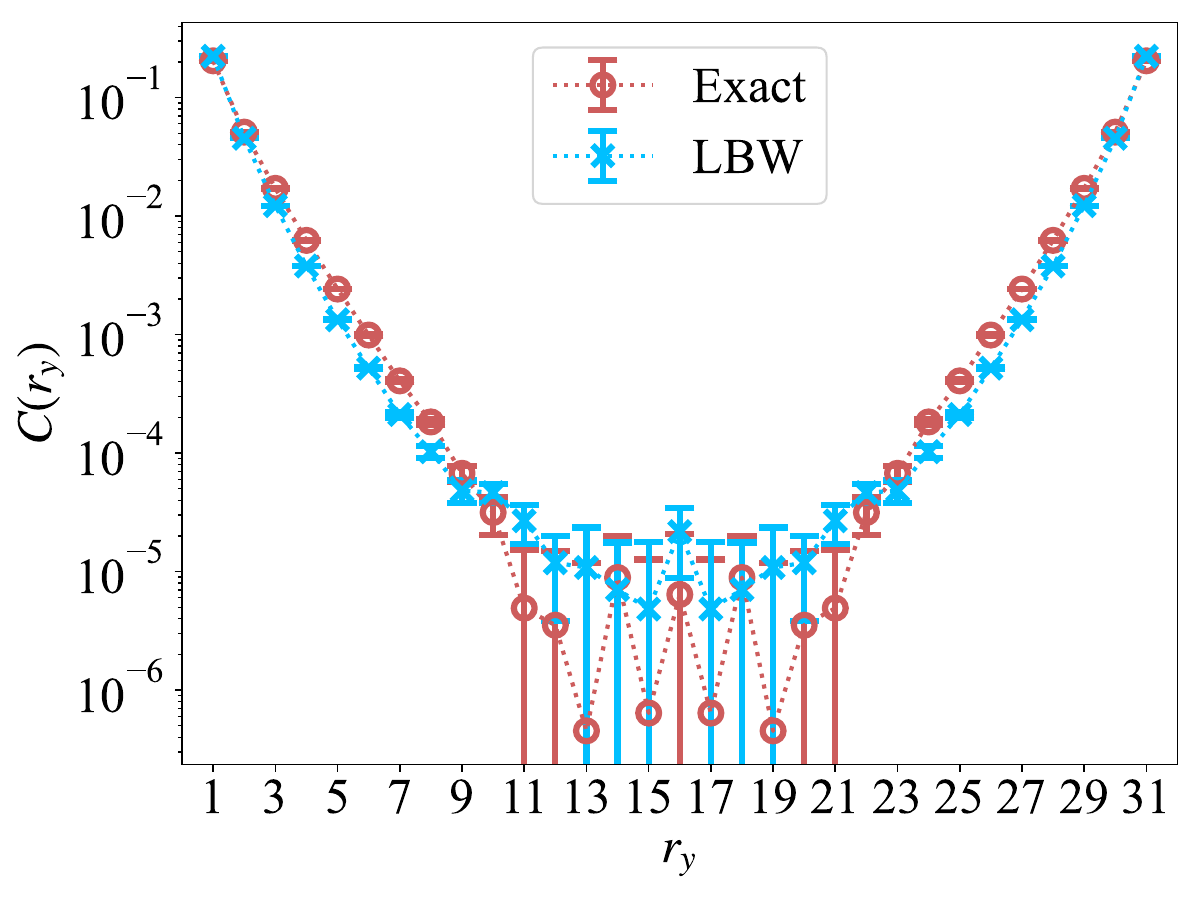}
\put (0,75) {{\textbf{(d)}}}
\end{overpic}
\caption{
Equal-time along-boundary correlators $C(r_y)$ at $L=32$ for the strong-bond bipartition shown in Fig.~\ref{fig:J1J2_config}(a), comparing the $32\times32$ LBW-EH and the $64\times32$ exact-EH at effective inverse temperature $\beta_A=1$. 
(a) Heisenberg limit $J_r=1$. 
(b) N\'eel phase $J_r=1.5$. 
(c) QCP $J_r=1.90951(1)$. 
(d) Dimer phase $J_r=3$.
}
\label{fig:L32_strong_appendixB}
\end{figure}

\begin{figure}[ht!]
\centering
\begin{overpic}[width=0.45\linewidth]{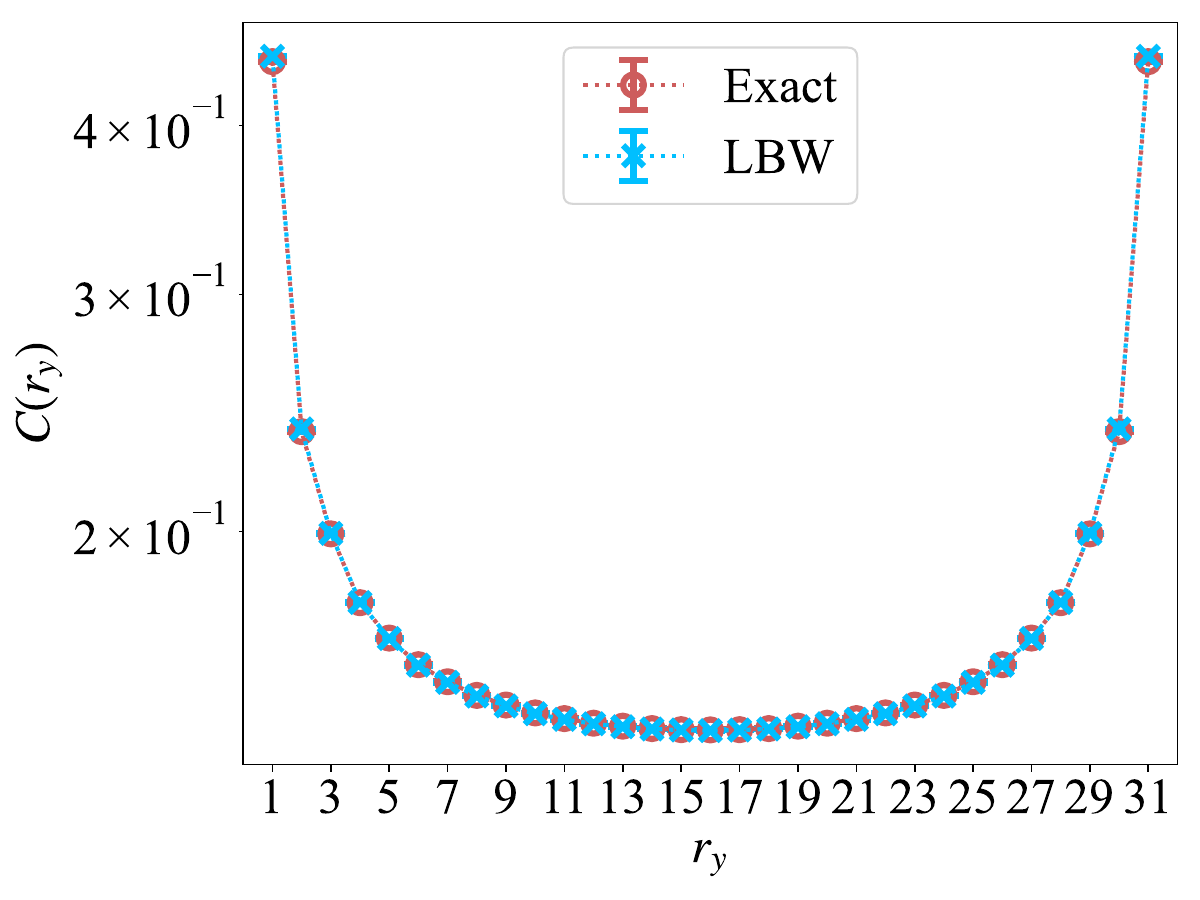}
\put (0,75) {{\textbf{(a)}}}
\end{overpic}
\begin{overpic}[width=0.45\linewidth]{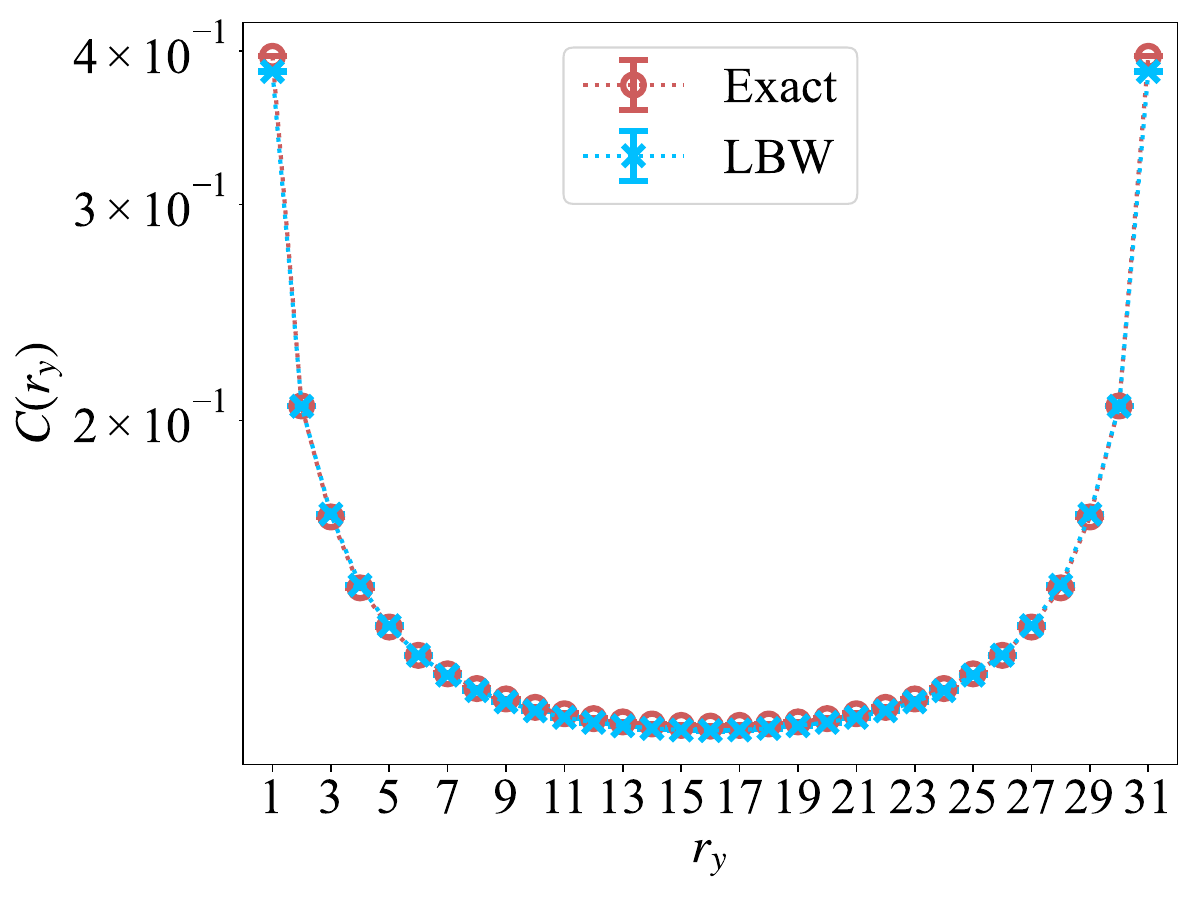}
\put (0,75) {{\textbf{(b)}}}
\end{overpic}
\begin{overpic}[width=0.45\linewidth]{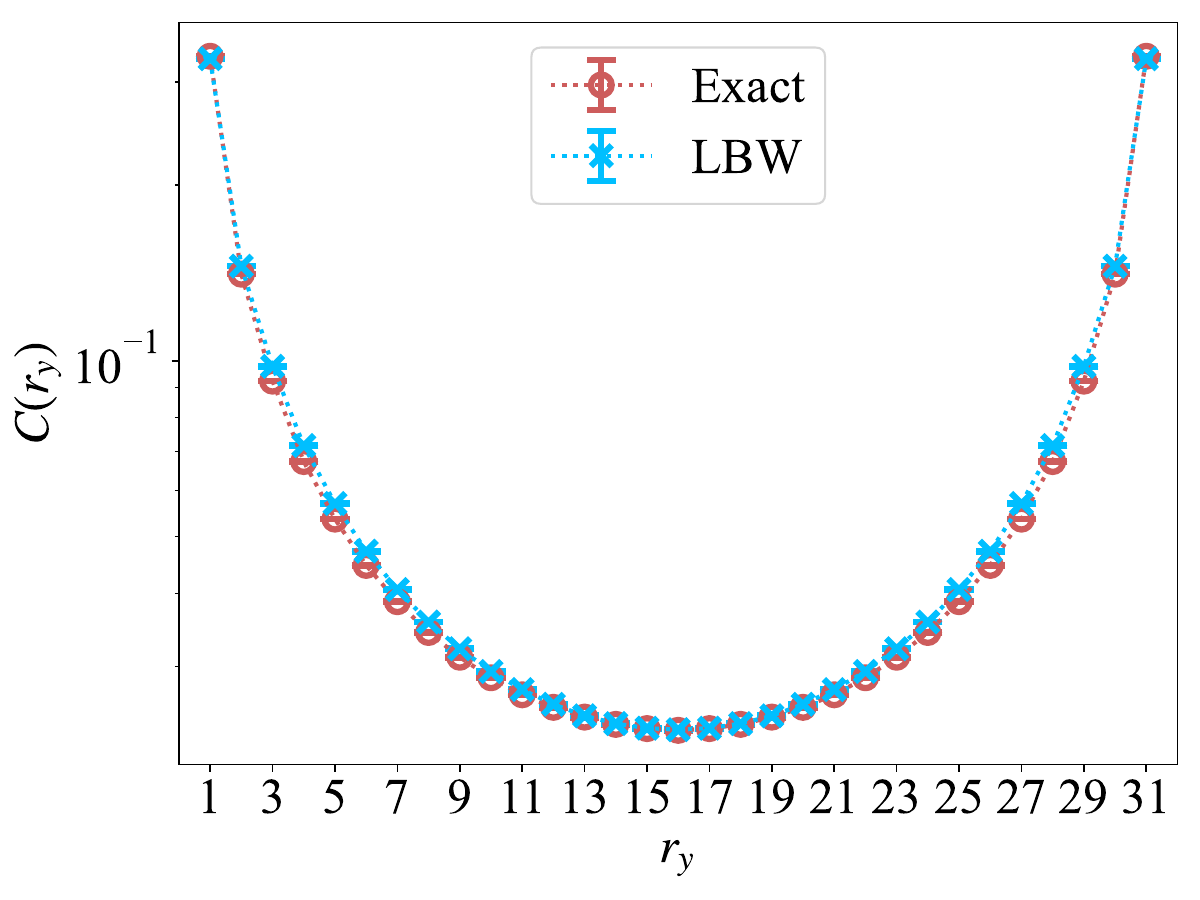}
\put (0,75) {{\textbf{(c)}}}
\end{overpic}
\begin{overpic}[width=0.45\linewidth]{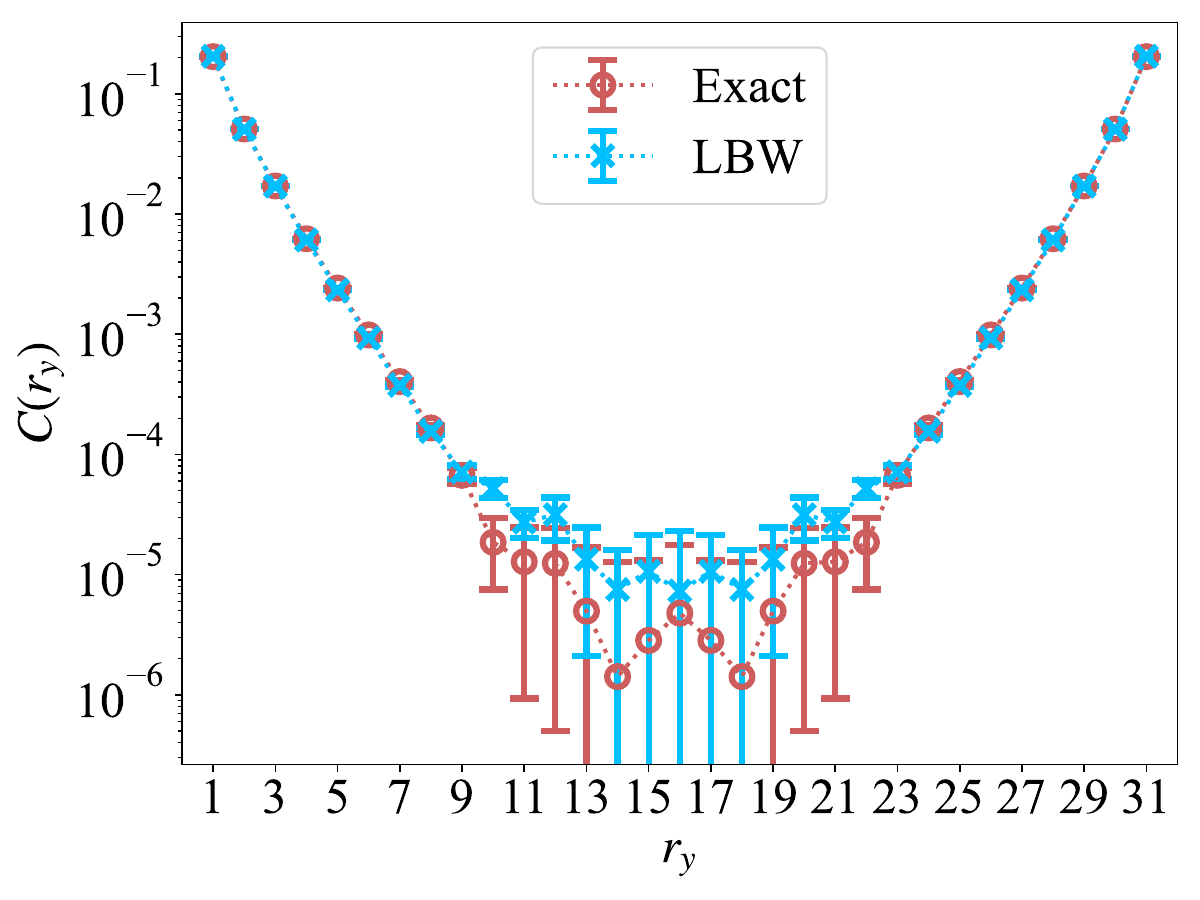}
\put (0,75) {{\textbf{(d)}}}
\end{overpic}
\caption{
Equal-time along-boundary correlators $C(r_y)$ at $L=32$ for the horizontal weak-bond bipartition shown in Fig.~\ref{fig:J1J2_config}(b), comparing the $32\times32$ LBW-EH and the $64\times32$ exact-EH at effective inverse temperature $\beta_A=1$. 
The correlator $C(r_y)$ is the same along-boundary correlator defined in Eq.~\eqref{eq:cry_def}, measured within subsystem $A$ on the boundary-adjacent line with separations taken along the periodic boundary direction. 
(a) Heisenberg limit $J_r=1$. 
(b) N\'eel phase $J_r=1.5$. 
(c) QCP $J_r=1.90951(1)$. 
(d) Dimer phase $J_r=3$.
}
\label{fig:L32_weak_appendixB}
\end{figure}

\begin{figure}[ht!]
\centering
\begin{overpic}[width=0.45\linewidth]{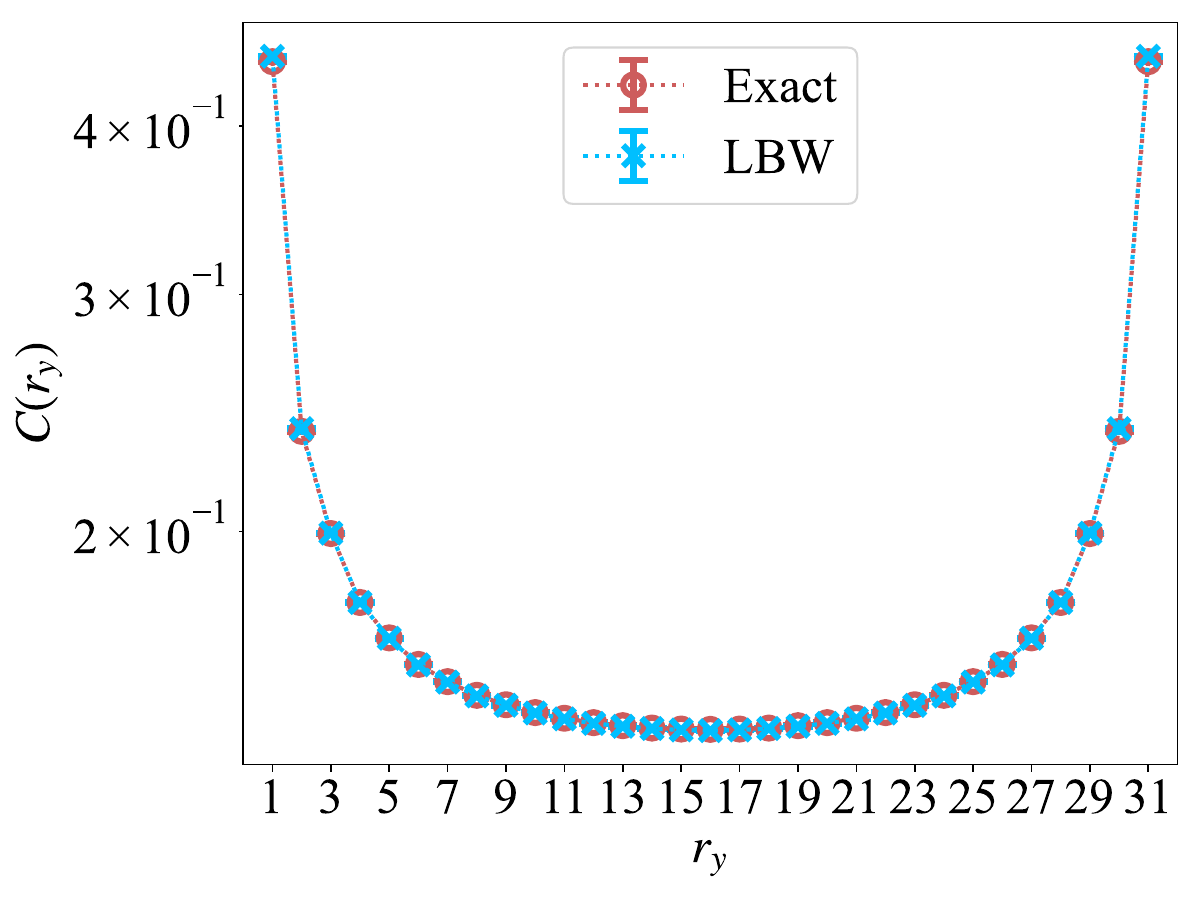}
\put (0,75) {{\textbf{(a)}}}
\end{overpic}
\begin{overpic}[width=0.45\linewidth]{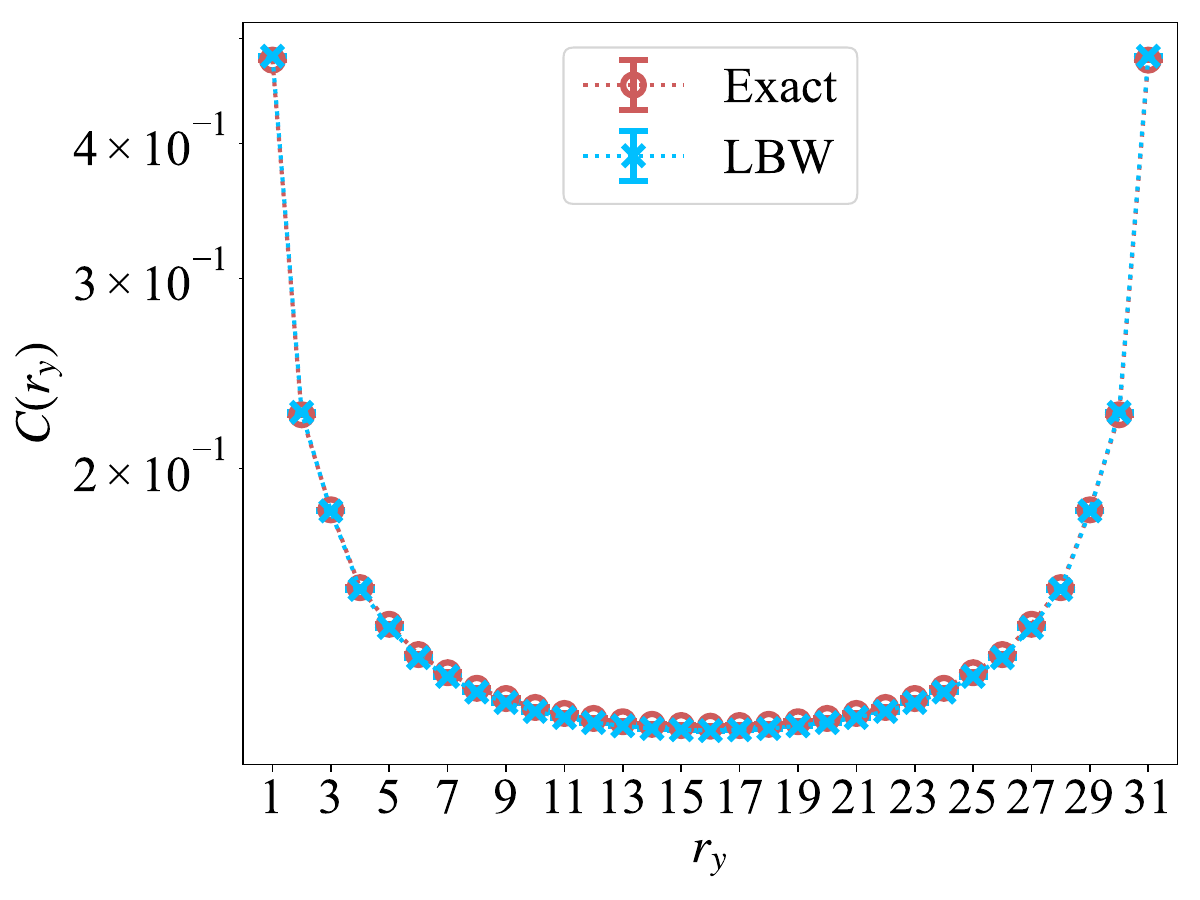}
\put (0,75) {{\textbf{(b)}}}
\end{overpic}
\begin{overpic}[width=0.45\linewidth]{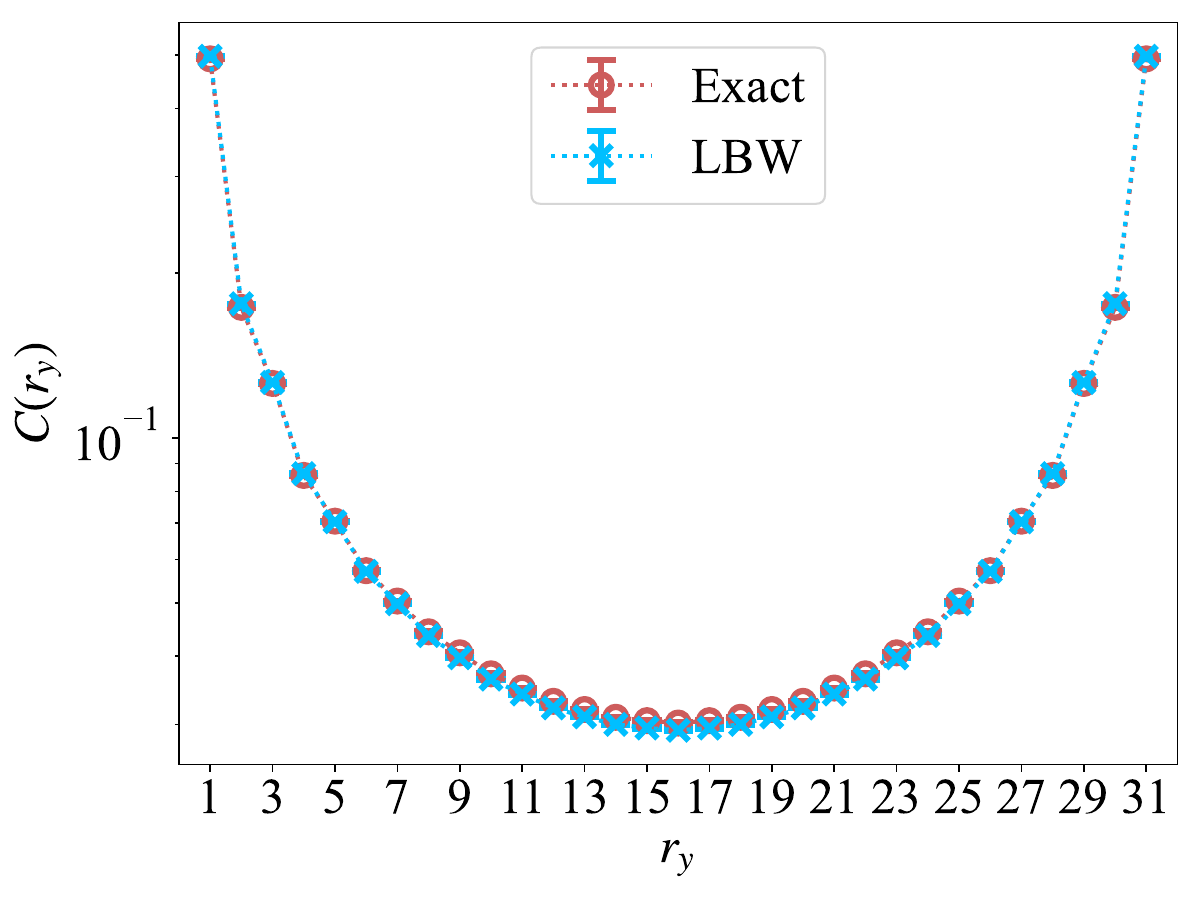}
\put (0,75) {{\textbf{(c)}}}
\end{overpic}
\begin{overpic}[width=0.45\linewidth]{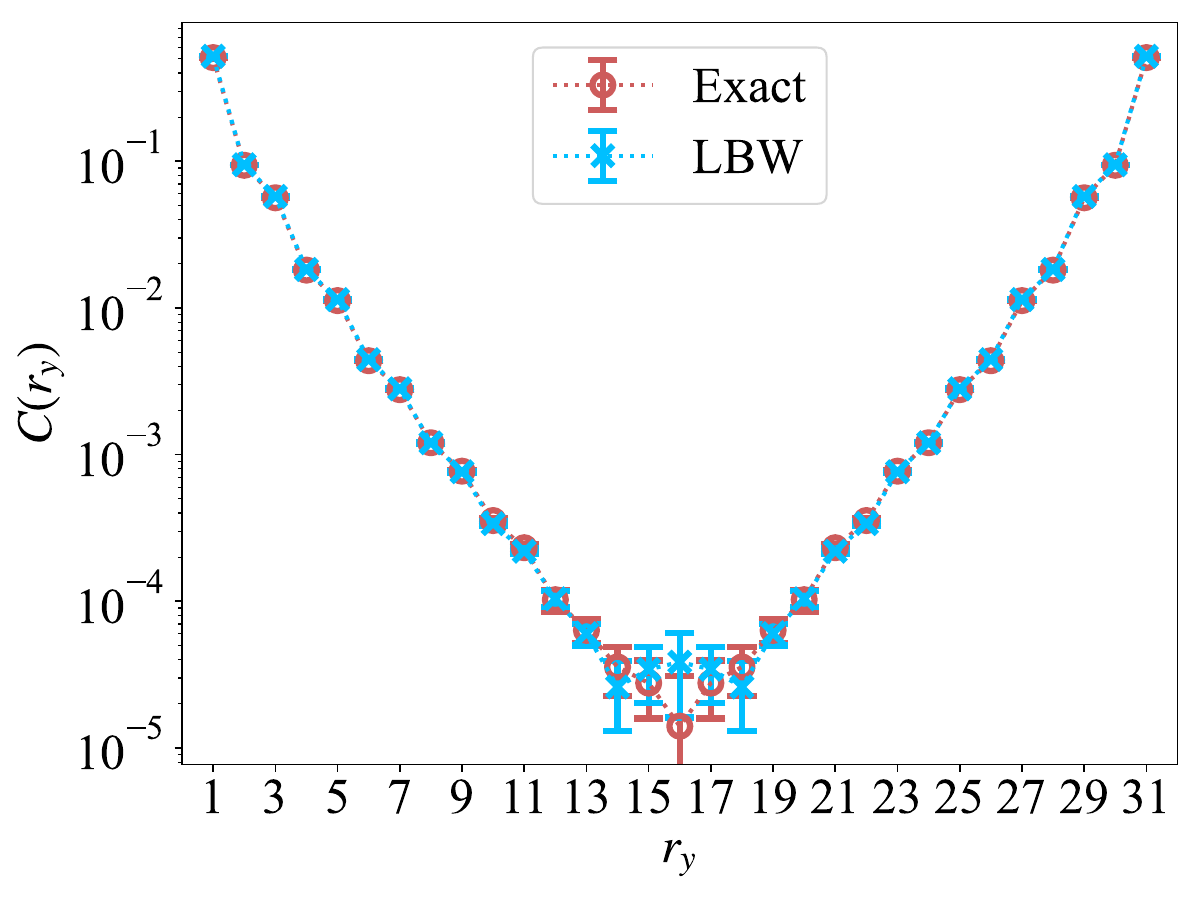}
\put (0,75) {{\textbf{(d)}}}
\end{overpic}
\caption{
Equal-time along-boundary correlators $C(r_y)$ at $L=32$ for the vertical weak-bond bipartition shown in Fig.~\ref{fig:J1J2_config}(c), comparing the $32\times32$ LBW-EH and the $64\times32$ exact-EH at effective inverse temperature $\beta_A=1$. 
The correlator $C(r_y)$ is the same along-boundary correlator defined in Eq.~\eqref{eq:cry_def}, measured within subsystem $A$ on the boundary-adjacent line with separations taken along the periodic boundary direction. 
(a) Heisenberg limit $J_r=1$. 
(b) N\'eel phase $J_r=1.5$. 
(c) QCP $J_r=1.90951(1)$. 
(d) Dimer phase $J_r=3$.
}
\label{fig:L32_weak_vertical_appendixB}
\end{figure}

A more important question concerns the cut dependence itself. 
At $L=16$, the equal-time along-boundary correlator $C(r_y)$ defined in Eq.~\eqref{eq:cry_def} already showed a clear distinction among the three bipartitions. 
For the strong-bond cut, the agreement between LBW-EH and exact-EH deteriorates away from the Heisenberg limit, whereas for the two weak-bond cuts the agreement remains excellent across phases. 
To determine whether this pattern survives at larger size, we evaluate the same equal-time along-boundary correlator at $L=32$ for all three inequivalent bipartitions in Fig.~\ref{fig:J1J2_config}. 
As in the main text, the correlator is measured within subsystem $A$ on the boundary-adjacent line and averaged along the periodic boundary direction. 
The corresponding results are shown in Figs.~\ref{fig:L32_strong_appendixB}--\ref{fig:L32_weak_vertical_appendixB}.

The large-size data display the same qualitative behavior as in the main text. 
For the strong-bond bipartition shown in Fig.~\ref{fig:L32_strong_appendixB}, the LBW-EH and exact-EH still agree very well at the Heisenberg limit, while visible discrepancies remain in the N\'eel phase, at the QCP, and in the dimer phase. 
Increasing the system size from $L=16$ to $L=32$ therefore does not remove the systematic mismatch associated with the strong-bond cut.

\vspace{0.3cm}

The situation is markedly different for the two weak-bond bipartitions. 
Fig.~\ref{fig:L32_weak_appendixB} and Fig.~\ref{fig:L32_weak_vertical_appendixB} show that the near-perfect agreement between LBW-EH and exact-EH persists at larger size across all four representative points. 
This confirms that the excellent performance of the LBW-EH ansatz for the ordinary cuts is not restricted to the smaller system studied in the main text.
We also note that the dimer phase at $J_r=3$ is statistically more demanding. 
Because this phase is gapped, the signal decays rapidly and the farthest-separation points under PBC are harder to resolve accurately. 
Although a few such points are less well converged, they do not affect the overall conclusion.

Taken together, these larger-size results show that the main cut-dependent conclusion of the present work remains stable as the system size is increased from $L=16$ to $L=32$, and therefore is not an artifact of the $L=16$ benchmark used in the main text. 
The larger-$L$ velocity extraction remains quantitatively consistent with the smaller-size results, while the equal-time along-boundary correlations continue to exhibit the same cut-dependent pattern across the three bipartitions. 
In particular, the strong-bond cut still exhibits systematic deviations between LBW-EH and exact-EH away from the Heisenberg limit, whereas the two weak-bond cuts continue to show near-perfect agreement across the Heisenberg limit, the N\'eel phase, the QCP, and the dimer phase. 
These results therefore reinforce the physical interpretation proposed in the main text, namely that the quality of the LBW-EH ansatz is controlled primarily by the cut geometry rather than by the particular finite size considered here. 
From this perspective, the strong-bond cut is associated with an anomalous boundary effect, while the two weak-bond cuts remain well described by the LBW-EH ansatz within the system sizes investigated.

\end{document}